\documentclass[11pt]{article}
\pdfoutput=1
\usepackage{amsmath, amssymb}

\usepackage{libertine} \usepackage[libertine]{newtxmath}

\usepackage[hang,flushmargin]{footmisc}
\usepackage[authoryear]{natbib}
\usepackage{booktabs}
\usepackage{enumitem}
\usepackage{multirow}

\usepackage{tikz}
\usetikzlibrary{arrows,positioning,decorations.pathreplacing}

\usepackage{graphicx}
\DeclareGraphicsExtensions{.pdf}
\usepackage{caption}

\usepackage[top=1in, bottom=1in, left=1.25in, right=1.25in]{geometry}
\usepackage{float}
\usepackage{array}
\usepackage{color}
\usepackage[justification=centering]{caption}
\usepackage[font={small}]{caption}
\usepackage{footmisc}
\DefineFNsymbols{mySymbols}{{}{} {\ensuremath\dagger}{\ensuremath\dagger}\S\P
   *{**}{\ensuremath{\dagger\dagger}}{\ensuremath{\ddagger\ddagger}}}
\setfnsymbol{mySymbols}
\usepackage{setspace}

\definecolor{NYUcolor}{RGB}{114.48, 29.16, 166.32}

\usepackage[pdftex,breaklinks]{hyperref}
\hypersetup{
    colorlinks=true,
    linkcolor=NYUcolor,
    filecolor=NYUcolor,
    urlcolor=NYUcolor,
    citecolor=NYUcolor,
    bookmarks=true,
    bookmarksopen=true,
    bookmarksnumbered=true,
    hypertexnames=true
}

\title{\textbf{The Testing Multiplier: Fear vs Containment}}

\author{\textbf{Francesco Furno}\footnote{\hspace*{-1.8em}Department of Economics, New York University. E-Mail: \href{mailto:francesco.furno@nyu.edu}{francesco.furno@nyu.edu}.}\thanks{\hspace*{-1.8em}I am particularly grateful to Olivier Blanchard, Jaroslav Borovi\u{c}ka, Tim Christensen and B\`{a}lint Sz\H{o}ke for their comments and suggestions. This work also benefited from conversations with  Valeria Ferraro, Sara Gerstner, Sebastian Hillenbrand, Pietro Monticone, Pietro Reggiani, Matheus Silva and Lena Song. I thank the NYU's High-Performance Computing Center for their support.} \\ NYU}

\begin{document}
\maketitle
\thispagestyle{empty}
\begin{abstract}
\noindent This paper studies the economic effects of testing during the outbreak of a novel epidemic disease. I propose a model where  testing permits isolation of the infected and provides agents with information about the prevalence and lethality of the disease. Additional testing reduces the perceived lethality of the disease, but \text{might} increase the perceived risk of infection. As a result, more testing \text{could} increase the perceived  risk of dying from the disease - i.e. ``stoke fear'' - and cause a fall in economic activity, despite improving health outcomes. Two main insights emerge. First, increased testing is beneficial to the economy and pays for itself if performed at a sufficiently large scale, but not necessarily otherwise. Second, heterogeneous risk perceptions across age-groups can have important aggregate consequences. For a SARS-CoV-2 calibration of the model,   heterogeneous risk perceptions across young and old individuals mitigate GDP losses by $50\%$ and reduce the death toll by $30\%$ relative to a scenario in which all individuals have the same perceptions of risk.\\
\\

\end{abstract}
\clearpage
\pagenumbering{arabic}

\newpage
\section{Introduction}
 There is widespread agreement that, during an epidemic outbreak, increased testing  can contain the spread of the disease, save lives and improve economic outcomes.\footnote{When one abstracts  from individuals' behavioral responses, testing's role is to selectively isolate the infected, which slows down epidemic transmission and benefits the economy - see \cite{Mongey2020} and \cite{droste2020economic} among others. When one assumes that testing affects behavior by resolving individual-level uncertainty, it can have unintended negative consequences  if not  followed by strictly enforced isolation - see \cite{EichenbaumTesting2020}.}  In this paper, I explore the economic effects of  testing during the outbreak of a novel disease in an epidemiological model where agents rely on testing data to form their perceptions of risk. Since more testing \textit{could} result in a higher number of detected cases - and thus increase the perceived risk of infection - it has the potential to  ``scare'' the population  and cause a fall in economic activity. Indeed, conditional on the latent size of the epidemic, more testing reveals more infections. However, thanks to the isolation of the infected, it also dynamically reduces the latent size of the epidemic. My results suggest that the perceived risk  decreases and the economy improves  when a sizeable share of the population is tested daily, but not necessarily otherwise. Interestingly, whenever more testing ``stokes fear'', the ensuing contraction of economic activity occurs despite improved public health outcomes.\\%

The model can be summarized as follows. I abstract from policy interventions  such as lockdowns or national mask mandates in order to focus exclusively on testing policies that mimic those adopted by health-care systems around the world. Economic activity is mainly a function of the perceived risk of dying from the epidemic disease, and risk perceptions are constructed  using testing data on total cases, active infections and deaths. Additional testing followed by (imperfect) isolation systematically improves health outcomes (i.e. it reduces infections and deaths), but has a non-monotone effect on risk perceptions.\footnote{If I assume  that agents  do not need to rely on testing data to form their perceptions because they can observe the true aggregate state of the epidemic, more testing  systematically improves  economic outcomes as well, in line with the existing literature.} Economic activity falls whenever additional testing increases the perceived risk of dying, and rises otherwise. The notion of a testing multiplier then naturally arises to summarize the economic effects of an additional dollar spent by the government on testing activity. \\

Risk perceptions  are introduced in the model as follows. Agents do not know the epidemiological process, and  testing data is the only source of information about the risk of dying from the epidemic disease. The risk of dying is given by the product of the probability of dying conditional on infection and the probability of infection. Following the epidemiological literature, I assume that the former is assessed with the case fatality rate, given by total reported deaths divided by total detected cases, while the latter is proportional to  detected active cases per capita. Since the disease is \textit{unknown}, agents do not know its true lethality and are  forced to resort to the case fatality rate. In \autoref{Section:Evidence}, I provide robust empirical evidence in favor of my specification.\\

Real-time risk perceptions in the model  can systematically differ from the truth, and depend on the level of testing activity. This is a general insight, and  it is useful to think about two wedges between perceptions and reality to see why. The first wedge captures the fact that the perceived probability of dying conditional on infection might exceed the true one. This happens because, without large-scale testing, real-world testing policies prioritize testing of individuals with severe symptoms, and infected individuals who are less likely to die are not tested.\footnote{See \cite{lipsitch2015potential} for a discussion of this issue.} In principle, this wedge can be eliminated over time  with  serological surveys and  other (natural) experiments, but, in practice, it takes time (or never completely occurs). Thus,  the perceived lethality of the disease heavily depends on the amount of testing performed, especially in the early stages of the outbreak. \\

The second wedge relates to the probability of infection, and arises because agents struggle to estimate the true number of active infections in real-time. This happens because many infections go undetected without large-scale testing, and because observable epidemiological variables are not enough to correctly estimate them in real-time. For instance, suppose that at some point during the outbreak a certain number of deaths is observed. Without knowledge of the true probability of dying conditional on infection, one is not able to estimate how many infections produced those deaths. A similar reasoning holds for hospitalizations.  Even when these probabilities are known, deaths and hospitalizations can only be used by agents to infer the  number of \textit{past} cases, since there are long lags between infection and death or hospitalization. Given that infection risk depends on how many active infections are \textit{currently} in the population,  agents are forced to rely on testing data for a real-time assessment of infection risk.\\

To further illustrate this point,  I consider an alternative  specification in \autoref{Section:Alternative_Beliefs} where agents learn over time about the true lethality of the disease, irrespective of testing. They  use this knowledge to correctly estimate the total number of past cases and construct an ascertainment bias factor that they use to scale up newly detected infections. Eventually, agents correctly assess both the lethality of the disease and the total number of cases, but they nonetheless fail to correctly estimate active infections in real-time without large-scale testing.\footnote{This approach to the estimation of active infections is inspired by and parallels what is proposed in the state-of-the-art work on the SARS-CoV-2 outbreak by \cite{chande2020real}. The authors construct an ascertainment bias factor given by the ratio of  total cases estimated  with serological surveys over detected cases through testing. They then use this factor to scale up newly detected infections.} This implies that my results are reproduced with this alternative specification as well.\\

Importantly, whether additional testing \textit{actually} increases  perceived risk is the result of three  forces that my epidemiological model is well-equipped to capture. The first two forces - described in the first paragraph - relate to the perceived risk of infection: additional testing dynamically reduces the true latent size of the epidemic, but  it also  unveils a larger share of it. The third force relates to the perceived risk of dying conditional on infection. More testing widens the  range of symptoms being tested, thereby mechanically decreasing the  case fatality rate. For a novel unknown disease, this reduces its perceived lethality. \\

Since the epidemic data produced by the health-care system through testing play a key role in the assessment of risk, I make sure that the testing policies in the model can match key features of the data, as I show in \autoref{Section:Testing_Policies_Validation}. This explains the choice of an agent-based epidemiological model with a second endemic disease and testing policies that prioritize severe symptoms. Economic considerations are then introduced only in a stylized way. Specifically,  agents in the model  mechanically reduce labor supply and enjoyment of leisure when their perceived risk of dying increases. Since contact rates and aggregate output are assumed to be a function of aggregate labor supply and leisure, perceived risk affects both aggregate economic activity and the spread of the epidemic disease in the population. \\

The first main result of the paper is that  additional testing is beneficial to the economy and (partially) pays for itself when performed at a large enough scale, but not necessarily otherwise. When a sizeable share of the population is tested every day and infected individuals are (imperfectly) isolated, epidemic containment succeeds and the perceived risk of dying decreases, improving economic outcomes. In this scenario, there is nothing to fear and the multiplier is positive. At a small scale, instead, testing might not succeed at containing the epidemic, but  would still detect a larger portion of it. Depending on a large set of parameters and on luck, this could increase the perceived risk of dying and cause a contraction of economic activity. When this happens, fear spreads in the population and the multiplier is negative.\\

The second main result clarifies  the aggregate importance of heterogeneous risk perceptions across age-groups. I extend the model to introduce young and old agents, and calibrate it to the U.S. and SARS-CoV-2, which features a steep risk-gradient across age-groups. As an extreme thought-experiment, I consider a scenario in which all individuals are assumed to have homogeneous perceptions of risk because the government releases only aggregate testing data - as it is often the case during epidemic outbreaks. I then compare it to another scenario in which age groups have heterogeneous risk perceptions because they can construct age-specific case fatality rates from disaggregated testing data. I find that, relative to the homogeneous case, heterogeneous risk perceptions vastly improve aggregate economic and health outcomes, because old agents - who are the most likely to die - protect themselves more while young agents protect themselves less and return to work.\footnote{The intuition behind this  parallels what is suggested in \cite{acemoglu2020multi} with respect to targeted lockdowns.}  \\

The paper is divided into the following sections. In \autoref{Section:Fear_Death}, I  discuss  risk perceptions. In \autoref{Section:Evidence}, I assess the empirical relevance of my proposed measure of risk. I present the model in \autoref{Section:Framework}, and provide extensive simulations to better understand its mechanisms in \autoref{Section:Understanding_Mechanism}.  I define  the testing multiplier in \autoref{Section:Testing_Multiplier}, and simulate it for various parameterizations. In \autoref{Section:Heterogeneity}, I explore the importance of heterogeneous risk perceptions across age-groups.

\subsection{\textcolor{black}{Relation to the Existing Literature}}
\label{Section:Literature}

This paper contributes to a fast-growing literature in economics that analyzes the interplay between epidemics and the economy. In terms of its methodological approach, this paper starts from an epidemiological model and extends it with a simple economic component, similarly in spirit to \cite{Mongey2020}, \cite{piguillem2020optimal}, \cite{taipale2020population}, and \cite{droste2020economic}  who also examine the economic and health benefits of testing. In none of these  papers, however, testing provides agents with information about the aggregate state of the epidemic.\footnote{Furthermore, these papers adopt a compartmental modeling strategy which results in better tractability  by permitting  the aggregation of  individuals into epidemiological compartments. I adopt an agent-based framework, which increases complexity but allows me to introduce more realistic testing policies and a more refined modeling of the epidemic disease. For a discussion of compartmental vs agent-based epidemiological models see \cite{murray2020epidemiology}, \cite{sukumar2012agent}, \cite{hunter2018comparison}, and \cite{gallagher2017comparing}.} \\

A complementary approach is to start from a macroeconomic model and expand it with a tractable stylized epidemiological component.\footnote{See the seminal contributions by  \cite{eichenbaum2020macroeconomics} and \cite{jones2020optimal}.} Within this strand of the literature, testing directly affects individual behavior in both \cite{brotherhood2020economic} and \cite{EichenbaumTesting2020} by resolving uncertainty regarding  an individual's health status, but does not provide information regarding the aggregate state of the epidemic.\\

In the epidemiological literature, there is a vast array of papers that attempt to endogenize behavior in response to `fear' of infection - see \cite{funk2010modelling}, \cite{verelst2016behavioural} and  \cite{wang2015coupled} for an overview. All these papers assume that agents' behavior reacts to information which could come from various sources, and could be `objective' or `subjective'.  To the best of my knowledge, no paper in this literature assumes that agents react to information coming from testing activity. \\

The present paper also relates to an emerging empirical literature that attempts to estimate the impact of the epidemic on economic activity. In particular, it echoes the findings in \cite{Goolsbee2020} that economic activity during an epidemic outbreak falls irrespective of non-pharmaceutical interventions by policy-makers, and the insight in \cite{eichenbaum2020people} that the probability of dying is a key determinants of individuals' behavioral responses.\\

Finally, this paper shares with \cite{droste2020economic} and \cite{cutler2020covid} the fundamental insight that large-scale testing can pay for itself, since its indirect benefits far outweigh its direct costs.

\newpage
\section{Perceptions of Risk}
\label{Section:Fear_Death}
Individuals fear deadly diseases and I argue that the relevant measure of `fear' of an epidemic disease is  the probability of dying from it, as opposed - for example - to the probability of contracting the disease.\footnote{Several epidemiological models with behavioral responses assume that individuals react to the prevalence rate, given by the ratio of cases to the population, a measure of infection risk - see \cite{funk2010modelling} for a review. In the  economics literature, for example,  \cite{kaplan2020great} and \cite{droste2020economic} assume that behavioral responses depend on the number of deaths.} It is insightful to  express the probability of death as follows:
\[ \textcolor{NYUcolor}{\text{Prob}(\text{Death})} = \text{Prob}(\text{Death} | \text{Infection}) \times \text{Prob}(\text{Infection}) \]
where the probability of death (`death risk') is given by the product between the conditional probability of death given infection (`disease lethality') and the probability of infection (`infection risk'). To be precise, the probability of death conditional on infection is labeled `infection fatality risk' (or IFR) in the epidemiological literature.\footnote{For example, see \cite{lipsitch2015potential}.}    \\

To see why the probability of dying is the relevant object, consider the following. Imagine first a widely spread epidemic disease which is completely harmless. This would result in a high infection risk but a null disease lethality, implying a null death risk, and thus no fear of the disease. Consider next a very deadly disease which is impossible to catch. This would imply a null infection risk and a null death risk, and thus no fear of the disease. \\

The main problem during an epidemic outbreak of a novel disease is that  key properties of the disease such as the infection fatality risk or the probability of developing severe symptoms are unknown and, without specific policy interventions,  remain unknown. As a result, the probabilities mentioned above are also unknown. This is especially true when the epidemic disease features a sizeable share of paucisymptomatic and asymptomatic individuals, which makes the estimation of  the number of total cases, active infections, asymptomatic infections and recovered individuals extremely hard in real-time. Despite the lack of reliable information, however, individuals will still try to perform a real-time assessment of the risk they face, and their behavioral responses will depend on this assessment. \\

In my theoretical analysis, I assume that individuals do not understand the epidemiological process and that they rely on testing data to form their perceptions of risk. Specifically,  agents look at the case fatality rate to estimate the risk of death conditional on infection:
\[ \textcolor{NYUcolor}{\text{Perceived Lethality}_t} = CFR_t \equiv \frac{D_t}{C_t} \]
where $\textcolor{NYUcolor}{CFR_t}$ is the case fatality rate, and $\textcolor{NYUcolor}{C_t}$ and $\textcolor{NYUcolor}{D_t}$ are cumulative cases  and deaths reported by the health-care system.  To assess the average infection risk in the population, I  assume that individuals follow a standard textbook epidemiological model - such as the SIR model - which posits that the probability of infection is proportional to the number of active infections over the population:
\[ \textcolor{NYUcolor}{\text{Perceived Infection Risk}_t} = IR_t = \beta \times \frac{I_t}{P_t} \]
where $\textcolor{NYUcolor}{\beta}$ is the transmission coefficient of the disease (which summarizes its contagiousness), $\textcolor{NYUcolor}{I_t}$ is the number of currently active infections detected by the health-care system, and $\textcolor{NYUcolor}{P_t}$ is the alive population.\footnote{In standard textbook epidemiological models, active infections - as opposed to cumulative infections -  are what matters for transmission because it is assumed that individuals who recover or die are no longer infectious.} The perceived risk of death - which from now on I will denote with the variable $\textcolor{NYUcolor}{\chi}$ - can then be re-constructed as follows:
\[ \textcolor{NYUcolor}{\chi_t} = \underbrace{CFR_t}_{\text{Perceived Lethality}} \times \underbrace{IR_t}_{\text{Perceived Infection Risk}}  \]
In the empirical analysis presented in \autoref{Section:Evidence}, I show that this proposed measure of perceived risk predicts precisely and robustly economic activity across U.S. states and counties during the SARS-CoV-2 outbreak.\\

\subsection{Discussion}
\label{Section:Perceptions_Discussion}
The proposed specification of beliefs implies that, without large-scale testing, perceptions of risk can systematically differ from the true latent risk of dying from the epidemic disease. To better understand this point, it is useful to decompose the probability of dying into its two components, and think about two wedges between perceptions and reality.\\

Let's start by considering the wedge between the perceived lethality of the disease and the true one. Since the lethality of a novel disease is unknown, individuals need to assess it in real-time using the information that is available. As explained in \cite{Wong2013} in relation to the 2009 influenza-A outbreak, testing data - i.e. laboratory-confirmed cases - produced by the health-care system are the most readily available source of information during an outbreak, and can be used by both experts and non-experts to assess the lethality of the emerging infectious disease. In particular, epidemiologists themselves construct a case fatality rate from testing data to assess in real-time the lethality of the disease, and I assume that individuals do that too.\footnote{See \cite{Ghani2005}.} \\

This  measure, unfortunately,  suffers from several shortcomings that are summarized, for example, in \cite{lipsitch2015potential}. Crucially, whenever the disease features a large share of sub-clinical infections (i.e. that do not require medical attention) that go undetected with narrow testing policies, the denominator of the case fatality rate is under-estimated, and therefore the  lethality of the disease is over-estimated. This issue is not easily solved even when combining available testing data with epidemiological theory, because of the identification problems outlined, for instance, in \cite{atkenson2020CFR} and \cite{korolev2020identification}. Large-scale testing is a way to solve the problem, since individuals  with non-severe infections who are less likely to die from the disease are  included in the total case count. Another way to solve the problem is to perform a  one-off large-scale random  experiment, with either a virological test (which detects an active infection) or a serological survey (which detects past infections). The problem is that, for a variety of reasons, this usually takes time to be performed - if it is ever performed. In principle, the wedge between the true lethality of the disease and the perceived one could be eliminated even without large-scale testing. In practice, this either takes time or never occurs, forcing individuals to rely on testing data to assess the lethality of the disease.\\

Let's now consider the second wedge, the one between the true and the perceived risk of infection. In a nutshell, my argument is that it is hard - if not impossible - to remove this wedge in real-time without large-scale testing. This is due to the fact that the risk of infection depends on the number of \textit{currently active} infections, which is more difficult to estimate than the total number of \textit{past} infections. For example, a large-scale serological survey can provide a very accurate estimate of the number of infections in the past, but has little to say about currently active infections. Similarly, observable epidemiological variables such as deaths or hospitalizations are not helpful because they contain information about past infections - as opposed to current infections. Indeed, deaths today are the results of infections days or weeks ago. These considerations suggest that the assessment of infection risk is heavily dependent on testing data, exactly as in the  specification of beliefs that I propose.\\

Further support to this argument comes from the recent state-of-the-art work by \cite{chande2020real}. In their paper on the SARS-CoV-2 outbreak in the U.S., the authors construct a location-specific real-time assessment of infection risk using \textit{``recent case reports multiplied by an ascertainment bias informed by serological surveys''}. In other words, they use testing data on new infections and they scale them up by a factor  given by the number of cases detected with serological surveys over the number of cases detected by the health-care system through testing. Since the serological surveys are conducted infrequently, daily variation in the estimated infection risk comes exclusively from testing activity. In \autoref{Section:Alternative_Beliefs}, I propose an alternative specification of beliefs that mimic this methodology. Agents eventually estimate correctly both the true lethality of the disease and the true number of total cases, but still fail to estimate the number of active infections in real-time without large-scale testing.

\newpage
\section{Fear and Economic Activity: Evidence from the U.S.}
\label{Section:Evidence}
I combine weekly data on economic activity with testing data on reported cases and deaths across U.S. states and counties during the first stages of the SARS-CoV-2 epidemic outbreak to investigate the relationship between  my proposed measure of perceived death risk and economic activity. My preferred proxies of economic activity are the Dallas FED's Mobility and Engagement Index (MEI) and the Google Workplace Mobility report because of their high-frequency availability. Perceived risk in location $\textcolor{NYUcolor}{i}$ during week $\textcolor{NYUcolor}{t}$ is given by:
\[ \textcolor{NYUcolor}{\chi_{i,t}} = \underbrace{\frac{D_{i, t}}{C_{i,t}}}_{\equiv CFR_{i,t}} \quad \times \quad \underbrace{\beta \cdot \frac{I_{i,t}}{P_{i,0}}}_{\equiv IR_{i,t}} \]

Data on reported total cases ($\textcolor{NYUcolor}{C_{i,t}}$), total deaths ($\textcolor{NYUcolor}{D_{i,t}}$) and population ($\textcolor{NYUcolor}{P_{i,0}}$) come from USA Facts, which are available at a county-level and can be easily aggregated up to the state-level. To estimate reported active infections ($\textcolor{NYUcolor}{I_{i,t}}$), I take new reported cases over 14-day horizon, although the results are robust to the time window considered. Finally, I assume that the transmission coefficient used to construct the perceived infection risk is $\textcolor{NYUcolor}{\beta = 0.30}$.\footnote{Given that the coefficient will be constant across time and space, this assumption will not affect the standardized estimated coefficients.} My dataset stretches from January 2020 to September 2020, and all the details can be found in \autoref{Appendix:Data_Sources}.\\

\begin{figure}[H]
\centerline{\includegraphics[scale=0.7, angle = 0]{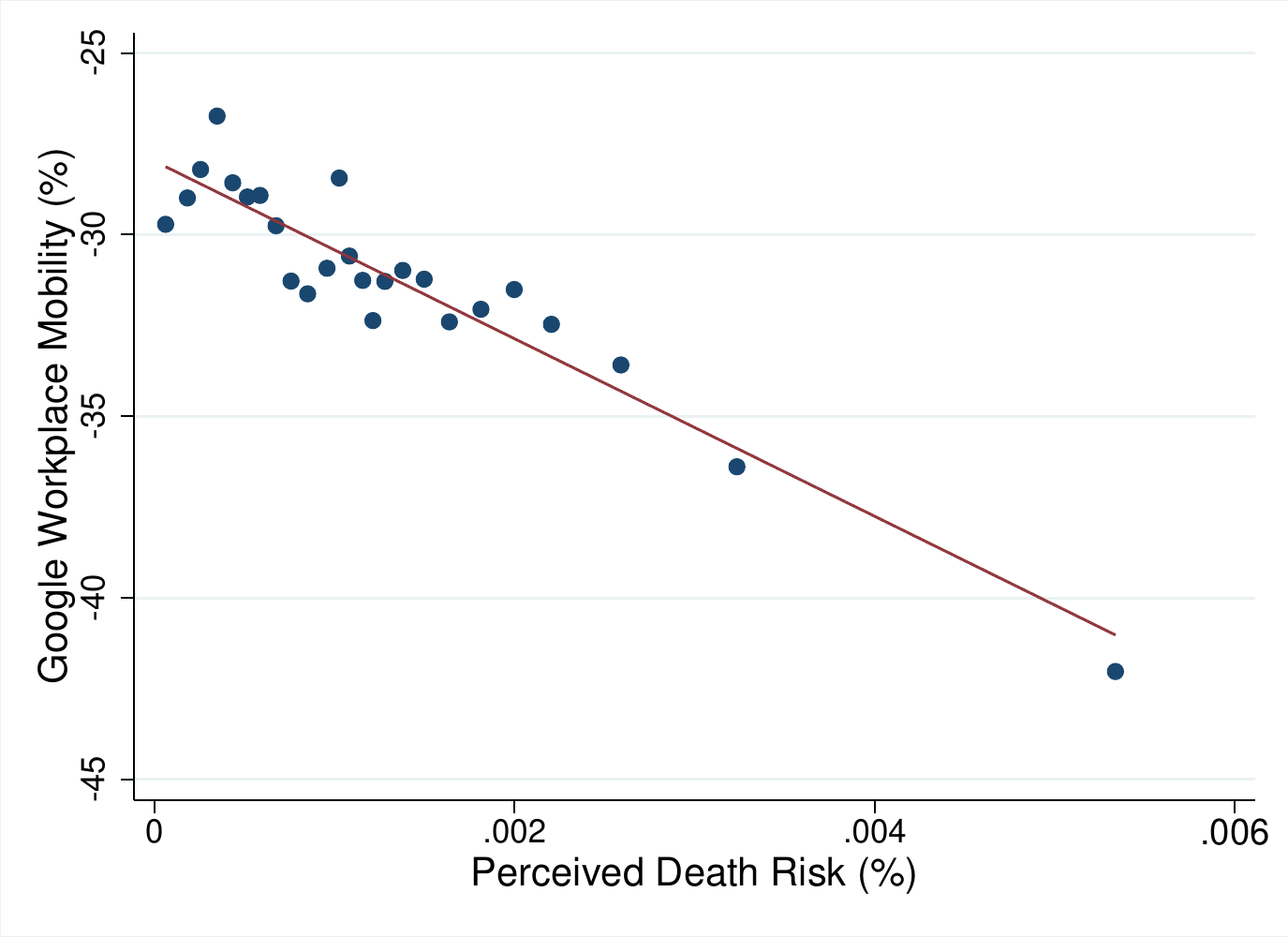}}
\caption{Correlation between Perceived Risk and Economic Activity during SARS-CoV-2 in the U.S. \\ \scriptsize \textbf{Notes}: The plot considers the period from 1 May 2020 to 1 September 2020 in order to leave out most of early lockdowns, business closures and similar interventions.}
\label{fig:Main_Summary_Correlation}
\end{figure}

\autoref{fig:Main_Summary_Correlation} reveals that a higher perceived risk of death is associated with  falls in economic activity across U.S. states. However,  this is not enough to establish causality for at least two reasons. First, reverse causality might be at play. Second, there might be omitted variable bias: a higher perceived death risk calls for lockdowns and similar non-pharmaceutical interventions, which produce a contraction of economic activity. \\

Reverse causality is unlikely to be an issue. SARS-CoV-2 is a disease characterized by lengthy lags between exposure and development of symptoms and/or hospitalization. Given the relatively narrow testing policies adopted in the U.S. during the first phase of the pandemic, the vast majority of detected infections were diagnosed after the appearance of symptoms or even after hospitalization, implying that a new infection was likely to be recorded by the health-care system with a sizeable delay. This implies that economic activity in a given week is likely to increase perceived death risk only in the future. Furthermore, reverse causality would suggest a positive relationship between economic activity and perceived risk, instead of a negative one.\\

Concerns about omitted variable bias are harder to dissipate. Policy-makers are likely to monitor epidemiological developments and respond to them promptly by implementing lockdowns and other containment policies. To control for such confounders, I employ fixed-effect regression models.  \\

\begin{table}[H]
\begin{center}
\begin{small}
\begin{tabular}{lccccccc}
  \hline \hline
                              & \multicolumn{3}{c}{\textbf{FED's MEI}}                                                                                                                                               & \multicolumn{1}{l}{} & \multicolumn{3}{c}{\textbf{Google's Workplace Mobility}}                                                                                                                             \\ \cline{2-4} \cline{6-8}
\multicolumn{1}{c}{\textbf{}} & \begin{tabular}[c]{@{}c@{}}(1)\\ OLS\end{tabular}          & \begin{tabular}[c]{@{}c@{}}(2)\\ FE\end{tabular}           & \begin{tabular}[c]{@{}c@{}}(3)\\ FE\end{tabular}           &                      & \begin{tabular}[c]{@{}c@{}}(1)\\ OLS\end{tabular}          & \begin{tabular}[c]{@{}c@{}}(2)\\ FE\end{tabular}            & \begin{tabular}[c]{@{}c@{}}(3)\\ FE\end{tabular}          \\ \hline
\textbf{Spec \#1}             &                                                            &                                                            &                                                            &                      &                                                            &                                                             & \multicolumn{1}{l}{}                                      \\
\textit{Death Risk ($\chi$)}  & \begin{tabular}[c]{@{}c@{}}-37.34***\\ (3.95)\end{tabular} & \begin{tabular}[c]{@{}c@{}}-38.19***\\ (3.49)\end{tabular} & \begin{tabular}[c]{@{}c@{}}-11.60***\\ (1.96)\end{tabular} &                      & \begin{tabular}[c]{@{}c@{}}-43.98***\\ (4.84)\end{tabular} & \begin{tabular}[c]{@{}c@{}}-49.46***\\ (5.20)\end{tabular}  & \begin{tabular}[c]{@{}c@{}}-9.88***\\ (1.50)\end{tabular} \\
\textit{}                     &                                                            &                                                            &                                                            &                      &                                                            &                                                             & \multicolumn{1}{l}{}                                      \\
\textbf{Spec \#2}             & \multicolumn{1}{l}{}                                       & \multicolumn{1}{l}{}                                       & \multicolumn{1}{l}{}                                       & \multicolumn{1}{l}{} & \multicolumn{1}{l}{}                                       & \multicolumn{1}{l}{}                                        & \multicolumn{1}{l}{}                                      \\
\textit{Lethality (CFR)}      & \begin{tabular}[c]{@{}c@{}}-36.41***\\ (6.10)\end{tabular} & \begin{tabular}[c]{@{}c@{}}-37.76***\\ (8.30)\end{tabular} & \begin{tabular}[c]{@{}c@{}}-3.38\\ (2.29)\end{tabular}     &                      & \begin{tabular}[c]{@{}c@{}}-45.60***\\ (7.48)\end{tabular} & \begin{tabular}[c]{@{}c@{}}-50.28***\\ (10.74)\end{tabular} & \begin{tabular}[c]{@{}c@{}}-3.44\\ (2.10)\end{tabular}    \\
\textit{Infection Risk (IR)}  & \begin{tabular}[c]{@{}c@{}}-18.81***\\ (5.16)\end{tabular} & \begin{tabular}[c]{@{}c@{}}-18.80***\\ (5.48)\end{tabular} & \begin{tabular}[c]{@{}c@{}}-10.39***\\ (1.80)\end{tabular} &                      & \begin{tabular}[c]{@{}c@{}}-31.05***\\ (3.92)\end{tabular} & \begin{tabular}[c]{@{}c@{}}-34.66***\\ (3.78)\end{tabular}  & \begin{tabular}[c]{@{}c@{}}-9.82***\\ (1.28)\end{tabular} \\
                              &                                                            &                                                            &                                                            &                      &                                                            &                                                             & \multicolumn{1}{l}{}                                      \\ \hline
State FE                      & N                                                          & Y                                                          & Y                                                          &                      & N                                                          & Y                                                           & Y                                                         \\
Time FE                       & N                                                          & N                                                          & Y                                                          &                      & N                                                          & N                                                           & Y                                                         \\
Adj. $R^2$ (Spec \#1)         & 0.15                                                       & 0.17                                                       & 0.96                                                       &                      & 0.20                                                       & 0.25                                                        & 0.97                                                      \\
Adj. $R^2$ (Spec \#2)         & 0.19                                                       & 0.21                                                       & 0.96                                                       &                      & 0.31                                                       & 0.37                                                        & 0.97                                                      \\
Obs                           & 1530                                                       & 1530                                                       & 1530                                                       &                      & 1479                                                       & 1479                                                        & 1479                                                      \\ \hline \hline
\multicolumn{8}{l}{\begin{tabular}[c]{@{}l@{}} \scriptsize \textbf{Notes:} Clustered standard errors at the state-level in parenthesis. *$p<0.10$, **$p<0.05$, ***$p<0.01$.\\ \scriptsize Standardized coefficients (\%) obtained by scaling variables by their standard deviation.\end{tabular}}
\end{tabular}
\caption{Main Regression Results at State-Level}
\label{Table:Main_Regression_Results_State}
\end{small}
\end{center}
\end{table}

\autoref{Table:Main_Regression_Results_State}  reports my regression results at the state-level.\footnote{All coefficients are standardized. The non-standardized estimates can be found in \autoref{Appendix:Empirical_Additional_Results}.} The first three columns report the regression results using the FED's MEI measure as dependent variable, while the last three using the Google Workplace Mobility measure. Furthermore, I consider two specifications: one in which I regress economic activity on  perceived death risk, and one in which I replace perceived death risk with its two components, namely  perceived lethality and  perceived infection risk.\\

 The estimates suggest  a strong negative relationship between economic activity and  perceived death risk, which holds also when the latter is decomposed into its two components. Importantly, the estimated effect is economically meaningful: perceived death risk explains between $35\%$ and $50\%$ of the fall in economic activity when time fixed-effects are excluded, and roughly $10\%$ of the relative fall in economic activity when time fixed-effects are included.  State fixed-effects ensure that unobserved heterogeneity in economic activity across states is properly accounted for. Time fixed-effects control for unobserved national developments common across states and countries, such as national containment guidelines, nation-wide communications from policy-makers and so on. The state-level estimates, however, might still suffer from omitted variable bias since they do not control for state-level developments that occur over time and might correlate with both economic activity and perceived risk.\\

\begin{table}[H]
\begin{center}
\begin{small}
\begin{tabular}{lccccc}
  \hline \hline
                              & \multicolumn{5}{c}{\textbf{FED's MEI}}                                                                                                                                                                                                                                                                      \\ \cline{2-6}
\multicolumn{1}{c}{\textbf{}} & \begin{tabular}[c]{@{}c@{}}(1)\\ OLS\end{tabular}          & \begin{tabular}[c]{@{}c@{}}(2)\\ FE\end{tabular}           & \begin{tabular}[c]{@{}c@{}}(3)\\ FE\end{tabular}          & \begin{tabular}[c]{@{}c@{}}(4)\\ FE\end{tabular}          & \begin{tabular}[c]{@{}c@{}}(5)\\ FE\end{tabular}          \\ \hline
\textbf{Spec \#1}             &                                                            &                                                            &                                                           &                                                           &                                                           \\
\textit{Death Risk ($\chi$)}  & \begin{tabular}[c]{@{}c@{}}-10.13***\\ (0.93)\end{tabular} & \begin{tabular}[c]{@{}c@{}}-9.95***\\ (0.79)\end{tabular}  & \begin{tabular}[c]{@{}c@{}}-4.67***\\ (0.40)\end{tabular} & \begin{tabular}[c]{@{}c@{}}-4.19***\\ (0.64)\end{tabular} & \begin{tabular}[c]{@{}c@{}}-4.19**\\ (1.65)\end{tabular}  \\
\textit{}                     &                                                            &                                                            &                                                           &                                                           &                                                           \\
\textbf{Spec \#2}             & \multicolumn{1}{l}{}                                       & \multicolumn{1}{l}{}                                       & \multicolumn{1}{l}{}                                      & \multicolumn{1}{l}{}                                      & \multicolumn{1}{l}{}                                      \\
\textit{Lethality (CFR)}      & \begin{tabular}[c]{@{}c@{}}-16.16***\\ (1.04)\end{tabular} & \begin{tabular}[c]{@{}c@{}}-18.67***\\ (1.13)\end{tabular} & \begin{tabular}[c]{@{}c@{}}-1.05***\\ (0.27)\end{tabular} & \begin{tabular}[c]{@{}c@{}}-1.91***\\ (0.51)\end{tabular} & \begin{tabular}[c]{@{}c@{}}-1.91***\\ (0.53)\end{tabular} \\
\textit{Infection Risk (IR)}  & \begin{tabular}[c]{@{}c@{}}-2.45***\\ (0.58)\end{tabular}  & \begin{tabular}[c]{@{}c@{}}-2.14***\\ (0.50)\end{tabular}  & \begin{tabular}[c]{@{}c@{}}-4.05***\\ (0.65)\end{tabular} & \begin{tabular}[c]{@{}c@{}}-3.05***\\ (0.67)\end{tabular} & \begin{tabular}[c]{@{}c@{}}-3.05**\\ (1.21)\end{tabular}  \\
                              &                                                            &                                                            &                                                           &                                                           &                                                           \\ \hline
County FE                     & N                                                          & Y                                                          & Y                                                         & N                                                         & N                                                         \\
Time FE                       & N                                                          & N                                                          & Y                                                         & N                                                         & N                                                         \\
State-Time FE                 & N                                                          & N                                                          & N                                                         & Y                                                         & Y                                                         \\
SE Clustering                 & County                                                     & County                                                     & County                                                    & County                                                    & State                                                     \\
Adj. $R^2$ (Spec \#1)         & 0.01                                                       & 0.14                                                       & 0.90                                                      & 0.81                                                      & 0.81                                                      \\
Adj. $R^2$ (Spec \#2)         & 0.03                                                       & 0.16                                                       & 0.90                                                      & 0.81                                                      & 0.81                                                      \\
Obs                           & 90599                                                      & 90599                                                      & 90599                                                     & 90599                                                     & 90599                                                     \\ \hline \hline
\multicolumn{6}{l}{\begin{tabular}[c]{@{}l@{}}\scriptsize \textbf{Notes:} Clustered standard errors in parenthesis. *$p<0.10$, **$p<0.05$, ***$p<0.01$.\\ \scriptsize Standardized coefficients (\%) obtained by scaling variables by their standard deviation.\end{tabular}}
\end{tabular}
\caption{Main Regression Results at County-Level}
\label{Table:Main_Regression_Results_County}
\end{small}
\end{center}
\end{table}
 \autoref{Table:Main_Regression_Results_County} reports my regression results at the more granular county-level. This allows me to introduce state-time fixed-effects which absorb state-level developments  over time. As the vast majority of lockdowns and containment policies during the first phase of the epidemic outbreak were enacted at a state-level, the state-time fixed-effects should be able to solve any omitted variable bias.\footnote{\cite{goolsbee2020covid} construct a dataset of stay-at-home and business closure orders for the first months of the epidemic outbreak. County-level lockdowns are highly correlated with state-level ones, although not perfectly. Moreover,  in late 2020 some states started to implement local lockdowns and stay-at-home orders, at a level as granular as the zip-code. This would invalidate the proposed identification for more recent data.} The county-level estimates remains negative and statistically significant across all specifications, and the same is true when perceived death risk is decomposed into its two components. The coefficients become smaller as fixed-effects are included, suggesting that the latter are successfully controlling for unobservables. Small coefficients could be due to the importance of local factors to explain local economic activity, but also to the fact that the proposed measure of perceived death risk is only an approximation - and this becomes clearer at a more granular level.\footnote{For example, different individuals will perceive death risk differently depending on a wide set of covariates, which could systematically differ across counties.}.  Nonetheless, the negative effect of perceived risk on economic activity appears clear.\\

\subsection{Additional Empirical Results}
\label{Section:Evidence_Additional_Results}
A question that naturally arises is how the proposed measure of death risk compares to alternative measures that have been adopted in the literature, such as total  or weekly cases and deaths. \autoref{Table:Additional_Regression_CasesDeaths} provides a tentative answer. \\

\begin{table}[H]
\begin{center}
\begin{small}
\begin{tabular}{lccccccccc}
  \hline \hline
                             & \multicolumn{4}{c}{\textbf{State-Level}}                                                                                                                                                                                                        & \multicolumn{1}{l}{} & \multicolumn{4}{c}{\textbf{County-Level}}                                                                                                                                                                                                     \\ \cline{2-5} \cline{7-10}
\textbf{FED's MEI}           & \begin{tabular}[c]{@{}c@{}}(1)\\ FE\end{tabular}          & \begin{tabular}[c]{@{}c@{}}(2)\\ FE\end{tabular}           & \begin{tabular}[c]{@{}c@{}}(3)\\ FE\end{tabular}          & \begin{tabular}[c]{@{}c@{}}(4)\\ FE\end{tabular}           &                      & \begin{tabular}[c]{@{}c@{}}(5)\\ FE\end{tabular}          & \begin{tabular}[c]{@{}c@{}}(6)\\ FE\end{tabular}          & \begin{tabular}[c]{@{}c@{}}(7)\\ FE\end{tabular}          & \begin{tabular}[c]{@{}c@{}}(8)\\ FE\end{tabular}          \\ \hline
\textbf{}                    &                                                           &                                                            & \multicolumn{1}{l}{}                                      &                                                            &                      &                                                           &                                                           & \multicolumn{1}{l}{}                                      & \multicolumn{1}{l}{}                                      \\
\textit{Death Risk ($\chi$)} &                                                           & \begin{tabular}[c]{@{}c@{}}-11.27***\\ (1.92)\end{tabular} &                                                           & \begin{tabular}[c]{@{}c@{}}-10.23***\\ (1.94)\end{tabular} &                      &                                                           & \begin{tabular}[c]{@{}c@{}}-3.07***\\ (0.56)\end{tabular} &                                                           & \begin{tabular}[c]{@{}c@{}}-2.47***\\ (0.56)\end{tabular} \\
\textit{Cases}         & \begin{tabular}[c]{@{}c@{}}-5.74***\\ (1.69)\end{tabular} & \begin{tabular}[c]{@{}c@{}}-1.55\\ (1.48)\end{tabular}     &                                                           &                                                            &                      & \begin{tabular}[c]{@{}c@{}}-7.45***\\ (2.84)\end{tabular} & \begin{tabular}[c]{@{}c@{}}-7.48***\\ (2.79)\end{tabular} &                                                           &                                                           \\
\textit{Deaths}        & \begin{tabular}[c]{@{}c@{}}-0.37\\ (1.16)\end{tabular}    & \begin{tabular}[c]{@{}c@{}}-3.43**\\ (1.40)\end{tabular}   &                                                           &                                                            &                      & \begin{tabular}[c]{@{}c@{}}-1.05\\ (1.65)\end{tabular}    & \begin{tabular}[c]{@{}c@{}}-0.61\\ (1.60)\end{tabular}    &                                                           &                                                           \\
\textit{$\Delta$Cases}    &                                                           &                                                            & \begin{tabular}[c]{@{}c@{}}-3.22***\\ (1.08)\end{tabular} & \begin{tabular}[c]{@{}c@{}}-2.02\\ (1.94)\end{tabular}     &                      &                                                           &                                                           & \begin{tabular}[c]{@{}c@{}}-6.85***\\ (1.93)\end{tabular} & \begin{tabular}[c]{@{}c@{}}-6.85***\\ (1.89)\end{tabular} \\
\textit{$\Delta$Deaths}   &                                                           &                                                            & \begin{tabular}[c]{@{}c@{}}-5.60*\\ (3.13)\end{tabular}   & \begin{tabular}[c]{@{}c@{}}-0.34\\ (2.84)\end{tabular}     &                      &                                                           &                                                           & \begin{tabular}[c]{@{}c@{}}-1.81*\\ (1.08)\end{tabular}   & \begin{tabular}[c]{@{}c@{}}1.24\\ (1.04)\end{tabular}     \\
                             &                                                           &                                                            & \multicolumn{1}{l}{}                                      &                                                            &                      &                                                           &                                                           & \multicolumn{1}{l}{}                                      & \multicolumn{1}{l}{}                                      \\ \hline
State FE                     & Y                                                         & Y                                                          & Y                                                         & Y                                                          &                      & N                                                         & N                                                         & N                                                         & N                                                         \\
Time FE                      & Y                                                         & Y                                                          & Y                                                         & Y                                                          &                      & N                                                         & N                                                         & N                                                         & N                                                         \\
State-Time FE                & N                                                         & N                                                          & N                                                         & N                                                          &                      & Y                                                         & Y                                                         & Y                                                         & Y                                                         \\
SE Clustering                & State                                                     & State                                                      & State                                                     & State                                                      &                      & County                                                    & County                                                    & County                                                    & County                                                    \\
Adj. $R^2$                   & 0.96                                                      & 0.96                                                       & 0.96                                                      & 0.96                                                       &                      & 0.82                                                      & 0.82                                                      & 0.82                                                      & 0.82                                                      \\
Obs                          & 1530                                                      & 1530                                                       & 1530                                                      & 1530                                                       &                      & 90569                                                     & 90569                                                     & 90569                                                     & 90569                                                     \\ \hline \hline
\multicolumn{10}{l}{\begin{tabular}[c]{@{}l@{}} \scriptsize \textbf{Notes:} Clustered standard errors in parenthesis. *$p<0.10$, **$p<0.05$, ***$p<0.01$. The dependent variable is the FED's Mobility \\ \scriptsize and Engagement Index (MEI). Standardized coefficients (\%) obtained  by scaling variables by their standard deviation. \end{tabular}}
\end{tabular}
\caption{Additional Regression Results on Reported Cases and Deaths}
\label{Table:Additional_Regression_CasesDeaths}
\end{small}
\end{center}
\end{table}
 The first four columns use state-level data and suggest that, while cases and deaths exhibit a negative relationship with economic activity, they tend to lose significance and become smaller when my measure of perceived risk is included in the regression. The last four columns replicate the exercise at the county-level. Overall, the estimates  suggest that the proposed measure of risk remains precise and robust even when detected cases and deaths are controlled for.\\

Another important question is whether the level of testing itself has any effects on economic activity. Indeed, one could argue that more testing reduces agents' uncertainty about the accuracy of reported cases and deaths, and that less uncertainty is beneficial to economic activity. In this respect, an indicator which is frequently monitored to assess how much testing is performed relative to the true latent epidemic is the test positivity rate.\footnote{Testing data come from the COVID Tracking Project, and the test positivity rate is defined as the number of new detected cases over the number of tests performed in a given time period. Due to data limitation, I limit myself to the construction of the test positivity rate at the state-level.} The results are reported in \autoref{Table:Additional_Regression_Positivity}.\\

\begin{table}[H]
\begin{center}
\begin{small}
\begin{tabular}{lccccc}
  \hline \hline
                              & \multicolumn{2}{c}{\textbf{FED's MEI}}                                                                               & \multicolumn{1}{l}{} & \multicolumn{2}{c}{\textbf{Workplace Mobility}}                                                             \\ \cline{2-3} \cline{5-6}
\multicolumn{1}{c}{\textbf{}} & \begin{tabular}[c]{@{}c@{}}(1)\\ FE\end{tabular}         & \begin{tabular}[c]{@{}c@{}}(2)\\ FE\end{tabular}          &                      & \begin{tabular}[c]{@{}c@{}}(3)\\ FE\end{tabular}         & \begin{tabular}[c]{@{}c@{}}(4)\\ FE\end{tabular}          \\ \hline
\textbf{}                     &                                                          &                                                           &                      &                                                          &                                                           \\
\textit{Death Risk ($\chi$)}  &                                                          & \begin{tabular}[c]{@{}c@{}}-8.96***\\ (1.64)\end{tabular} &                      &                                                          & \begin{tabular}[c]{@{}c@{}}-8.53***\\ (1.15)\end{tabular} \\
\textit{Test Positivity Rate} & \begin{tabular}[c]{@{}c@{}}-2.24**\\ (1.09)\end{tabular} & \begin{tabular}[c]{@{}c@{}}-1.49\\ (0.95)\end{tabular}    &                      & \begin{tabular}[c]{@{}c@{}}-2.24**\\ (1.09)\end{tabular} & \begin{tabular}[c]{@{}c@{}}-0.82\\ (0.95)\end{tabular}    \\
                              &                                                          &                                                           &                      &                                                          &                                                           \\ \hline
State FE                      & Y                                                        & Y                                                         &                      & Y                                                        & Y                                                         \\
Time FE                       & Y                                                        & Y                                                         &                      & Y                                                        & Y                                                         \\
Adj. $R^2$                    & 0.96                                                     & 0.96                                                      &                      & 0.97                                                     & 0.97                                                      \\
Obs                           & 1318                                                     & 1318                                                      &                      & 1318                                                     & 1318                                                      \\ \hline \hline
\multicolumn{6}{l}{\begin{tabular}[c]{@{}l@{}} \scriptsize \textbf{Notes:} Clustered standard errors at the state-level. *$p<0.10$, **$p<0.05$, ***$p<0.01$.\\ \scriptsize Standardized coefficients (\%) obtained by scaling variables by their standard deviation.\end{tabular}}
\end{tabular}
\caption{Additional Regression Results on Test Positivity Rate}
\label{Table:Additional_Regression_Positivity}
\end{small}
\end{center}
\end{table}

The test positivity rate exhibits a significant negative relationship with economic activity,  in line with the previous conjecture. Interestingly, however, when the proposed measure of perceived risk is included in the regression, the estimated effect of the test positivity rate becomes indistinguishable from zero, and the magnitude of the estimated standardized coefficient is almost an order of magnitude lower than that of the perceived death risk.

\newpage
\section{A Stochastic Epidemiological Model with Fear}
\label{Section:Framework}
This section presents a stochastic epidemiological model with symptoms-based testing policies and reduced-form behavioral responses driven by fear of the epidemic disease.\footnote{From a technical viewpoint, the model presented in this paper is an extension of standard epidemiological models, and nests the most common ones as special cases. For the sake of illustration, I show in \autoref{Appendix:Deterministic_SIR} how to recover the textbook deterministic SIR model.} The model is agent-based, i.e. each agent is modeled individually, and features two diseases: a novel emerging epidemic disease and an endemic confounding disease. The role of the confounding disease is literally to confound the diagnosis of the epidemic disease, since individuals exhibiting symptoms might be infected with either disease.\\

Testing policies in the model mimic real-world ones which prioritize testing of severe symptomatic individuals, and play two important roles. First, detected active infections are put into (imperfect) isolation, allowing the government to slow down epidemic transmission.  Second, testing provides agents with information about the latent epidemic disease. Because the true epidemic is unobservable, agents base their behavior on the data produced by the health-care system through testing. More precisely, they use reported cases, active infections, and deaths to construct a measure of death risk which embodies the familiar notion of fear: a higher perceived risk of death stokes fear and prompts a reduction in labor supply, causing a fall in economic activity - consistently with the empirical  part of the paper. Agents' behavior is introduced in a reduced-form manner: labor supply and leisure respond to the perceived risk of death according to a fixed elasticity parameter which can be estimated in the data.\\

The level of testing activity in the model is partially, but not fully, under the control of the government. Indeed, I assume that all severe symptomatic individuals are always tested by the health-care system, which implies that the government is left with the possibility to test non-severe symptomatic individuals, i.e. those with mild symptoms or no symptoms at all. This is also referred to as  ``screening'' of the population for infections.\\

The overall effect of increased testing on perceived risk, and thus on economic activity, is non-monotone. More testing results in better epidemic containment - thanks to the targeted isolation of the infected and to stronger behavioral responses - but it also uncovers a larger portion of the true latent epidemic. Furthermore, it reduces the perceived lethality of the disease. The ultimate goal of the model is to analyze these forces and to understand which one prevails.\\

\subsection{Aggregate Epidemic Dynamics}
\label{Subsection:General_Environment}
Time is discrete, each time period is interpreted as a day, and the population will be studied over an horizon $\textcolor{NYUcolor}{T}$. Consider a homogeneous population of ex-ante identical individuals with initial size $\textcolor{NYUcolor}{P_0}$, and suppose that no individual is added to the population (e.g. no births, no immigration). There are two diseases circulating in the population: the \textcolor{NYUcolor}{\text{epidemic disease}} and a \textcolor{NYUcolor}{\text{confounding disease}}. The latter is an \textit{endemic disease} which circulates in the population irrespective of the epidemic disease and is named `confounding' because it confounds the  diagnosis of the epidemic diseases due to the fact that infected individuals share similar symptoms across the two diseases. I assume the following:
\begin{itemize}[noitemsep]
  \item[\textcolor{NYUcolor}{\textbf{E1:}}] \textit{For each disease, individuals who recover obtain immunity.}
  \item[\textcolor{NYUcolor}{\textbf{E2:}}] \textit{Each individual can catch only one of the two diseases.}
\end{itemize}
Assumption \textcolor{NYUcolor}{\textbf{E1}} is often adopted in epidemiological models - since most epidemic diseases feature at least a temporary immunity -  and simplifies the problem from a modeling perspective. Assumption \textcolor{NYUcolor}{\textbf{E2}} is a simplification that allows to abstract from what happens when an agent catches both diseases.\\

From now on, latent variables will be denoted with an asterisk and observable ones without. At any point in time, each individual $\textcolor{NYUcolor}{j}$ can find themselves in one of three states:
\[ \textcolor{NYUcolor}{x^*_t(j)} + \textcolor{NYUcolor}{c^*_t(j)} + \textcolor{NYUcolor}{c^{f*}_T(j)} = 1 \]
where $\textcolor{NYUcolor}{x^*_t(j)}$ denotes susceptibility to the epidemic disease and takes value $1$ if individual $\textcolor{NYUcolor}{j}$ has never contracted any disease at time $\textcolor{NYUcolor}{t}$ and will never contract the confounding disease over the time horizon considered. The variable $\textcolor{NYUcolor}{c^*_t(j)}$  takes value $1$ if the individual has contracted the epidemic disease at time $t$ or before, while $\textcolor{NYUcolor}{c^{f*}_T(j)}$ takes value $1$ if the individual has ever contracted the confounding disease over the time horizon considered.\\

Given that the confounding disease is an endemic disease, I will model new aggregate cases each day as an exogenous stationary process:
\[ \textcolor{NYUcolor}{\Delta C_t^{f*}} \sim \textit{Normal}\Bigg(\frac{\omega^f \cdot P_0}{T}, \Big(\sigma^f \cdot \frac{\omega^f \cdot P_0}{T} \Big)^2 \Bigg) \]
where realizations are rounded to the nearest integer. Notice that $\textcolor{NYUcolor}{\omega^{f}}$ is the share of the population  that on average contracts the confounding disease over the time horizon $\textcolor{NYUcolor}{T}$. So for instance, if $\textcolor{NYUcolor}{T} = 90$ and $\textcolor{NYUcolor}{\omega^{f}} = 0.20$, then on average twenty percent of the initial population contracts the infection over a 90 day period. Moreoever,  $\textcolor{NYUcolor}{\sigma^f}$ is the coefficient of variation of new daily infections.\\

Turning to the epidemic disease, I assume that the event that a susceptible individual catches the epidemic disease follows a bernoulli random variable:

\[ \textcolor{NYUcolor}{\Delta c^*_{t+1}(j)| x_t^*(j) = 1} \sim \textit{Bernoulli}\Big(IR_t^*\Big) \]
where $\textcolor{NYUcolor}{IR_t^*}$ is the true latent infection risk and will be defined shortly. Assuming that individual infection events are independent, and aggregating across individuals one gets new daily aggregate infections

\[ \textcolor{NYUcolor}{\Delta C_{t+1}^{*}} \sim \textit{Binomial}\Big(X_{t}^{*}, IR_t^* \Big) \]
where $\textcolor{NYUcolor}{X_t^*}$ is the (latent) number of susceptible individuals. Importantly, the true latent infection risk in the model will be assumed to be the following:
\[ \textcolor{NYUcolor}{IR_t^*} = \underbrace{\beta}_{\text{Transmission Coefficient}} \times \underbrace{\rho_{t}}_{\text{Contact Rate}} \times \underbrace{\frac{ I_{I}^{*} - \theta \cdot I_{t}}{P_{t} - \theta \cdot I_{t}}}_{\text{Probability of Meeting an Infected}} \]
where $\textcolor{NYUcolor}{\beta}$ is the exogenous transmission coefficient, which is the product of the transmission risk upon contact with an infected and the average number of pre-epidemic contacts;  $\textcolor{NYUcolor}{\rho_t}$ is the contact rate, which is normalized to one absent the epidemic;  $\textcolor{NYUcolor}{I^*_t}$ is the true latent number of  active infections; $\textcolor{NYUcolor}{I_t}$ is the number of detected active infections by the health-care system; $\textcolor{NYUcolor}{\theta}$ is a parameter summarizing the degree of enforcement  of the isolation policy adopted by the health care system;  and  $\textcolor{NYUcolor}{P_t}$ is population.\footnote{Notice that when $\textcolor{NYUcolor}{\theta} = 1$ each detected active infection is put into full isolation and when $\textcolor{NYUcolor}{\theta} = 0$ none is. Imperfect isolation obtains when $\textcolor{NYUcolor}{\theta}\in(0,1)$.}
Notice that, throughout the paper, variables in capital letters will denote aggregate time-series and are recovered as follows:
\[  \textcolor{NYUcolor}{W_t} =  \sum_{j = 1}^{P_0} w_t(j) \]
where $\textcolor{NYUcolor}{W_t}$ denotes a generic time-series variable and  $\textcolor{NYUcolor}{w_t(j)}$ denotes the individual-level counterpart.\\

While isolation of infected individuals directly affects the probability of meeting an infected, behavioral responses directly affect the endogeneous contact rate:
\[ \textcolor{NYUcolor}{\rho_t} = \pi \cdot \bar{N}_t + (1 - \pi) \cdot \bar{L}_t \]
where $\textcolor{NYUcolor}{\pi}$ is the (exogenous) share of contacts due to work (as opposed to leisure), $\textcolor{NYUcolor}{\bar{N}_t}$ is average labor supply across agents, and $\textcolor{NYUcolor}{\bar{L}_t}$ is average leisure. More precisely:
\begin{align*}
\textcolor{NYUcolor}{\bar{N}_t} &= P_t^{-1} \cdot \sum_{j = 1}^{P_0} n_t(j) \\
\textcolor{NYUcolor}{\bar{L}_t} &= P_t^{-1} \cdot \sum_{j = 1}^{P_0} l_t(j)
\end{align*}
Implicitly, the idea is that labor supply, leisure and interactions across agents are all sides of the same coin. A fall in labor supply and/or leisure therefore reduces interactions among agents, which in turn reduces the true infection risk. The next section turns to individuals and describes how agents' behavior works in the model.\\

\subsection{Individuals}
Individuals supply labor for production, enjoy leisure and can be infected by either disease. I first describe the reduced-form behavior of labor supply and leisure, and then turn to the evolution of each disease conditional on infection.\\

\subsubsection{Work and Leisure}
Individuals achieve a daily production $\textcolor{NYUcolor}{y_t(j)}$ by supplying labor:
\[ \textcolor{NYUcolor}{y_t(j)} = A \cdot n_t(j) \]
where $\textcolor{NYUcolor}{A}$ captures the daily average productivity of an individual, and $\textcolor{NYUcolor}{n_t(j)}$ denotes the individual's labor supply. I assume that labor supply depends on health status, fear of the epidemic, and whether the individual is subject to mandatory isolation. In a reduced-form way, I posit that:
\[ \textcolor{NYUcolor}{n_t(j)} =
\begin{cases}
   n_0 \cdot (1 + \chi_t)^{-\varepsilon_n} & \text{if $j$ has no or mild symptoms and not isolated} \\
  (1 - \theta) \cdot n_0 & \text{if $j$ has no or mild symptoms and isolated} \\
  0 & \text{if $j$ is dead or has severe symptoms}
\end{cases}
\]
where $\textcolor{NYUcolor}{n_0}$ is labor  for a healthy individual in normal times (i.e. absent the epidemic), $\textcolor{NYUcolor}{\chi_t}$ is the perceived risk of death from the epidemic disease, and $\textcolor{NYUcolor}{\varepsilon_n}$ is the (approximate) elasticity of labor supply with respect to the perceived risk of death. The equation above  basically says that dead individuals and those with severe symptoms cannot work, while what those alive do depends on whether they have been tested. An individual with no or mild symptoms that has not been tested, will not be isolated and will not know whether he is infected with the epidemic disease. As a result, she will be assumed to be capable of working, but will protect herself from the epidemic disease. Individuals who have tested positively are put under (imperfect) isolation. Given that they are currently infected, they have no reason to `protect' themselves from the epidemic disease, and their labor supply will depend exclusively on the strictness of the isolation policy.\footnote{ Individuals in the model engage in a `selfish' behavior similar to what described in \cite{EichenbaumTesting2020}.} However, once the isolation is over and they are no longer infected, they keep `protecting' themselves because they are not sure as to whether past infections guarantee immunity from the epidemic disease. \\

Consistently with the empirical analysis in \autoref{Section:Evidence}, I assume that the test positivity rate does not affect behavior and I model perceived death risk as:
\begin{gather*}
 \textcolor{NYUcolor}{\chi_t} = \underbrace{\frac{D_t}{C_t}}_{\text{Case Fatality Rate}}  \times \underbrace{\beta \cdot \frac{I_t}{P_t}}_{\text{Perceived Infection Risk}}
\end{gather*}
where I assume for simplicity that individuals use the true transmission coefficient $\textcolor{NYUcolor}{\beta}$ when forming their perceptions. The reduced-form behavior of leisure is symmetric to that of labor supply:
\[ \textcolor{NYUcolor}{l_t(j)} =
\begin{cases}
   l_0 \cdot (1 + \chi_t)^{-\varepsilon_l} & \text{if $j$ has no or mild symptoms and not isolated} \\
  (1 - \theta) \cdot l_0 & \text{if $j$ has no or mild symptoms and isolated} \\
  0 & \text{if $j$ is dead or has severe symptoms}
\end{cases}
\]
Together, labor supply and leisure determine the level of interactions between agents.\\

\subsubsection{Epidemic Disease}
\label{Subsection:Epidemic_Disease}
 \autoref{fig:Epidemic_Disease_TimeLine} provides a visual summary of the assumed evolution of the disease upon infection:\\

\tikzstyle{level 1}=[level distance=3.5cm, sibling distance=0cm]
\tikzstyle{level 2}=[level distance=3.5cm, sibling distance=2cm]
\tikzstyle{level 3}=[level distance=4.5cm, sibling distance=1.25cm]

\tikzstyle{bagg} = [text width=5em, text centered]
\tikzstyle{bag} = [text width=9em, text centered]
\tikzstyle{end} = [circle, minimum width=3pt,fill, inner sep=0pt]

\begin{figure}[H]
\begin{center}
\begin{footnotesize}
\begin{tikzpicture}[grow=right, sloped]
\node[bagg] {Infection}
    child {
      node[bagg]{Incubation}
    child {
        node[bag] {Severe Symptoms ($s$)}
        child {
                node[bag] {Recovery ($1 - \phi_s$)}
            }
            child {
                node[bag] {Death ($\phi_s$)}
            }
    }
    child {
        node[bag] {Mild Symptoms ($m$)}
            child {
                node[bag] {Recovery ($1 - \phi_m$)}
            }
            child {
                node[bag] {Death ($\phi_m$)}
            }
    }
    child {
        node[bag] {Asymptomatic ($a$)}
            child {
                node[bag] {Recovery ($1 - \phi_a$)}
            }
            child {
                node[bag] {Death ($\phi_a$)}
            }
    }
    };
\end{tikzpicture}
\end{footnotesize}
\end{center}
\caption{Evolution of the Epidemic Disease upon Infection}
\label{fig:Epidemic_Disease_TimeLine}
\end{figure}
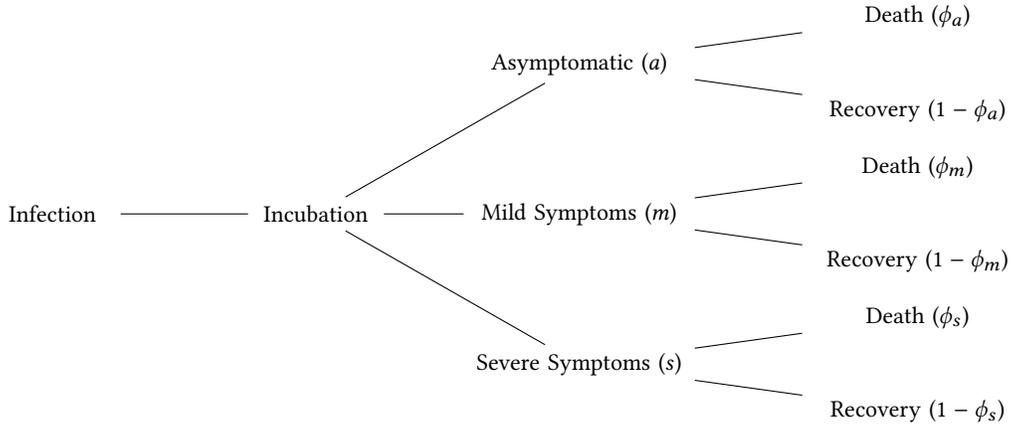
Conditional on infection, the epidemic disease evolves as follows: first, the individual enters a \textit{pre-symptomatic period} (a.k.a. \textit{incubation period}), during which they are infected (and can infect others), but do not manifest any symptoms. What happens next is the result of two random events. The first random event determines what type of symptoms the individual will display. I will assume three types of symptoms: severe symptoms, mild symptoms and no symptoms. The second random event determines the terminal outcome of the disease, i.e. whether the individual recovers or dies.\\

Let's now introduce a more formal modeling of the disease. For a generic  individual $\textcolor{NYUcolor}{j}$ who has been infected by the epidemic disease at a generic time $\tilde{t}$, we have that

\[ \textcolor{NYUcolor}{c_t^*(j)} = \begin{cases}
1 & \text{if } t \geq \tilde{t} \\
0 & \text{if } t < \tilde{t}
\end{cases} \]

To model the type of symptoms developed and the terminal outcome for a generic individual $\textcolor{NYUcolor}{j}$, I will introduce two random variables: $\textcolor{NYUcolor}{\text{symptoms}^*(j)}$ describing the type of symptoms developed, and $\textcolor{NYUcolor}{\text{death}^*(j)}$ to denote the terminal outcome of the disease. Their \textit{joint probability distribution} is given by:

\begin{table}[H]
\begin{center}
  \begin{tabular}{c|cc|c}
                  & recovers                           & dies                           &                                   \\ \hline
  severe symptoms & $\textcolor{NYUcolor}{s \cdot (1 - \phi_s)}$ & $\textcolor{NYUcolor}{s \cdot \phi_s}$ & $\textcolor{NYUcolor}{s}$         \\
  mild symptoms   & $\textcolor{NYUcolor}{m \cdot (1 - \phi_m)}$ & $\textcolor{NYUcolor}{m \cdot \phi_m}$ & $\textcolor{NYUcolor}{m}$         \\
  asymptomatic    & $\textcolor{NYUcolor}{a \cdot  (1 - \phi_a)}$ & $\textcolor{NYUcolor}{a \cdot \phi_a}$ & $\textcolor{NYUcolor}{a}$ \\ \hline
                  & $\textcolor{NYUcolor}{1 - \phi}$   & $\textcolor{NYUcolor}{\phi}$   &
  \end{tabular}
\end{center}
\end{table}

Notice that $\textcolor{NYUcolor}{\phi}$ is the \textit{unconditional infection fatality risk}, i.e. the probability that an individual  who contracts the epidemic disease dies, while  $\textcolor{NYUcolor}{\phi_s}$, $\textcolor{NYUcolor}{\phi_m}$ and $\textcolor{NYUcolor}{\phi_a}$ denote  the \textit{conditional infection fatality risks}, i.e. the probability that an individual who contracts the epidemic disease and exhibits a certain type of symptoms dies.\\

The timing of these random events is random itself. In particular, for each individual $\textcolor{NYUcolor}{j}$, the random variable $\textcolor{NYUcolor}{p^*(j)}$ represents the length of the incubation period or, equivalently, the number of days the individual spends in the pre-symptomatic state; $\textcolor{NYUcolor}{\tilde{k}^*(j)}$ represents the number of days between the onset of symptoms and the terminal outcome death; $\textcolor{NYUcolor}{\tilde{q}^*(j)}$ represents the number of days between the onset of symptoms and the terminal outcome recovery. I will assume that these lags do not depend on the type of symptoms developed, nor on the terminal outcome, and that they are distributed as Poisson random variables:\footnote{Standard stochastic SIR-type models assume that these timings are exponentially distributed, as this assumption allows the aggregation of individuals into compartments, resulting in a noticeable simplification of the problem thanks to the memorylessness property of the exponential random variables. See \cite{Feng2007} and \cite{Feng2007a} for more details.}
\begin{align*}
  \textcolor{NYUcolor}{p^*(j), \tilde{k}^*(j), \tilde{q}^*(j)} &\perp  \text{symptoms}^*(j), \text{death}^*(j) \\
   \textcolor{NYUcolor}{p^*(j)} &\sim \text{Poisson}(p - 1) + 1\\
   \textcolor{NYUcolor}{\tilde{k}^*(j)}  &\sim \text{Poisson}(\tilde{k})  \\
   \textcolor{NYUcolor}{\tilde{q}^*(j)}  &\sim \text{Poisson}(\tilde{q})
\end{align*}
I opt for a shifted-Poisson distribution of $\textcolor{NYUcolor}{p^*(j)}$ so that it takes at least one period - i.e. one day - between infection and the terminal outcome. The number of days between infection and  terminal outcomes are therefore given by:
\begin{align*}
  \textcolor{NYUcolor}{k^*(j)} &= p^*(j) + \tilde{k}^*(j) \sim  \text{Poisson}(p + \tilde{k} - 1) + 1 \\
  \textcolor{NYUcolor}{q^*(j)} &= p^*(j) + \tilde{q}^*(j) \sim  \text{Poisson}(p + \tilde{q} - 1) + 1
\end{align*}

Analytically, the dynamic evolution of the disease for a generic individual $j$  can  be  expressed as follows:
\begin{align*}
  \textcolor{NYUcolor}{u_t^*(j)} &= c_t^*(j) - c_{t-p^*(j)}^*(j) \\
  \textcolor{NYUcolor}{d_t^*(j)} &= \text{death}^*(j) \cdot c_{t-k^*(j)}^*(j) \\
  \textcolor{NYUcolor}{r_t^*(j)} &= \big[1 - \text{death}^*(j) \big] \cdot c^*_{t-q^*(j)}(j) \\
  \textcolor{NYUcolor}{s_t^*(j)} &= \text{severe}^*(j) \cdot [c^*_{t-p^*(j)}(j) - d^*_t(j) - r^*_t(j)] \\
  \textcolor{NYUcolor}{m_t^*(j)} &= \text{mild}^*(j) \cdot [c^*_{t-p^*(j)}(j) - d^*_t(j) - r^*_t(j)] \\
  \textcolor{NYUcolor}{a_t^*(j)} &= \text{asymptomatic}^*(j) \cdot [c^*_{t-p^*(j)}(j) - d^*_t(j) - r^*_t(j)] \\
  \textcolor{NYUcolor}{i_t^*(j)} &= c^*_t(j) - r^*_t(j) - d^*_t(j) = u^*_t(j) + s^*_t(j) + m^*_t(j) + a^*_t(j)
\end{align*}

where  $\textcolor{NYUcolor}{u_t^*(j)}$ is one when the individual in the incubation period (and their symptoms are still Unknown), $\textcolor{NYUcolor}{d_t^*(j)}$ and $\textcolor{NYUcolor}{r_t^*(j)}$ are, respectively, one when the individual dies or recover, $\textcolor{NYUcolor}{s_t^*(j)}$ is one when the individual displays severe symptoms, $\textcolor{NYUcolor}{m_t^*(j)}$ is one when the individual displays mild symptoms and $\textcolor{NYUcolor}{a_t^*(j)}$ is one when the individual is asymptomatic. Furthermore, $\textcolor{NYUcolor}{i_t^*(j)}$ is one when the individual has an active infection, and becomes zero again when the infection is no longer active (either because of recovery or death). At the same time, an active infection can manifest itself in four forms: incubation period, severe symptoms, mild symptoms or lack of symptoms. Finally, notice that $\textcolor{NYUcolor}{c^*_t(j)}$,  $\textcolor{NYUcolor}{d^*_t(j)}$,  $\textcolor{NYUcolor}{r^*_t(j)}$ are \textit{absorbing states} that, if reached, are never left, while $\textcolor{NYUcolor}{u^*_t(j)}$,  $\textcolor{NYUcolor}{s^*_t(j)}$,  $\textcolor{NYUcolor}{m^*_t(j)}$,  $\textcolor{NYUcolor}{a^*_t(j)}$,  $\textcolor{NYUcolor}{i^*_t(j)}$ are \textit{transient states}.\\

\subsubsection{Confounding Disease}
\label{Subsection:Confounding_Disease}
Since the confounding disease is not the main object of investigation, its characterization will be simplified as much as possible. I will assume no incubation period and two types of symptoms: severe symptoms and mild symptoms.\footnote{Notice that the confounding disease can also be thought of as two different diseases which differ in the type of symptoms they induce. For instance, the mild-symptom state can be thought of as seasonal flu, and the severe-symptom state can be thought of as pneumonia.} Importantly, symptoms induced by the confounding disease are similar to those arising from the epidemic disease, so that the former literally acts as a confounder in the diagnostic process of the latter. Similarly to the epidemic disease, two random events determine the type of symptoms and the final outcome:

\[ \textcolor{NYUcolor}{\text{severe}^{f*}(j)} = \begin{cases} \text{severe}(\equiv 1) \quad &\text{wp $s^f$} \\ \text{mild}(\equiv 0) \quad &\text{wp $1 - s^f$}
\end{cases} \quad \quad \textcolor{NYUcolor}{\text{death}^{f*}(j)} = \begin{cases} 1 \quad \text{wp $\phi^f$} \\ 0 \quad \text{wp $1 - \phi^f$}
\end{cases}
\]
The timing of the terminal outcome is described by $\textcolor{NYUcolor}{k^{f*}(j)}$, which represents the number of periods between infection and death, and by $\textcolor{NYUcolor}{q^{f*}(j)}$, which represents the number of periods between infection and recovery. For simplicity, I will assume that these lags are degenerate and independent of the type of symptoms. I will also assume that the terminal outcome is independent of the type of symptoms.  \autoref{fig:Confounding_Disease_TimeLine}  provides a visual summary of the evolution of the disease upon infection.

\tikzstyle{level 1}=[level distance=4cm, sibling distance=2.5cm]
\tikzstyle{level 2}=[level distance=5cm, sibling distance=1.25cm]

\tikzstyle{bag} = [text width=11em, text centered]
\tikzstyle{end} = [circle, minimum width=3pt,fill, inner sep=0pt]

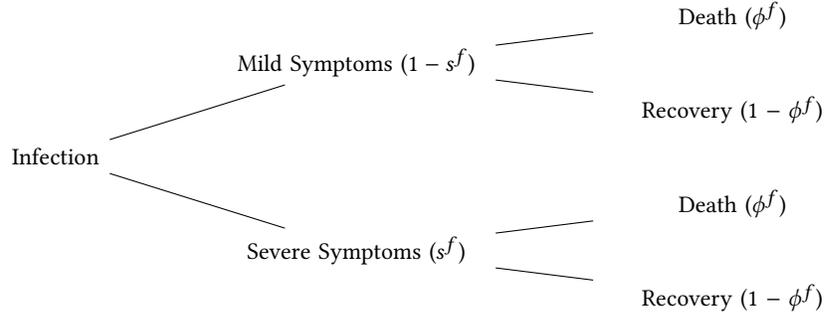
\begin{figure}[H]
\begin{center}
\begin{footnotesize}
\begin{tikzpicture}[grow=right, sloped]
\node[bag] {Infection}
    child {
        node[bag] {Severe Symptoms ($s^f$)}
        child {
                node[bag] {Recovery ($1 - \phi^f$)}
            }
            child {
                node[bag] {Death ($\phi^f$)}
            }
    }
    child {
        node[bag] {Mild Symptoms ($1 - s^f$)}
            child {
                node[bag] {Recovery ($1 - \phi^f$)}
            }
            child {
                node[bag] {Death ($\phi^f$)}
            }
    };
\end{tikzpicture}
\end{footnotesize}
\end{center}
\caption{Evolution of the Confounding Disease upon Infection}
\label{fig:Confounding_Disease_TimeLine}
\end{figure}

More formally, the evolution of the confounding disease at the individual level is given by:
\begin{align*}
  \textcolor{NYUcolor}{d^{f*}_t(j)} &= \text{death}^{f*}(j) \cdot c_{t-k^{f*}}^{f*}(j) \\
  \textcolor{NYUcolor}{r^{f*}_t(j)} &= \big[1 - \text{death}^{f*}(j) \big] \cdot c_{t-q^{f*}}^{f*}(j) \\
  \textcolor{NYUcolor}{s^{f*}_t(j)} &=  \text{severe}^{f*}(j) \cdot \big[ c^{f*}_{t}(j) - d^{f*}_t(j) - r^{f*}_t(j) \big] \\
  \textcolor{NYUcolor}{m^{f*}_t(j)} &=  \big[1 -  \text{severe}^{f*}(j) \big] \cdot \big[ c^{f*}_{t}(j) - d^{f*}_t(j) - r^{f*}_t(j) \big] \\
  \textcolor{NYUcolor}{i^{f*}_t(j)} &= c^{f*}_t(j) - r^{f*}_t(j) - d^{f*}_t(j) = s^{f*}_t(j) + m^{f*}_t(j)\\
\end{align*}

\subsection{The Government}
The government plays two roles in the model. First, it performs testing activity through the health-care system. Second, it collects revenues and engages in health-care spending.\\

\subsubsection{Symptoms-Based Testing Policies}
I assume that the health-care system in the model adopts symptoms-based testing policies that mimic real-world ones. As the World Health Organization puts it, \textit{``the decision to test should be based on clinical and epidemiological factors and linked to an assessment of the likelihood of infection"}, and there are few better indicators of infections than symptoms, especially at the early stages of an epidemic outbreak when the characteristics of the disease are still unknown.\footnote{See \href{https://www.who.int/publications/i/item/10665-331501}{https://www.who.int/publications/i/item/10665-331501} for the testing guidelines as of March 19 2020 during the COVID-19 outbreak. As of January 31 2020, right at the start of the outbreak, a suspected case to be tested was required to display either ``severe acute respiratory infection requiring admission to hospital'' or ``any acute respiratory illness''. See  \href{https://www.who.int/publications/i/item/laboratory-testing-of-2019-novel-coronavirus-(-2019-ncov)-in-suspected-human-cases-interim-guidance-17-january-2020}{https://www.who.int/publications/i/item/laboratory-testing-of-2019-novel-coronavirus-(-2019-ncov)-in-suspected-human-cases-interim-guidance-17-january-2020}.  } These considerations are introduced  into the model by assuming that testing activity is \textit{prioritized} based on the severity of symptoms displayed by individuals: severe symptomatic individuals are always tested first, then individuals with mild symptoms, and finally asymptomatic ones.\\

As opposed to standard compartmental epidemiological models, which assume that a (constant) share of some compartment is tested each period, the proposed framework allows for the specification of a daily testing capacity in terms of the number of tests to be performed. This implies, for example, that a certain testing capacity permits testing of a large share of symptomatic individuals at the beginning of the epidemic, but of a very small share during its peak. Importantly, I am going to assume the following:
\begin{itemize}[noitemsep]
 \item[\textcolor{NYUcolor}{\textbf{T1:}}] \textit{There is always enough daily testing capacity to test severe symptomatic individuals.}
\end{itemize}
Assumption $\textcolor{NYUcolor}{\textbf{T1}}$ is motivated by two considerations. First, it reflects testing priority of individuals exhibiting severe symptoms, which is justified by the need to determine the underlying disease in order to decide appropriate medical treatment. Second, it reflects the fact that individuals with severe symptoms are more likely to show up at the hospital, be hospitalized, and tested.\\

What the government can therefore choose is whether to perform additional tests $\textcolor{NYUcolor}{T_t^{NS}}$ on individuals who do not display severe symptoms (NS stands for `Non-Severe'). Consistently with the idea of symptoms-based policies, these additional tests are administered to mild symptomatic individuals first, and to asymptomatic individuals only if there is any remaining testing capacity.\footnote{
Notice that there is no modeling of voluntary testing. However, since a large portion of voluntary testing arguably arises due to the appearance of symptoms, testing policies that prioritize testing of symptomatic individuals should implicitly account - at least partially - for voluntary testing.}

\subsubsection{Implementation of Testing Policies}
 Depending on the technical characteristics of the existing testing technology and on the properties of the epidemic disease, one can make slightly different assumptions about how testing activity is actually implemented in the model. I will assume the following:
\begin{itemize}[noitemsep]
  \item[\textcolor{NYUcolor}{\textbf{T2:}}] \textit{Individuals who have tested positively are not tested again. }
  \item[\textcolor{NYUcolor}{\textbf{T3:}}] \textit{Tests detect only active infections, with a false negativity rate $\textcolor{NYUcolor}{\alpha}$.}
  \item[\textcolor{NYUcolor}{\textbf{T4:}}] \textit{The outcome of the test is known with a fixed delay $\textcolor{NYUcolor}{d}$, and an individual is not tested again until the outcome of the previous test is known.}
\end{itemize}
Assumption $\textcolor{NYUcolor}{\textbf{T2}}$  is  justified when immunity is obtained after recovery from the infection.\footnote{It also implicitly assumes that the health care system has a way to detect recovery that does not require the use of an additional test, or that there is another testing capacity dedicated for this purpose.} Assumption  $\textcolor{NYUcolor}{\textbf{T3}}$ implies that a test (imprecisely) detects the infection during the incubation period and irrespective of the type of symptoms while infection is active, but not after death or recovery. The delay  $\textcolor{NYUcolor}{d}$ in assumption $\textcolor{NYUcolor}{\textbf{T4}}$ could reflect both technological and organizational constraints that create a fixed lag between the time a test is administered and the time its outcome is known. \\

The health-care system's testing policy is implemented using set theory. First, by assumption $\textcolor{NYUcolor}{\textbf{T1}}$, all severe symptomatic individuals that need to get tested are tested. The set of severe symptomatic individuals tested at time $t$ is given by
\[ \textcolor{NYUcolor}{\mathcal{T}^S_t} = \mathbf{\Sigma}^S_t \setminus  (\mathcal{C}_{t-1} \cup \mathcal{T}^p_t )  \]
and consists of the individuals displaying severe symptoms  ($\textcolor{NYUcolor}{\mathbf{\Sigma}^S_t}$), minus those that have been diagnosed with the disease in the past ($\textcolor{NYUcolor}{\mathcal{C}_{t-1}}$), minus those whose test result is still pending ($\textcolor{NYUcolor}{\mathcal{T}^p_t}$).\\

 When the government mandates additional testing capacity (i.e. $\textcolor{NYUcolor}{T_t^{NS}>0}$), the health-care system expands testing to mild symptomatic individuals. The set of individuals that it would like to test is given by
\[ \textcolor{NYUcolor}{\mathcal{G}^M_t} = \mathbf{\Sigma}^M_t \setminus  (\mathcal{C}_{t-1} \cup \mathcal{T}^p_t )  \]
where $\textcolor{NYUcolor}{\mathbf{\Sigma}^M_t}$ is the set of individuals displaying mild symptoms. A random subset $\textcolor{NYUcolor}{\mathcal{T}_t^M \subseteq \mathcal{G}^M_t}$ of size equal to $\textcolor{NYUcolor}{T_t^M = |\mathcal{T}_t^M| = \min\{T_t^{NS}, |\mathcal{G}^M_t| \}}$ is tested. \\

After individuals with mild symptoms are tested,  the health-care system starts testing asymptomatic individuals if there is additional testing capacity, i.e.  $\textcolor{NYUcolor}{T_t^{NS} - T_t^M>0}$. The set of individuals that it would like to test is given by
\[ \textcolor{NYUcolor}{\mathcal{G}^A_t} = \Big(\mathcal{P}_t \setminus (\mathbf{\Sigma}^S_t \cup  \mathbf{\Sigma}^M_t)\Big) \setminus  (\mathcal{C}_{t-1} \cup \mathcal{T}^p_t )  \]
where  $\textcolor{NYUcolor}{\mathcal{P}_t}$ is the set of alive individuals.  A random subset $\textcolor{NYUcolor}{\mathcal{T}_t^A \subseteq \mathcal{G}^A_t}$ is tested, and its size is equal to $\textcolor{NYUcolor}{T_t^A = |\mathcal{T}_t^A| = \min\{T_t^{NS} - T_t^M, |\mathcal{G}^A_t| \}}$. \\

The set of all individuals tested at a generic time $t$ is given by
\[ \textcolor{NYUcolor}{\mathcal{T}_t} = \mathcal{T}^S_t \cup \mathcal{T}^M_t \cup \mathcal{T}^A_t \]
and the total number of tests performed is given by $\textcolor{NYUcolor}{T_t = |\mathcal{T}_t|}$.  Tests can turn out to be either positive or negative, i.e. $\textcolor{NYUcolor}{\mathcal{T}_t = \mathcal{T}_t^+ \cup \mathcal{T}_t^-}$, and positive tests are given by:\footnote{I implicitly assume that all tests administered share the same technological characteristics. This can be easily relaxed, for example, by assuming that severe symptomatic individuals receive a different type of test from the rest of the population, as done in \cite{droste2020economic}.}
\[ \textcolor{NYUcolor}{\mathcal{T}^+_t} \subseteq  \mathcal{I}^*_t \cap \mathcal{T}_t \]
where $\textcolor{NYUcolor}{x\in \mathcal{I}^*_t \cap \mathcal{T}_t}$ also belongs to $ \textcolor{NYUcolor}{\mathcal{T}^+_t} $ with probability $\textcolor{NYUcolor}{1 - \alpha}$, where $\textcolor{NYUcolor}{\alpha}$ is the false negativity rate of the testing technology. Because of the delay in test results, positive cases are known only with a delay:
\[    \textcolor{NYUcolor}{\mathcal{C}_t} = \mathcal{C}_{t-1} \cup \mathcal{T}_{t-d}^+  \]
At time $t$, the list of pending test results is given by
\[ \textcolor{NYUcolor}{\mathcal{T}^p_t} = \bigcup\limits_{index=1}^{d} \mathcal{T}_{t - index} \]
 with $\textcolor{NYUcolor}{\mathcal{T}^p_t = \emptyset}$ if $\textcolor{NYUcolor}{d = 0}$. Finally, given the list of detected cases $\textcolor{NYUcolor}{\mathcal{C}_t}$, one can recover any detected set $\textcolor{NYUcolor}{\mathcal{Z}_t}$ as follows:\footnote{To be clear, what I refer to as `detected' cases are laboratory-confirmed cases and are not indirect estimates obtained in other ways.}
\[ \textcolor{NYUcolor}{\mathcal{Z}_t} = \mathcal{C}_t \cap   \mathcal{Z}_t^* \]
where $\textcolor{NYUcolor}{\mathcal{Z}_t^*}$ is its latent counterpart. Reported  epidemic time-series are then given by $\textcolor{NYUcolor}{Z_t = |\mathcal{Z}_t |}$. In other words, once an individual enters the list of positive cases, her health-status is perfectly known by the health-care system.\\

\subsubsection{The Government's Budget}
Government expenditure is thus given by:
\begin{align*}
  \textcolor{NYUcolor}{Exp_t} &= c_T \cdot T_t + c_s \cdot |\mathbf{\Sigma}^S_t|
\end{align*}
where $\textcolor{NYUcolor}{c_T}$ is the cost of each test performed, $\textcolor{NYUcolor}{T_t}$ is the number of tests performed,  $\textcolor{NYUcolor}{c_S}$ is the cost of treating a severe infection, and  $\textcolor{NYUcolor}{|\mathbf{\Sigma}^S_t|}$ is the number of individuals displaying severe symptoms.\footnote{I implicitly assume that individuals  with severe symptoms from any disease require costly medical treatment.} The government also collects revenues by taxing economic activity:
\[   \textcolor{NYUcolor}{Rev_t}= \tau \cdot Y_t \]
where $\textcolor{NYUcolor}{\tau}$ is a uniform tax rate and  $\textcolor{NYUcolor}{Y_t}$ is daily GDP, which is obtained aggregating individuals' daily production. For simplicity, I assume that the government can run budget deficits or surpluses which are freely rolled over.

\newpage
\section{Understanding the Mechanism}
\label{Section:Understanding_Mechanism}
To build intuition, I will first simulate the model under a coronavirus-like parameterization  to analyze the dynamic behavior of both economic and epidemiological variables.\footnote{Coronavirus diseases, such as SARS-CoV-2, are usually deemed as `influenza-like' diseases, because of their similarity with influenza. The 2002-2004 SARS epidemic, the 2009 swine flu pandemic, and the ongoing MERS  are all examples of influenza-type diseases. On the other hand, measles and ebola are epidemic disease which are not influenza-type.} I will then validate the testing policies in the model using Northern Italy as a case study.\\

\subsection{A Coronavirus-Like Influenza Disease}
\autoref{Table:Parameterization} below reports the parameterization for a generic coronavirus-disease, which I will refer to as the `baseline parameterization'.

\begin{table}[H]
\begin{center}
\begin{small}
  \begin{tabular}{lccllcc}
    \hline \hline
  \multicolumn{3}{c}{\textbf{Individuals}}                                                            &  & \multicolumn{3}{c}{\textbf{Government}}                                                                 \\ \cline{1-3} \cline{5-7}
  \textit{Baseline Labor Supply}                    & $\textcolor{NYUcolor}{n_0}$           & $1$     &  & \textit{Test Cost}                          & $\textcolor{NYUcolor}{c_T}$        & $25$                 \\
  \textit{Baseline Leisure}                         & $\textcolor{NYUcolor}{l_0}$           & $1$     &  & \textit{Treatment Cost}                     & $\textcolor{NYUcolor}{c_S}$        & $300$                \\
  \textit{Daily Productivity}                       & $\textcolor{NYUcolor}{A}$             & $175$   &  & \textit{Tax Rate}                           & $\textcolor{NYUcolor}{\tau}$       & $0.30$               \\
  \textit{Elasticity of Labor to Perceived Death Risk}              & $\textcolor{NYUcolor}{\varepsilon_n}$ & $1000$  &  & \textit{Test Outcome Delay}                 & $\textcolor{NYUcolor}{d}$          & $1$                  \\
  \textit{Elasticity of Leisure to Perceived Death Risk}            & $\textcolor{NYUcolor}{\varepsilon_l}$ & $1000$  &  & \textit{Test False Negative Rate}           & $\textcolor{NYUcolor}{\alpha}$     & $0.25$               \\
  \textit{Contact Share from Work}                  & $\textcolor{NYUcolor}{\pi}$           & $0.5$   &  & \textit{Isolation Effectiveness}            & $\textcolor{NYUcolor}{\theta}$     & $0.9$                \\
                                                    &                                       &         &  &                                             & \multicolumn{1}{l}{}               & \multicolumn{1}{l}{} \\
  \multicolumn{3}{c}{\textbf{Epidemic Disease}}                                                       &  & \multicolumn{3}{c}{\textbf{Confounding Disease}}                                                        \\ \cline{1-3} \cline{5-7}
  \textit{Transmission Coefficient}                        & $\textcolor{NYUcolor}{\beta}$         & $0.275$ &  & \textit{Probability of Severe Symptoms}     & $\textcolor{NYUcolor}{s^f}$        & $0.10$               \\
  \textit{Probability of Severe Symptoms}           & $\textcolor{NYUcolor}{s}$             & $0.30$  &  & \textit{Infection Fatality Risk}            & $\textcolor{NYUcolor}{\phi^f}$     & $0.02$               \\
  \textit{Probability of Mild Symptoms}             & $\textcolor{NYUcolor}{m}$             & $0.40$  &  & \textit{Time from Infection to Recovery}    & $\textcolor{NYUcolor}{q^f}$        & $7$                  \\
  \textit{Probability of No Symptoms}               & $\textcolor{NYUcolor}{a}$             & $0.30$  &  & \textit{Time from Infection to Death}       & $\textcolor{NYUcolor}{k^f}$        & $7$                  \\
  \textit{Infection Fatality Risk for Severe}       & $\textcolor{NYUcolor}{\phi_s}$        & $0.15$  &  & \textit{Share of Population Infected}       & $\textcolor{NYUcolor}{\omega^f}$   & $0.20$               \\
  \textit{Infection Fatality Risk for Mild}         & $\textcolor{NYUcolor}{\phi_m}$        & $0$     &  & \textit{Volatility of New Daily Infections} & $\textcolor{NYUcolor}{\sigma^{f}}$ & $0.10$               \\
  \textit{Infection Fatality Risk for Asymptomatic} & $\textcolor{NYUcolor}{\phi_a}$        & $0$     &  & \textit{}                                   &                                    &                      \\
  \textit{Unconditional Infection Fatality Risk}    & $\textcolor{NYUcolor}{\phi}$          & $0.045$ &  & \multicolumn{3}{c}{\textbf{General}}                                                                    \\ \cline{5-7}
  \textit{Incubation Period}                        & $\textcolor{NYUcolor}{p}$             & $3$     &  & \textit{Initial Population}                 & $\textcolor{NYUcolor}{P_0}$        & $5e4$                \\
  \textit{Average Time from Symptoms to Recovery}   & $\textcolor{NYUcolor}{\tilde{q}}$     & $11$    &  & \textit{Time Horizon}                       & $\textcolor{NYUcolor}{T}$          & $350$                \\
  \textit{Average Time from Symptoms to Death}      & $\textcolor{NYUcolor}{\tilde{k}}$     & $5$     &  &                                             & \multicolumn{1}{l}{}               & \multicolumn{1}{l}{} \\
  \textit{Initial Infections}                       & $\textcolor{NYUcolor}{C_0^*}$             & $50$    &  &                                             & \multicolumn{1}{l}{}               & \multicolumn{1}{l}{} \\ \hline \hline
  \end{tabular}
\caption{Parameterization of a Coronavirus-Like Disease}
\label{Table:Parameterization}
\end{small}
\end{center}
\end{table}
In this benchmark parameterization, the epidemic disease features a large share of mild symptomatic and asymptomatic infections, an incubation period, and a relatively high probability of death for severe infections, but a null one for non-severe infections. The average time from infection to death is $8$ days, and from infection to recovery is $14$ days. The confounding disease, instead, is meant to resemble a seasonal flu. Most infections are mild symptomatic, and the infection fatality risk for this disease is relatively low. The time from infection to any terminal outcome is assumed to be a week, and  $20\%$ of the population is infected over the course of roughly a year.\\

Individual labor supply and leisure in normal times are normalized to $1$. Daily productivity is chosen as to roughly match daily GDP per capita in the U.S., while the elasticities of labor and leisure to the perceived death risk are set to $1000$, which  generates a sizeable fall in GDP over a 1-year horizon. Half of contacts arise from work, while the other half from leisure activities. I set the cost of a test to $25$ dollars, the cost of daily treatment of a severe infection to $300$ dollars, and the tax rate to $30\%$. I assume that it takes one day to learn the outcome of a test, that the probability of a false negative is $25\%$ and that the compliance rate of mandatory isolation is $90\%$. The population size is set to strike a balance between computational speed and uncertainty of outcomes.\footnote{All the simulations in the paper report $68\%$ confidence bands. A larger population size reduces variance of outcomes, but increases the computational costs. Notice that, unlike standard epidemiological models, even when the population size increases to infinity, uncertainty remains due to the testing activity.}\\

\begin{figure}[H]
\centerline{\includegraphics[scale=0.575, angle=0]{Baseline_Epidemic}}
\caption{Epidemic Dynamics for a Coronavirus-Type Epidemic Disease}
\label{fig:DGP_baseline_epidemic}
\end{figure}

 \autoref{fig:DGP_baseline_epidemic} displays the epidemic dynamics under the baseline parameterization  when the government does not mandate any additional testing on non-severe individuals, i.e. $\textcolor{NYUcolor}{T_t^{NS} = 0, \forall t}$. This implies that the health-care system tests only individuals who exhibit severe symptoms. The top row of the figure summarizes the dynamic evolution of the true latent epidemic. First, most infections result in recovery. Second, active infections can be in the incubation period, asymptomatic, mild symptomatic or severe symptomatic. Third, new daily cases are asymmetric because the spread of the disease slows down after peak, as a result of agents' endogenous behavior.\\

The middle row of the figure displays, instead, what is detected by the health-care system.  Only a small portion of true cases is detected, and, since only individuals displaying severe-symptoms are tested, the health-care system does not detect any infection in the incubation period, nor any infection who displays mild or no symptoms. In other words, detected infections are not representative of overall infections and  are the most likely to die from the disease.\footnote{Under the baseline parameterization, the health-care system correctly detects all the deaths due to the epidemic disease.}\\

The bottom row of the figure reports the test positivity rate and the composition of symptomatic individuals. Agents exhibiting symptoms can be infected by the epidemic disease or by the confounding disease, and it is not possible to tell them apart without testing. The test positivity rate takes the shape of an inverted parabola, as it peaks when true active infections peak, but it is low in the initial and final stages of the outbreak.

\begin{figure}[H]
\centerline{\includegraphics[scale=0.55, angle=0]{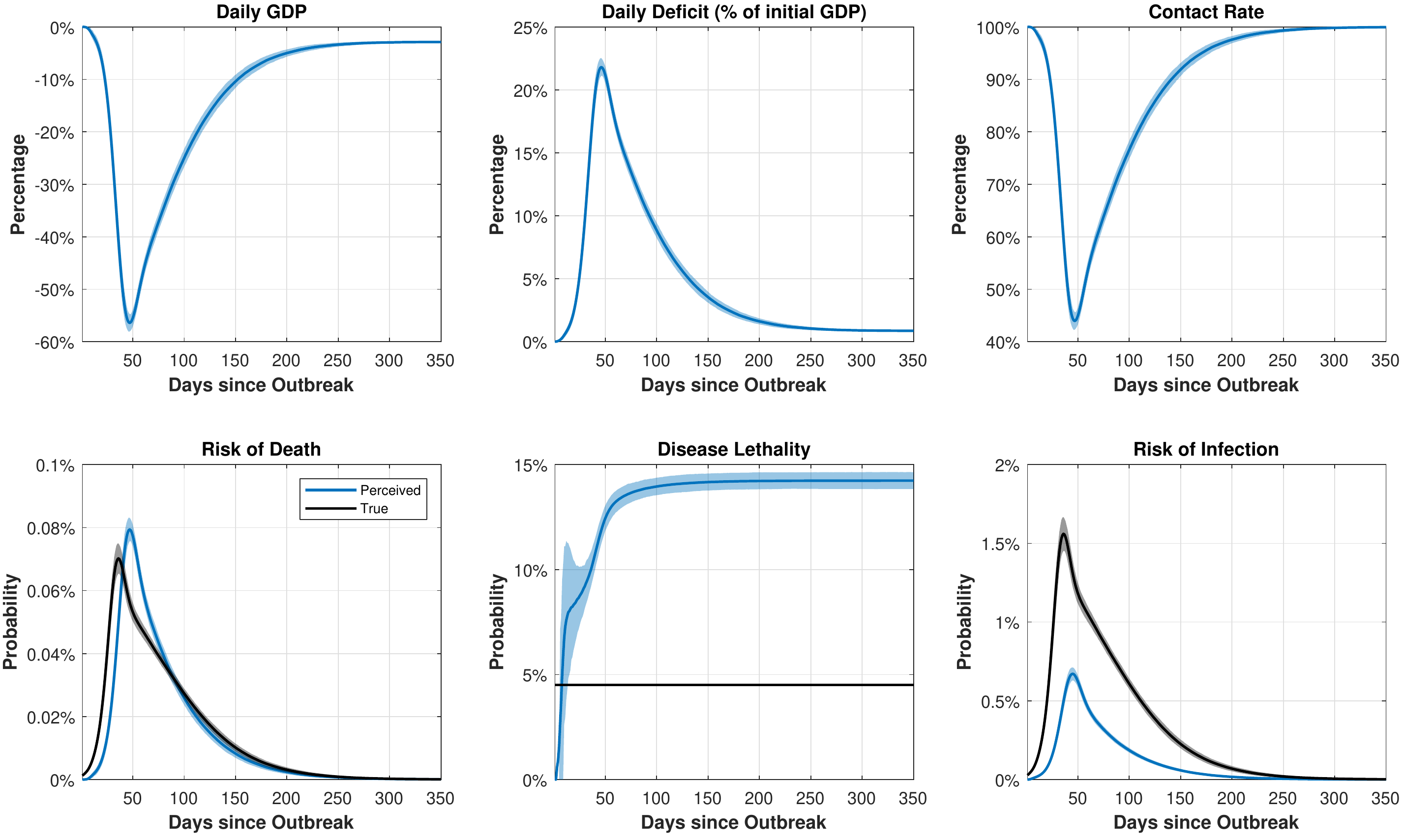}}
\caption{Economic Dynamics for a Coronavirus-Type Epidemic Disease}
\label{fig:DGP_baseline_economics}
\end{figure}

\autoref{fig:DGP_baseline_economics} summarizes what happens to the economy. Output contracts, interactions among agents fall, and budget deficits soar, mainly as a result of agents' responses to the perceived risk of death -  reported in the second row of the figure. As the outbreak unfolds, the health-care system produces time-series of testing data on cases, deaths and active infections which are used  by agents to assess death risk. As the latter increases, agents cut on both work and leisure activities, causing output to fall, deficits to rise and the number of interactions to fall.\footnote{Under the proposed parameterization, the cumulative GDP loss over the course of one year is around $15\%$, while the public deficit rises by around $5\%$ of pre-epidemic GDP.} In \autoref{Appendix:Behavioral_responses}, I further clarify the role of behavioral responses.\\

The second row of the figure clarifies what happens to agents' perceptions of risk, and compares them to the true latent risk. For this specification, the perceived death risk (in blue) is similar to the true death risk (in black), although this is not a general property of the model. In fact,  agents' perceptions about the two determinants of overall risk are incorrect. The perception of disease lethality is off because the case fatality rate substantially over-estimates the true infection fatality risk, as a result of a `narrow' testing policy that focuses on severe symptomatic individuals, who have the  highest conditional infection fatality risk. The perception of infection risk is also off because  the health-care system sizably under-estimates the number of true active infections.\\

\begin{figure}[H]
\centerline{\includegraphics[scale=0.6, angle=0]{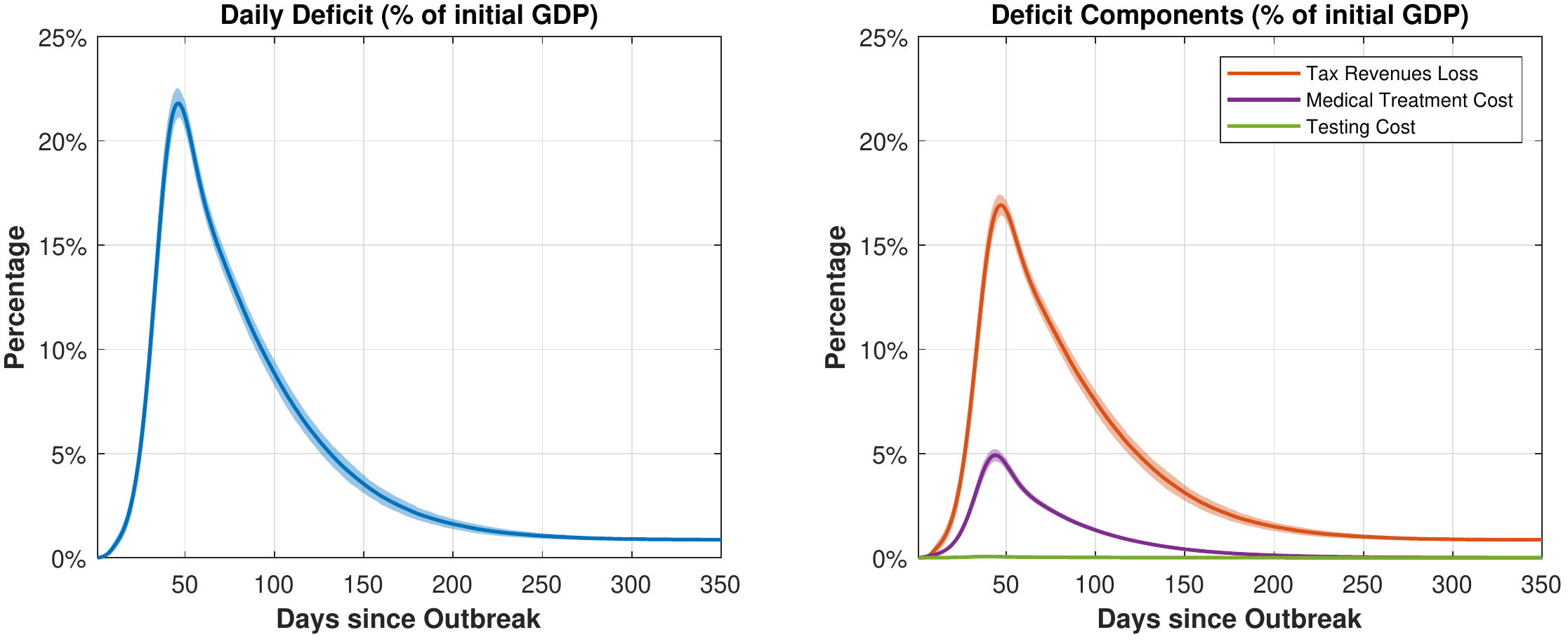}}
\caption{The Evolution of Government's Finances}
\label{fig:DGP_baseline_budget}
\end{figure}

Finally,  soaring deficits arise from the combination of several forces, as summarized in \autoref{fig:DGP_baseline_budget}. The first and main reason behind rising deficits is the loss of tax revenues resulting from the fall in economic activity. The second is the increase in expenditure on medical treatments: each severe symptomatic individual requires costly medical attention and, as a result, health-care expenditure rises. Last, there is the cost of testing. Given that the government is not mandating any additional testing on non-severe individuals, expenditure on testing is modest because of the small number of tests performed.\\

\newpage
\subsection{Validating Testing Policies: The Case of Northern Italy}
\label{Section:Testing_Policies_Validation}
In this section, I compare the dynamic evolution of key testing variables in the model with their empirical counterparts, using the Italian regions of Lombardy and Veneto  during the first SARS-CoV-2 outbreak in early 2020 as a case study. These two regions were the first two territories who experienced the SARS-CoV-2 epidemic outbreak in the West. They border each other, have a similar GDP per capita, similar infrastructures, are both highly populated, and were both among the most hit regions during the epidemic outbreak in Spring 2020. \\

A key difference between the two regions, however, lies in their testing policies. At the beginning of the outbreak, Lombardy followed the Italian government's testing guideline, which in turn followed the WHO's initial instructions to limit testing to severe symptomatic individuals. Veneto's approach, instead, was shaped by Professor Andrea Crisanti and consisted  in implementing a wider testing policy right away. As Science Magazine reports: \textit{``Crisanti persuaded the regional government of Veneto to test anyone with even the mildest of symptoms, and to trace and test their contacts as well''}, while \textit{``[g]uidelines from the World Health Organization and Italy’s National Institute of Health said to test only patients with symptoms''}.\footnote{See \href{https://www.sciencemag.org/news/2020/08/how-italy-s-father-swabs-fought-coronavirus}{https://www.sciencemag.org/news/2020/08/how-italy-s-father-swabs-fought-coronavirus} for the full article.}\\

Testing data from the two regions therefore provide a useful case study to validate the testing policies in the model. Specifically, I parallel testing data from Lombardy to those generated under the baseline parameterization, where the health-care system tests only severe symptomatic individuals. Testing data from Veneto, instead, are paralleled to those generated from the baseline parameterization with the twist that  both severe and mild symptomatic individuals are tested daily.\\

\begin{figure}[H]
\centerline{\includegraphics[scale=0.575, angle=0]{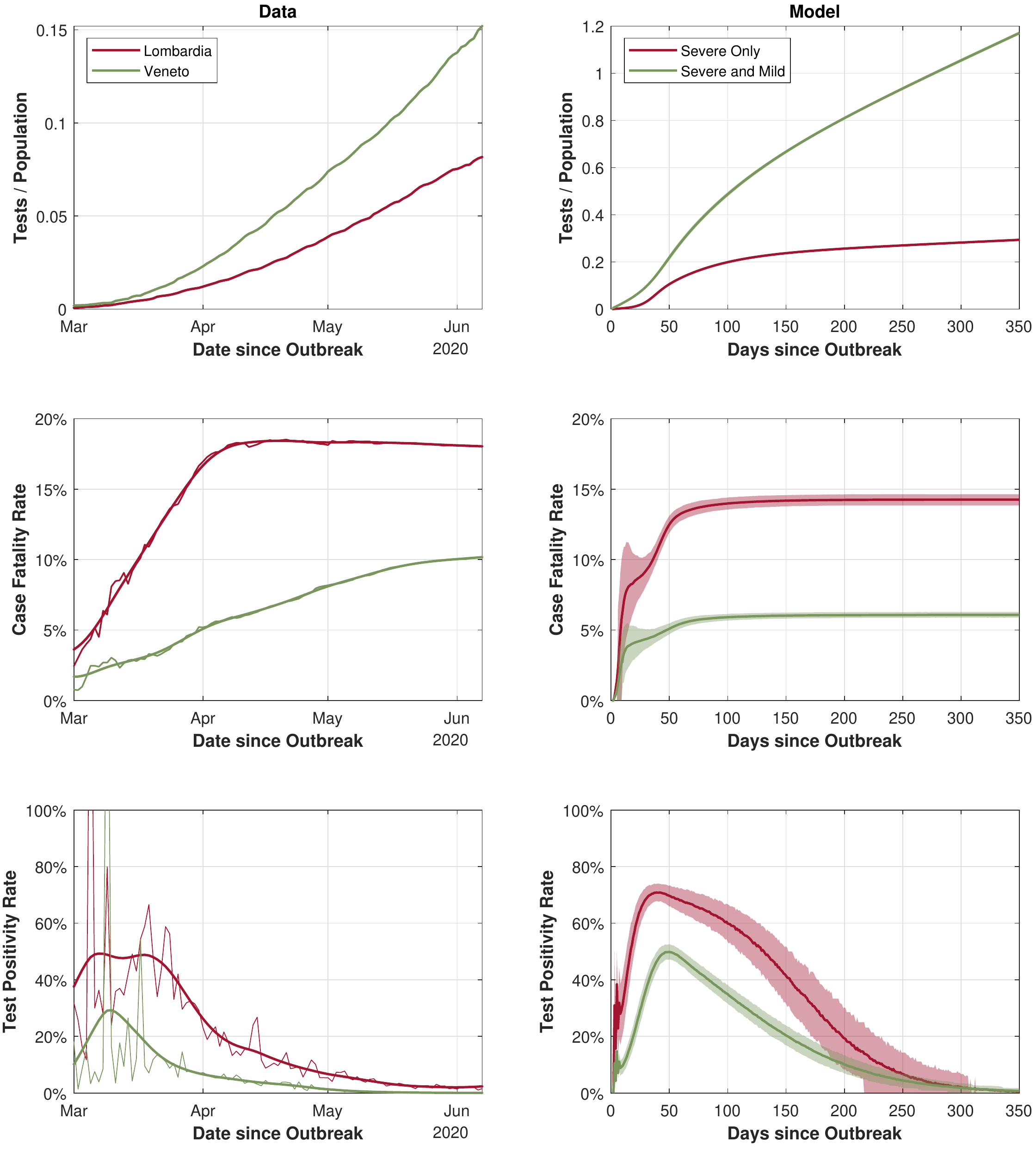}}
\caption{Validation of Model's Testing Policies with Data from Northern Italy \\
\scriptsize \textbf{Source:} Protezione Civile. Data are smoothed with an HP-Filter with smoothing parameter set to 200.}
\label{fig:Model_Validation}
\end{figure}

The results are presented in \autoref{fig:Model_Validation}. The top row of the figure reports the number of total tests performed over time, scaled by population, and serves as a sanity check. A higher level of per-capita testing in Veneto confirms the narrative that the region enacted a wider testing policy than Lombardy, and the  model is able to reproduce the same fact. \\

The middle row looks at the case fatality rate. Under a wider testing policy, the case fatality rate decreases since individuals with a lower probability of dying are included in the case count. This is generated by the model and confirmed in the data.\\

\vfill
\newpage

Finally, the bottom row shows the test positivity rate. The test positivity rate is generally expected to decrease with a wider testing policy, since the probability of being infected is usually increasing in the severity of symptoms. Under a wider testing policy, it becomes harder to find infected individuals and the test positivity rate falls. The model can reproduce this fact and match the data. Furthermore, the model is also able to replicate the dynamic evolution of the test positivity rate observed in the data.\footnote{Standard epidemiological models struggle to generate the dynamic behavior of the test positivity rate observed in the data. In fact, with symptoms-based testing, standard models would generate a constant positivity rate equal to one (assuming that the testing technology is precise). The ability of the proposed model to match the data is due to the presence of a stationary confounding disease. The latter makes detection of epidemic infections hard when there are few of them,  resulting in a low test positivity rate. But when epidemic infections peak, the test positivity rate rises.}

\newpage
\section{The Testing Multiplier}
\label{Section:Testing_Multiplier}
Given that the ultimate goal of this paper is to understand the impact of increased testing on economic activity, I conveniently introduce the notion of the `testing multiplier' which summarizes the average effect on economic activity of an additional dollar spent on testing.

\subsection{Definitions}
Let's assume that the government mandates a constant daily capacity $\textcolor{NYUcolor}{\bar{T}}$ for testing of mild symptomatic and asymptomatic individuals.\footnote{The government policy does not have to be constant over time, but assuming so is a way to restrict the space of possible government's actions and simplify the comparison of alternative policies.} Let's define cumulative output over the course of one-year since the epidemic outbreak for a generic testing policy $\textcolor{NYUcolor}{\bar{T}}$ as follows:
\[ \textcolor{NYUcolor}{\mathcal{Y}(\bar{T})} = \sum_{t = 1}^T Y_t \quad \text{with} \quad \{T_t^{NS} = \bar{T} \}_{t=1}^{T} \]
where $\textcolor{NYUcolor}{Y_t}$ is daily GDP. Let's also define the direct cumulative cost of this testing policy:
\[ \textcolor{NYUcolor}{\mathcal{E}(\bar{T})} = \sum_{t = 1}^T c_T \cdot T_t \quad \text{with} \quad \{T_t^{NS} = \bar{T} \}_{t=1}^{T} \]
where $\textcolor{NYUcolor}{T_t}$ is the total number of tests performed at time $\textcolor{NYUcolor}{t}$. Notice that both $ \textcolor{NYUcolor}{\mathcal{Y}(\bar{T})} $ and $\textcolor{NYUcolor}{ \mathcal{E}(\bar{T})}$ are random variables, since the evolution of the epidemic is random. It is then possible to define the testing multiplier for output:
\begin{align*}
\textcolor{NYUcolor}{\text{GDP-Multiplier}} &= \frac{\mathcal{Y}(\bar{T}) - \mathcal{Y}(0)}{\mathcal{E}(\bar{T}) - \mathcal{E}(0)}
\end{align*}
which is an \textit{average multiplier} because it summarizes the effect on GDP of an additional dollar spent on testing \textit{relative} to the case where there is no additional spending on testing mandated by the government. Also, notice that the multiplier is a random variable itself.\\

Similarly, one can also define the testing multiplier for the budget surplus:
\begin{align*}
\textcolor{NYUcolor}{\text{Surplus-Multiplier}} &= \frac{\mathcal{B}(\bar{T}) - \mathcal{B}(0)}{\mathcal{E}(\bar{T}) - \mathcal{E}(0)}
\end{align*}
where $\textcolor{NYUcolor}{\mathcal{B}(\bar{T})}$ represents the cumulative budget-surplus over the course of roughly one-year:
\[ \textcolor{NYUcolor}{\mathcal{B}(\bar{T})} = - \sum_{t = 1}^T Def_t \quad \text{with} \quad \{T_t^{NS} = \bar{T} \}_{t=1}^{T} \]
where $\textcolor{NYUcolor}{Def_t}$ is daily deficit. Similarly to the GDP-Multiplier, the Surplus-Multiplier summarizes the effect on the budget of an additional dollar spent on testing. When the Surplus-Multiplier is positive, an additional dollar spent on testing not only does not add to the deficit, but it even improves it. In the model, this could (and does) happen because testing expenditure reduces the fall in tax revenues and curbs the rise in costly medical treatments. When the Surplus-Multiplier is zero, an additional dollar spent on testing leaves public finances untouched, meaning that testing expenditure fully repays itself. When the Surplus-Multiplier is negative, but less than $1$ in absolute value, an additional dollar spent on testing adds to the deficit, but less than one-for one, implying that testing partially repays itself.\\

\subsection{The Multiplier is Positive for the Baseline Parameterization...}
\begin{figure}[H]
\centerline{\includegraphics[scale=0.6, angle=0]{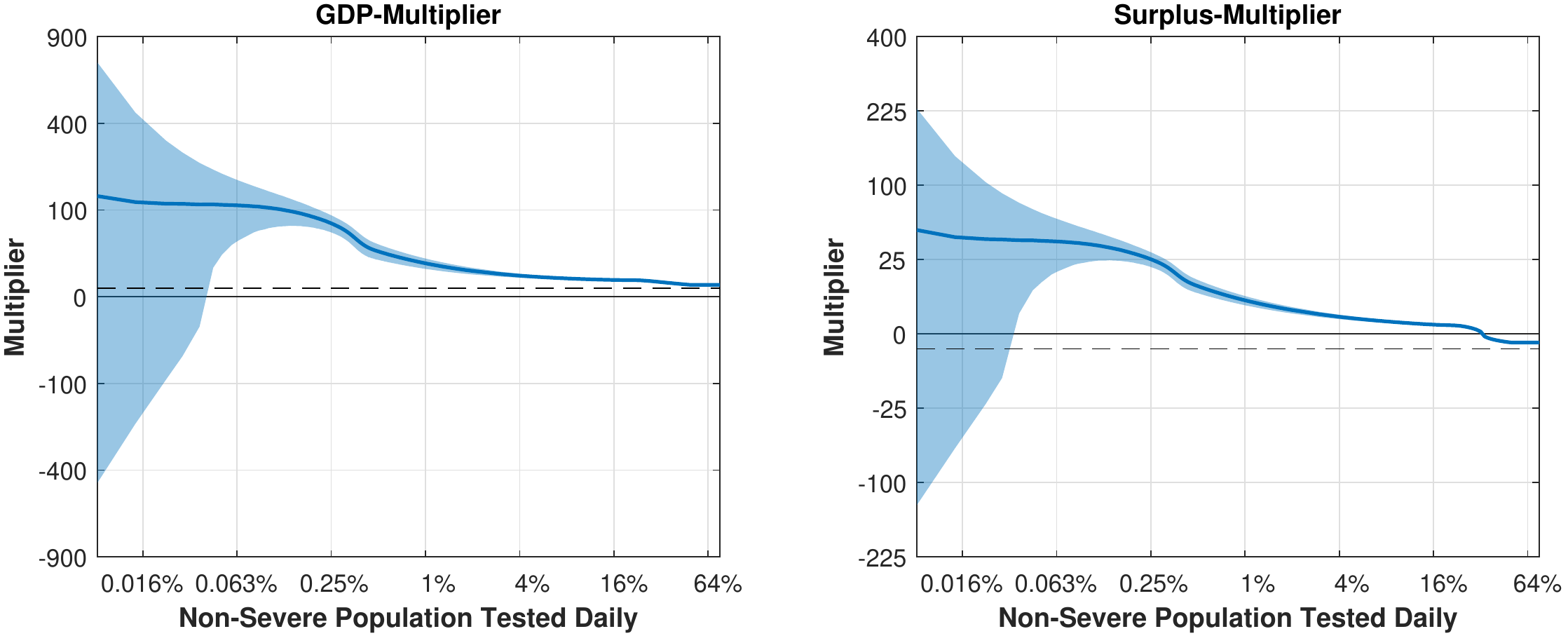}}
\caption{The Testing Multiplier under the Baseline Parameterization}
\label{fig:Multiplier_Baseline}
\end{figure}

\autoref{fig:Multiplier_Baseline} shows the GDP-Multiplier and Surplus-Multiplier for the baseline parameterization. The GDP-Multiplier is on average always positive and above one. Since luck plays an important role  at  low testing levels, however,  there are realizations of the process where the multiplier takes negative values.\footnote{Given the highly non-linear nature of the testing multiplier, I report it on a non-linear scale. The x-axis is on a $log_2$ scale, while the y-axis on a square-root scale.} With a negative multiplier, additional testing causes GDP to fall.\\

The Surplus-Multiplier is positive for most testing levels, although it turns negative at very high ones. In any case, it is always greater than minus one, meaning that testing at least partly pays for itself at all testing levels. Similarly to the multiplier for output, luck plays an important role at low testing levels.\\

\begin{figure}[H]
\centerline{\includegraphics[scale=0.55, angle=0]{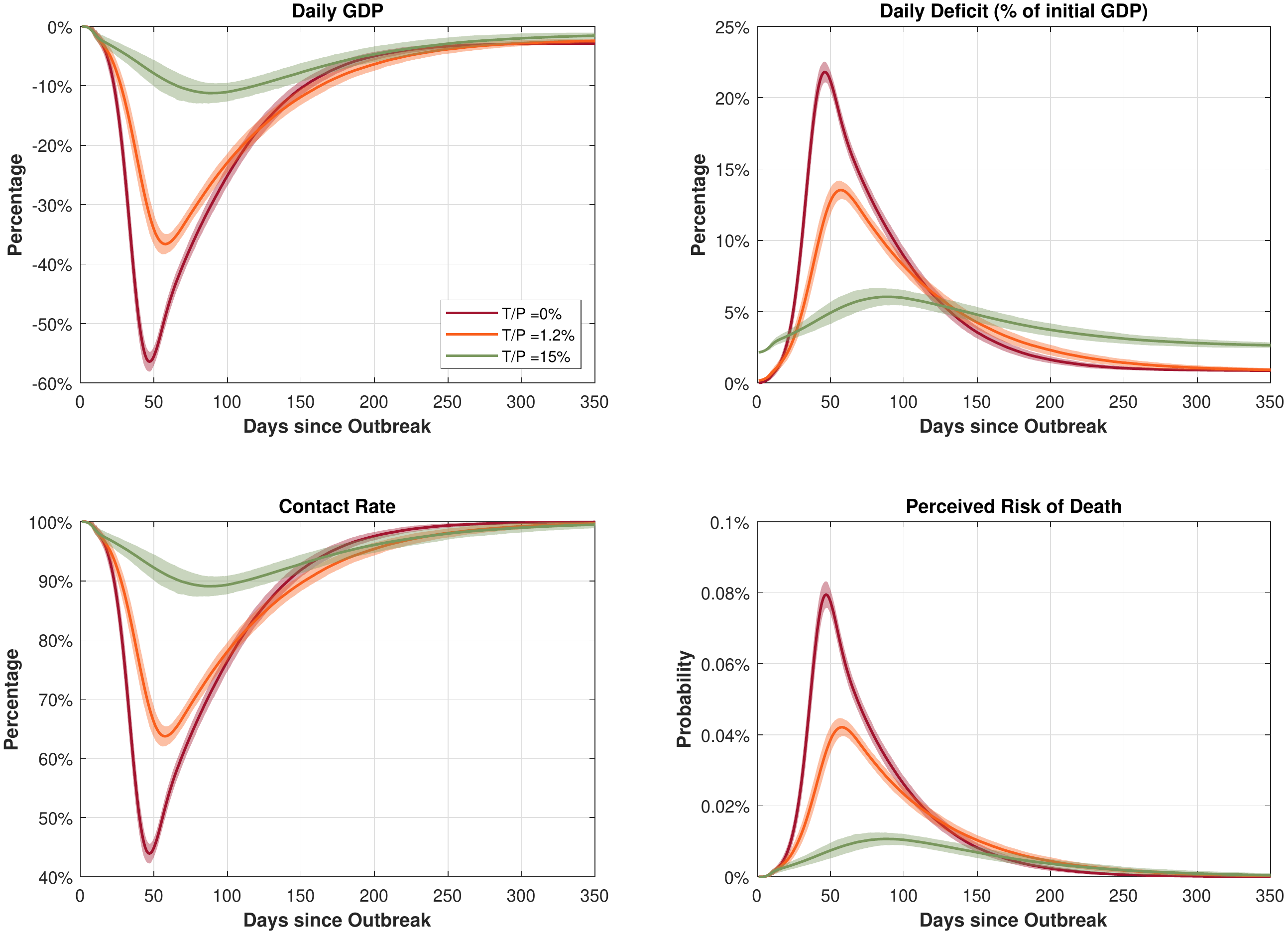}}
\caption{Economic Dynamics under Different Testing Levels}
\label{fig:Multipliers_channels_main_one}
\end{figure}

While the testing multiplier is a great summary of the effects of increased testing on the economy, it is not immediate to understand the channels through which it operates. \autoref{fig:Multipliers_channels_main_one} helps clarify what happens by illustrating the dynamic evolution of the disease under three arbitrary  testing levels: low (in red), medium (in orange), and high (in green). Under the low testing level, the overall perceived  risk of death is  highest, which results in lower labor supply, thus higher GDP loss and deficit increase. By expanding testing more and more, the government slows down epidemic transmission thanks to isolation of the infected and behavioral responses. This, in turn, reduces agents' perceived risk,  thereby mitigating the fall in GDP and curbing the rising deficit. \\

\begin{figure}[H]
\centerline{\includegraphics[scale=0.55, angle=0]{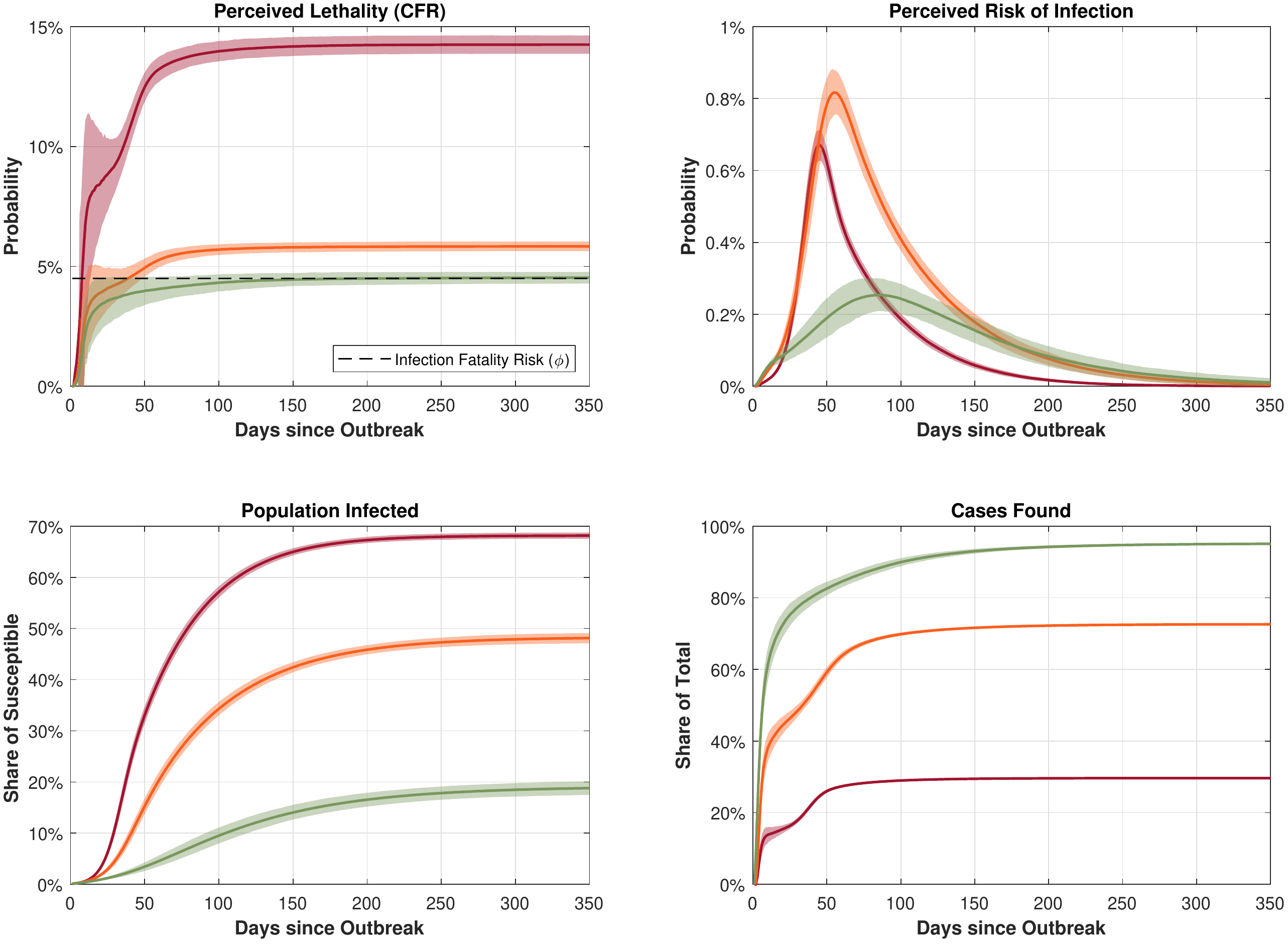}}
\caption{Perceptions under Different Testing Levels}
\label{fig:Multipliers_channels_main_two}
\end{figure}

What happens to the perceived risk of death is illustrated in \autoref{fig:Multipliers_channels_main_two}.  As previously explained, because of symptoms-based testing policies who prioritize testing according to the severity of symptoms,  individuals with a much lower risk of death are tested when the government mandates additional testing capacity. As a result, the case fatality rate falls reducing the perceived lethality of the disease, as shown in the top-left panel.\footnote{With more and more testing, the case fatality rate keeps falling to the point where it converges to the true infection fatality risk.}\\

What happens to the perceived infection risk is more complicated. Whenever the government decides to expand testing, the overall number of true latent infections fall, but the share that is detected rises. Which of these two forces prevail is not obvious. In \autoref{fig:Multipliers_channels_main_two}, for example, the perceived infection risk rises when ones moves from low to medium testing, but falls when one moves from medium to high testing. \\

In this simulation, even when additional testing increases the perceived  risk of infection, the fall in the case fatality rate prevails so that the overall perceived death risk falls.\\

\subsection{... But Can Be Negative for Alternative Diseases}
The testing multiplier is a complicated object which does not need to be positive on average.  To illustrate this point, I introduce other three influenza-like diseases, each one of them departing from the coronavirus-disease of the baseline parameterization under a specific aspect, as summarized in \autoref{Table:Parameterization_Others}. All other parameters of the model stay untouched.\\

 Disease B is `unstoppable' because its transmission coefficient  is so high that moderate levels of testing and isolation might not be enough to slow down its spread. Disease C is `less-lethal' because its infection fatality risk for individuals who develop severe symptoms is lower than in the baseline. This implies that there is little room to reduce the perceived lethality of the disease with additional testing.  Finally, disease D is `never-ending' because the length of infection is longer than in the baseline. This implies that an infected individual remains contagious for more days, increasing the probability of infection for others.  \\

\begin{table}[H]
\begin{small}
\begin{center}
\begin{tabular}{c|cccc}
  \hline \hline
\text{}                & \textbf{\begin{tabular}[c]{@{}c@{}}Disease A\\ Baseline\end{tabular}} & \textbf{\begin{tabular}[c]{@{}c@{}}Disease B\\ ``Unstoppable''\end{tabular}} & \textbf{\begin{tabular}[c]{@{}c@{}}Disease C\\ ``Less-Lethal''\end{tabular}} & \textbf{\begin{tabular}[c]{@{}c@{}}Disease D\\ ``Never-Ending''\end{tabular}} \\ \hline
$\textcolor{NYUcolor}{\beta}$     & $0.275$                                                               & $\textcolor{NYUcolor}{0.475}$                                                & $0.275$                                                                    & $0.275$                                                                       \\
$\textcolor{NYUcolor}{\phi_s}$    & $0.15$                                                                & $0.15$                                                                       & $\textcolor{NYUcolor}{0.01}$                                               & $0.15$                                                                        \\
$\textcolor{NYUcolor}{\tilde{k}}$ & $5$                                                                   & $5$                                                                          & $5$                                                                        & $\textcolor{NYUcolor}{20}$                                                    \\
$\textcolor{NYUcolor}{\tilde{q}}$ & $11$                                                                  & $11$                                                                         & $11$                                                                       & $\textcolor{NYUcolor}{26}$  \\ \hline \hline
\end{tabular}
\caption{Parameterization of Alternative Influenza-Like Diseases}
\label{Table:Parameterization_Others}
\end{center}
\end{small}
\end{table}

\begin{figure}[H]
\centerline{\includegraphics[scale=0.55, angle=0]{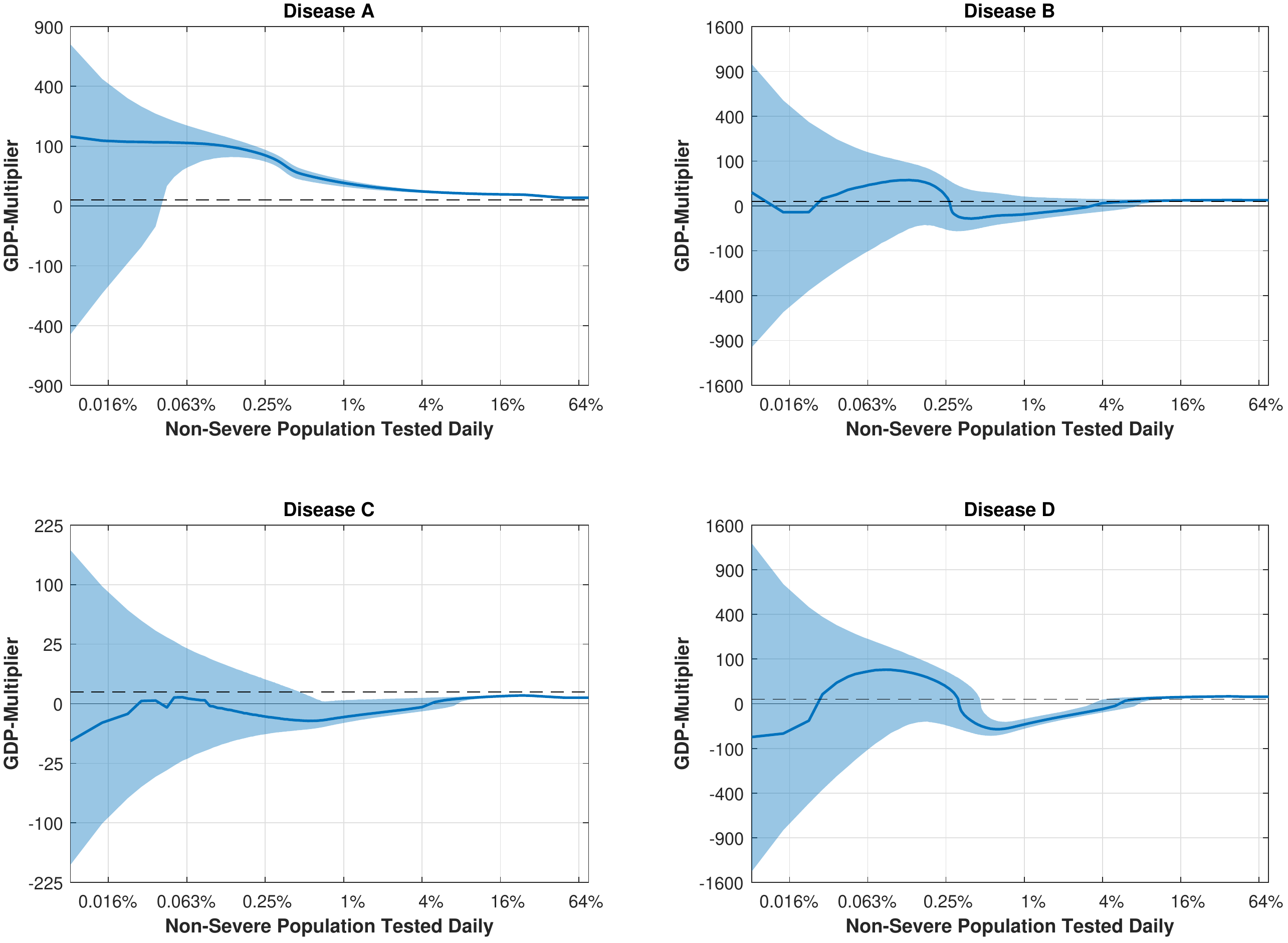}}
\caption{The GDP-Multiplier for Alternative Diseases}
\label{fig:Multiplier_Other_Diseases}
\end{figure}

\autoref{fig:Multiplier_Other_Diseases} shows the GDP-multiplier across these alternative diseases, while the results for the Surplus-Multiplier can be found in \autoref{Appendix:Additional_Results}. \\

Across all diseases, the GDP-multiplier is on average positive when a sizeable share of the population is tested every day, but not necessarily otherwise. The multiplier becomes negative  when additional testing increases the perceived risk of death, leading to a fall in economic activity.
The main reason why this happens is that, at a small scale, additional testing fails to contain the epidemic and results in a higher number of detected cases, increasing the perceived  risk of infection.\\

\begin{figure}[H]
\centerline{\includegraphics[scale=0.60, angle=0]{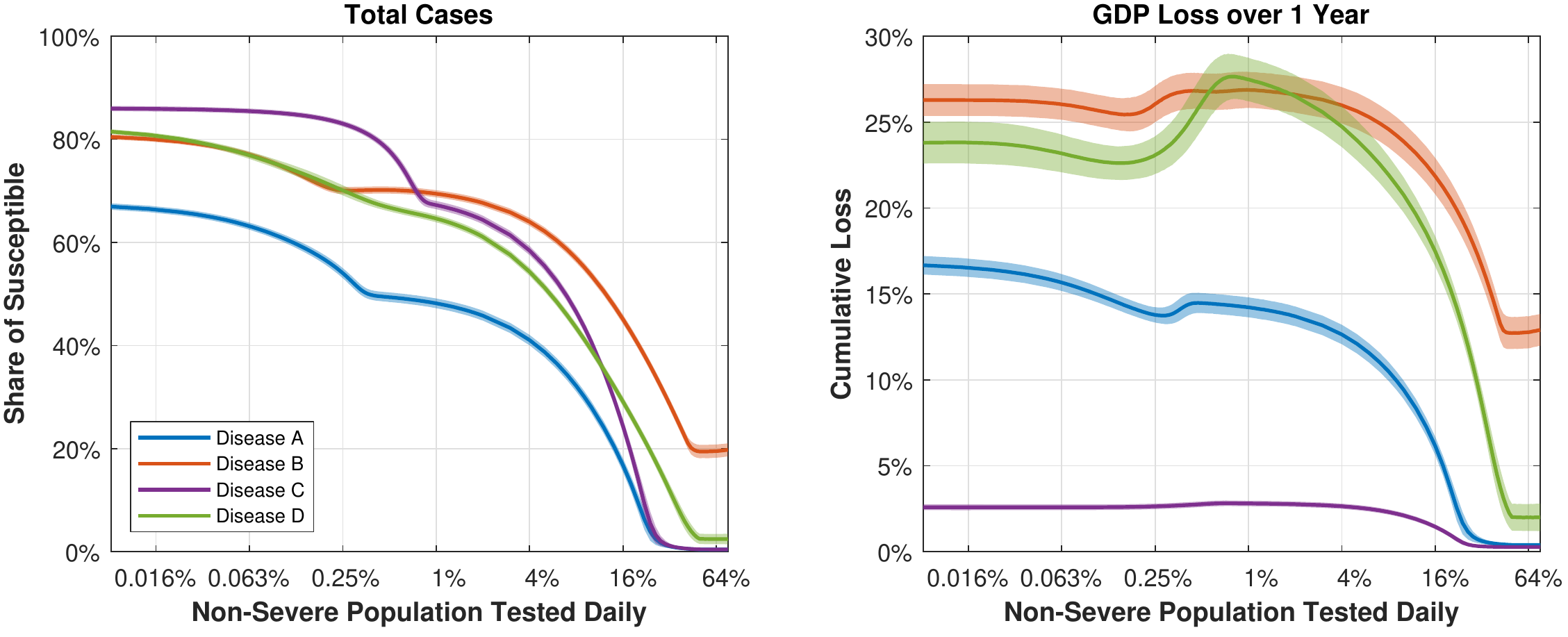}}
\caption{Economic and Public Health Outcomes for Alternative Diseases}
\label{fig:Public_health_outcomes}
\end{figure}

Interestingly, the contraction of economic activity occurs in spite of a systematic improvement in health outcomes, as \autoref{fig:Public_health_outcomes} shows.\footnote{The number of deaths is on average proportional to the number of total infections, which implies that the former would also monotonically decrease in the number of individuals tested daily.}\\

\subsection{Technological Determinants of the Testing Multiplier}
\begin{figure}[H]
\centerline{\includegraphics[scale=0.55, angle=0]{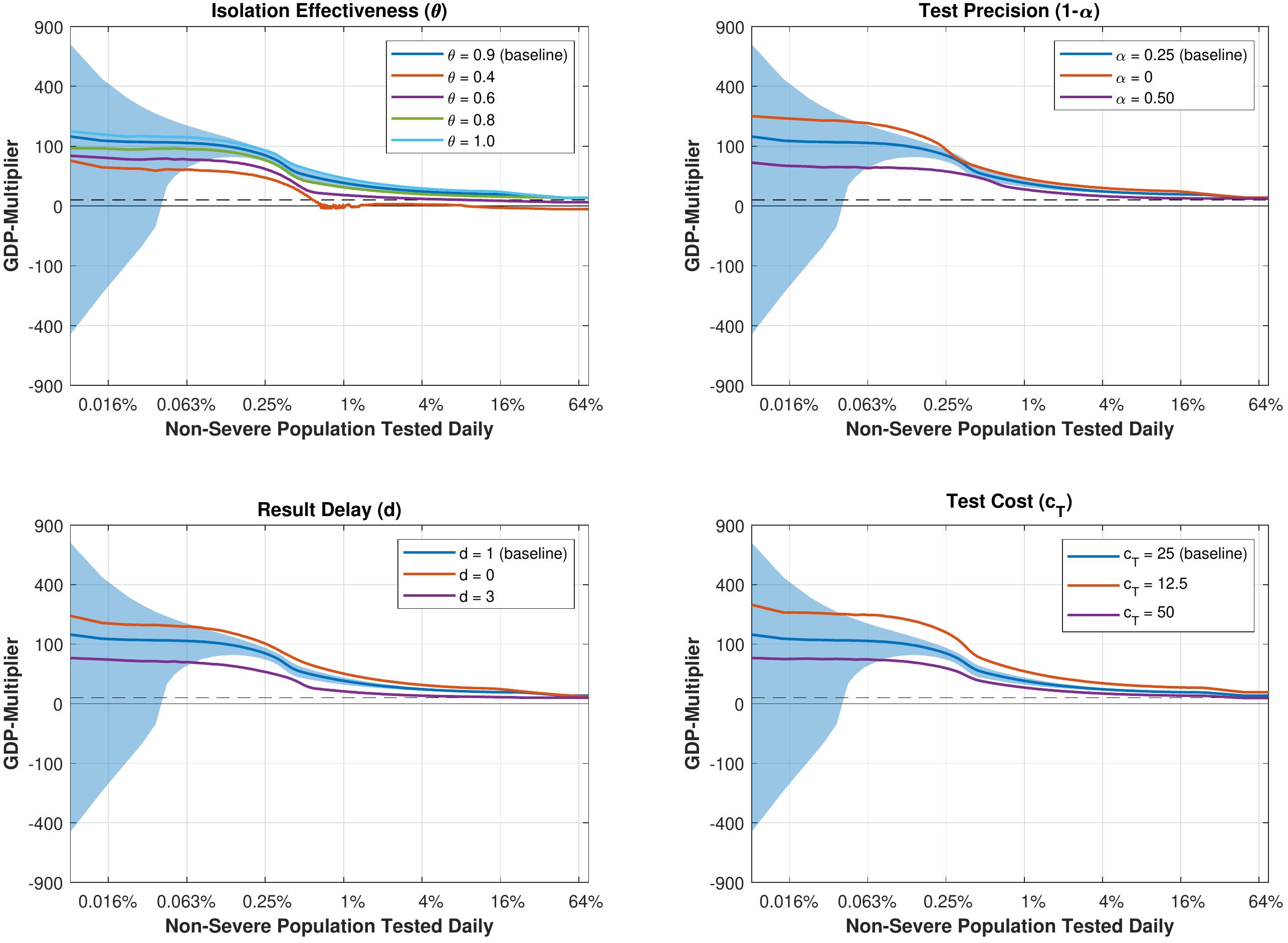}}
\caption{Technological Determinants of the GDP-Multiplier}
\label{fig:Multiplier_Technological_Characteristics}
\end{figure}

Intuitively, the testing multiplier depends on the  characteristics of the testing and isolation technology. Under the baseline parameterization, when the testing technology is more precise (i.e. it has a lower false negative rate),  cheaper, and more timely (i.e.  the lag between the day the test is administered and the day of the result is lower), the multiplier is higher. The multiplier is also higher when isolation of the infected is more rigorously enforced.\footnote{These findings are in line with what found by the existing literature. Importantly, my theoretical analysis assumes that all the tests administered share the same technological characteristics, an assumption that can be easily relaxed. For example, one could allocate very accurate tests to severe symptomatic individuals who require medical attention, and less accurate  but faster and cheaper ones to screen the rest of the population. This dual approach is advocated in \cite{mina2020rethinking}, \cite{larremore2020test}, and \cite{droste2020economic}.}   \autoref{fig:Multiplier_Technological_Characteristics} reports the results for the GDP-Multiplier, while the results for the Surplus-Multiplier can be found in \autoref{fig:Multiplier_Surplus_Technological_Characteristics} in the appendix.\\

\vfill
\newpage

It is important to realize that these sensitivity results are relative to a scenario in which the multiplier is always positive. In scenarios where the multiplier takes negative values, the effect of better technology can actually be detrimental to economic activity - at least until testing reaches a scale large enough that the multiplier becomes positive. To see this, think for example about the cost of each test kit. Around a testing level where the multiplier is negative, an additional dollar spent on testing will produce higher harm to economic activity when the testing technology is cheaper, because the same dollar with translate in more testing being performed. Once again, this highlights the complex non-linearities involved in the analysis.\\

\subsection{An Alternative Specification of Beliefs}
\label{Section:Alternative_Beliefs}
\begin{figure}[H]
\centerline{\includegraphics[scale=0.60, angle=0]{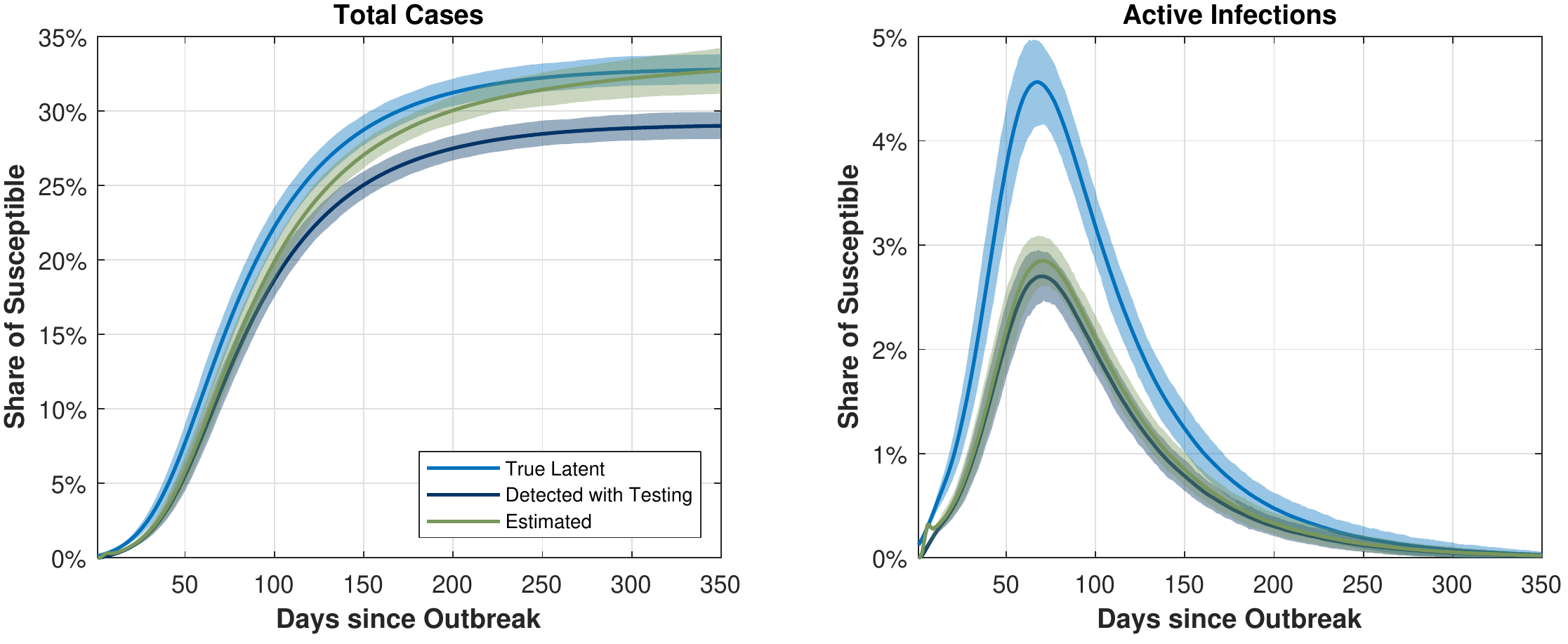}}
\caption{Alternative Specification of Beliefs}
\label{fig:Alternative_Beliefs}
\end{figure}

The insight that additional testing has the potential to increase risk perceptions is more general than it may appear. To show this, I twist the original specification of beliefs in the following way. Agents are assumed to learn about the true lethality of the disease over time as follows:
\[ \textcolor{NYUcolor}{\text{Perceived Lethality}_t} = (1 - \lambda_t) \cdot CFR_t + \lambda_t \cdot \phi  \]
where $\textcolor{NYUcolor}{\lambda_t} = \frac{t}{T}$, where $\textcolor{NYUcolor}{T}$ is the time horizon considered. The perceived lethality of the disease is therefore an average between the case fatality rate and the true infection fatality risk, where the weight of the latter linearly increases over time.\footnote{This `exogenous learning' is meant to  capture the idea that agents gradually learn about the true lethality of the disease over time from sources other than testing.} Agents then use the total number of deaths from the epidemic disease to construct an estimate of the total number of cases:
\[ \textcolor{NYUcolor}{\hat{C}_t} = \frac{D_t}{\text{Perceived Lethality}_t}  \]
and compare this estimate with the detected number of cases in order to construct an ascertainment bias factor:
\[ \textcolor{NYUcolor}{\text{Ascertainment Bias}_t} = \frac{\hat{C}_t}{C_t}  \]
which provides an estimate of the degree to which testing under-estimates the total number of infections. Finally, they estimate the number of currently active infections by scaling up the detected number of active infections:
\[ \textcolor{NYUcolor}{\hat{I}_t} = I_t \times \text{Ascertainment Bias}_t   \]
Finally, they form their perceived risk of death as in the original specification:
\[ \textcolor{NYUcolor}{\chi_t} = \text{Perceived Lethality}_t \ \ \times \ \  \beta \cdot \frac{\hat{I}_t}{P_t} \]
The key property of this specification is that, irrespective of testing, agents correctly learn over time the total number of infections. Yet, they still fail to learn the number of active infections in real-time, exactly as in the original specification. \autoref{fig:Alternative_Beliefs} illustrates this point for the baseline parameterization when a sizeable share (namely $8\%$) of the non-severe population is tested daily:\\

Since agents need to rely on testing data to form their perceptions of infection risk, the testing multiplier can still be negative, as reported in \autoref{fig:Multiplier_Alternative_Beliefs} and \autoref{fig:Multiplier_Alternative_Beliefs_Surplus} in the appendix.

\newpage
\section{Age-Heterogeneity and Risk Perceptions}
\label{Section:Heterogeneity}
In this section, I explore the aggregate effects of heterogeneous risk perceptions across age groups. To this end, I divide the population into two groups: the young and the old. I then  calibrate the model to the U.S. and  SARS-CoV-2, which features sharp age-heterogeneity in the infection fatality risk. In turn, this implies that the two groups are subject to a different risk of dying from the epidemic disease. \\

I  consider two extreme scenarios. In the first, the government releases only testing data that are aggregated across age-groups - as it often happens during epidemic outbreaks. Because of this, I assume that the two groups share the same perceptions of   risk. In the second scenario,  the government provides them with disaggregated testing  data, i.e. group-specific data on cases, deaths and so on. For simplicity, I assume that agents still share the same perceived infection risk, but they are now able to construct age-specific case fatality rates. As a result, risk perceptions are now different across the two age groups.\\

While the assumption that heterogeneous individuals could  share the same risk perceptions is certainly a stretch, it provides a useful thought-experiment to assess the importance of heterogeneous perceptions.  Referring to the U.S. response to the SARS-CoV-2 outbreak in 2020, Jay Bhattacharya, Professor of Medicine at Stanford University, points out that  \textit{``[...] a major public health message that we failed at is describing the [...] age-gradient in the risk. Older people think that they are at lower risk than they actually are, and younger people think they are at higher risk than they actually are. I think that is an enormous public health mistake''}.\footnote{The full interview can be watched at \href{https://youtu.be/2tsUTAWBJ9M}{https://youtu.be/2tsUTAWBJ9M}. The quote can be found at minute 24:20.}  \\

This section proceeds as follows. First, I extend the model to introduce age-heterogeneity. Then, I calibrate it to the U.S. and SARS-CoV-2 and analyze the economic and health outcomes under heterogeneous vs homogeneous risk perceptions across age groups.\\

\subsection{Heterogeneous-Agent Framework}
This section generalizes the homogeneous population model by introducing ex-ante heterogeneous groups. The generalized model nests the homogeneous case when $\textcolor{NYUcolor}{G = 1}$ or when the various groups are parameterized to be identical. In what follows, I will assume $\textcolor{NYUcolor}{G=2}$ and that initial population is comprised by two groups, young and old:
\begin{align*}
   \textcolor{NYUcolor}{P_0} &= P_0^y + P_0^o  \\
   \textcolor{NYUcolor}{P^y_0} &= \omega^y \cdot P_0  \\
   \textcolor{NYUcolor}{P^o_0} &= (1 - \omega^y) \cdot P_0
\end{align*}
where $\textcolor{NYUcolor}{\omega^y}$ is the share of young agents in the initial period. In any time period, new true latent epidemic infections for each group are given by:
\begin{align*}
  \textcolor{NYUcolor}{\Delta C_{t+1}^{y*}} &\sim Binomial(X_t^{y*}, IR_t^{y*}) \\
  \textcolor{NYUcolor}{\Delta C_{t+1}^{o*}} &\sim Binomial(X_t^{o*}, IR_t^{o*})
\end{align*}
where $\textcolor{NYUcolor}{X_t^{g*}}$ is the number of susceptible individuals in group $\textcolor{NYUcolor}{g}$, and  $\textcolor{NYUcolor}{IR_t^{g*}}$ is the latent infection risk for group $\textcolor{NYUcolor}{g}$. The two groups interact with each other and these interactions will determine the group-specific infection risk as follows:\footnote{This specification is similar in spirit to \cite{acemoglu2020multi}, and to what is used in the epidemiological literature with heterogeneous age groups. See for example \cite{mistry2020inferring}.}
\begin{align*}
  \textcolor{NYUcolor}{IR_{t}^{y*}} &= \beta \times \Bigg[\rho^{yy}_t \times \frac{ I_{t-1}^{y*} - \theta \cdot I_{t-1}^{y}}{P_{t-1}^{y} - \theta \cdot I_{t-1}^{y}}  + \rho^{yo}_t \times \frac{ I_{t-1}^{o*} - \theta \cdot I_{t-1}^{o}}{P_{t-1}^{o} - \theta \cdot I_{t-1}^{o}}\Bigg] \\
  \textcolor{NYUcolor}{IR_{t}^{o*}} &= \beta \times \Bigg[\rho^{yo}_t \times \frac{ I_{t-1}^{y*} - \theta \cdot I_{t-1}^{y}}{P_{t-1}^{y} - \theta \cdot I_{t-1}^{y}}  + \rho^{oo}_t \times \frac{ I_{t-1}^{o*} - \theta \cdot I_{t-1}^{o}}{P_{t-1}^{o} - \theta \cdot I_{t-1}^{o}}\Bigg]
\end{align*}
which assumes that the infection risk depends on the transmission coefficient (which is assumed to be homogeneous across groups), on the number of interactions within and between groups, and the probability of meeting an infected individual within each group. Pre-epidemic contact rates between groups are given by:
\[ \textcolor{NYUcolor}{\rho_0} = \begin{bmatrix}
\rho^{yy}_0 & \rho^{yo}_0 \\
\rho^{oy}_0 & \rho^{oo}_0
\end{bmatrix} \]
where the rows of the matrix sum to $1$, and each entry represents the share of contacts that a group entertains with another. Importantly, as the epidemic unfolds, behavioral responses make the matrix of contact rates become endogenous as follows:
\begin{align*}
  \textcolor{NYUcolor}{\rho^{yy}_t} &= \rho_0^{yy} \cdot \Big[ \pi \cdot \bar{N}^y_t + (1 - \pi) \cdot \bar{L}^y_t   \Big] \\
  \textcolor{NYUcolor}{\rho^{oo}_t} &= \rho_0^{oo} \cdot \Big[ \pi \cdot \bar{N}^o_t + (1 - \pi) \cdot \bar{L}^o_t   \Big] \\
  \textcolor{NYUcolor}{\rho^{yo}_t} &= \rho_0^{yo} \cdot \Big[ \pi \cdot \bar{N}_t + (1 - \pi) \cdot \bar{L}_t   \Big] \\
  \textcolor{NYUcolor}{\rho^{oy}_t} &= \rho_0^{oy} \cdot \Big[ \pi \cdot \bar{N}_t + (1 - \pi) \cdot \bar{L}_t   \Big]
\end{align*}
where $\textcolor{NYUcolor}{\rho_t^{gg'}}$ is the contact rate between group $\textcolor{NYUcolor}{g}$ and $\textcolor{NYUcolor}{g'}$, $\textcolor{NYUcolor}{\bar{N}^g_t}$ is average labor supply in group $\textcolor{NYUcolor}{g}$, and $\textcolor{NYUcolor}{\bar{L}^g_t}$ is average leisure in group $\textcolor{NYUcolor}{g}$, $\textcolor{NYUcolor}{\bar{N}_t}$ is average labor supply in the population, and $\textcolor{NYUcolor}{\bar{L}_t}$ is average leisure in the population. The underlying assumption  is that interactions within a group depend on labor supply and leisure of that group, while between-group interactions are a population-weighted average of labor supply and leisure. As a result, the model features a reduced-form infection externality between groups.\footnote{The infection externality arises because  the labor supply and enjoyment of leisure of one group affects the number of interactions of the other, and thus their infection risk.} Labor supply for a generic individual $\textcolor{NYUcolor}{j}$ in group $\textcolor{NYUcolor}{g}$ is given by:
\[ \textcolor{NYUcolor}{n_t(j, g)} =
\begin{cases}
   n_0 \cdot (1 + \chi^g_t)^{-\varepsilon_n} & \text{if $j$ is alive, without symptoms and not isolated} \\
  (1 - \theta) \cdot n_0 & \text{if $j$ is alive, with no/mild symptoms and  isolated} \\
  0 & \text{if $j$ is dead or has severe symptoms from any disease}
\end{cases}
\]
where the crucial difference is that  $\textcolor{NYUcolor}{\chi^g_t}$ is the perceived risk of death for group $\textcolor{NYUcolor}{g}$. Leisure is symmetrically given. Daily production is given by:
\[ \textcolor{NYUcolor}{y_t(j, g)} = A^g \cdot n_t(j, g) \]
where $\textcolor{NYUcolor}{A^g}$ is the productivity of an individual belonging to group $\textcolor{NYUcolor}{g}$. As a result of these assumptions, the contribution of each group to GDP will depend on its size, average labor supply and productivity.\\

To illustrate the importance of risk perceptions across groups, I will consider two scenarios. In the first scenario, the government releases only aggregate testing data and  the perceived death risk is given by:
\[ \textcolor{NYUcolor}{\chi_t} = \underbrace{\frac{D_t}{C_t}}_{\text{Aggregate Case Fatality Rate}} \times \underbrace{\beta \cdot \frac{I_t}{P_t}}_{\text{Perceived Infection Risk}} \]
and is the same across groups, i.e. $\textcolor{NYUcolor}{\chi_t = \chi_t^y = \chi_t^o}$. In the second scenario, the government releases both group-specific and aggregate testing  data, and the perceived risk of death for each group is given by:
\begin{align*}
  \textcolor{NYUcolor}{\chi^y_t} &= \underbrace{\frac{D^y_t}{C^y_t}}_{\text{Youngs' Case Fatality Rate}} \times \underbrace{\beta \cdot \frac{I_t}{P_t}}_{\text{Perceived Infection Risk}} \\
  \textcolor{NYUcolor}{\chi^o_t} &= \underbrace{\frac{D^o_t}{C^o_t}}_{\text{Olds' Case Fatality Rate}} \times \underbrace{\beta \cdot \frac{I_t}{P_t}}_{\text{Perceived Infection Risk}}
\end{align*}
Again, the assumption that the perceived infection risk is the same across groups is made for simplicity, and  isolates the importance of heterogeneity in the perceived lethality of the disease.\\

\subsection{A SARS-CoV-2 Calibration}
\label{Section:COVID19}
I calibrate the model to the U.S. and SARS-CoV-2, and the parameters are reported in \autoref{Table:COVID19_parameterization}. Several parameters are omitted since they are the same as in the baseline parameterization of \autoref{Table:Parameterization}.\\

\begin{table}[H]
\begin{center}
\begin{small}
\begin{tabular}{lccc}
\hline \hline
\multicolumn{4}{c}{\textbf{SARS-CoV-2}}                                                                                                                                           \\ \hline
\multicolumn{1}{c}{\textbf{Description}}                       & \textbf{Parameter}                               & \textbf{Value} & \textbf{Source}                              \\
\textit{Young Population Share}                                & $\textcolor{NYUcolor}{\omega^y}$                 & $0.835$        & \textcolor{NYUcolor}{U.S. Census (2019)}                                  \\
\textit{Initial Young Infections}                              & $\textcolor{NYUcolor}{C_0^{y*}}$                 & $42$           &                                              \\
\textit{Initial Old Infections}                                & $\textcolor{NYUcolor}{C_0^{o*}}$                 & $8$            &                                              \\
\textit{Infection Fatality Risk for Severe Young}              & $\textcolor{NYUcolor}{\phi^y_s}$                 & $0.005$        & \cite{levin2020assessing}               \\
\textit{Infection Fatality Risk for Severe Old}                & $\textcolor{NYUcolor}{\phi^o_s}$                 & $0.248$         & \cite{levin2020assessing}               \\
\textit{Share of Asymptomatic Infections}                      & $\textcolor{NYUcolor}{a}$                        & $0.4$          & \cite{oran2020asymptomatic, jung2020clinical} \\
\textit{Share of Mild Infections}                              & $\textcolor{NYUcolor}{m}$                        & $0.3$          & \cite{oran2020asymptomatic, jung2020clinical} \\
\textit{Share of Severe Infections}                            & $\textcolor{NYUcolor}{s}$                        & $0.3$          & \cite{oran2020asymptomatic, jung2020clinical} \\
\textit{Implied Infection Fatality Risk for Young}              & $\textcolor{NYUcolor}{\phi^y}$                 & $0.002$        & \cite{levin2020assessing}               \\
\textit{Implied Infection Fatality Risk for Old}                & $\textcolor{NYUcolor}{\phi^o}$                 & $0.074$         & \cite{levin2020assessing}               \\
\textit{Incubation Period}                                     & $\textcolor{NYUcolor}{p}$                        & $7$            & \cite{qin2020incubation}    \\
\textit{Symptoms to Death}                   & $\textcolor{NYUcolor}{\tilde{k}_s}$              & $10$           & \textcolor{NYUcolor}{U.S. CDC (2020)}  \\
\textit{Symptoms to Recovery for Severe}     & $\textcolor{NYUcolor}{\tilde{q}_s}$              & $10$           & \textcolor{NYUcolor}{U.S. CDC (2020)}  \\
\textit{Symptoms to Recovery for Non-Severe} & $\textcolor{NYUcolor}{\tilde{q}_s, \tilde{q}_a}$ & $7$           & \textcolor{NYUcolor}{U.S. CDC (2020)}  \\
\textit{Transmission Coefficient}                                     & $\textcolor{NYUcolor}{\beta}$                    & $0.20$         &                                              \\
\textit{Daily Productivity for Young}                          & $\textcolor{NYUcolor}{A^y}$                      & $230$          & \textcolor{NYUcolor}{U.S. BLS (2019)} \\
\textit{Daily Productivity for Old}                            & $\textcolor{NYUcolor}{A^o}$                      & $46$           &  \textcolor{NYUcolor}{U.S. BLS (2019)} \\
\textit{Pre-Epidemic Contact Rate Young-Young}                 & $\textcolor{NYUcolor}{\rho_0^{yy}}$              & $0.95$         & \cite{prem2017projecting}   \\
\textit{Pre-Epidemic Contact Rate Young-Old}                   & $\textcolor{NYUcolor}{\rho_0^{yo}}$              & $0.05$         & \cite{prem2017projecting}   \\
\textit{Pre-Epidemic Contact Rate Old-Old}                     & $\textcolor{NYUcolor}{\rho_0^{oo}}$              & $0.24$         & \cite{prem2017projecting}   \\
\textit{Pre-Epidemic Contact Rate Old-Young}                   & $\textcolor{NYUcolor}{\rho_0^{oy}}$              & $0.76$         & \cite{prem2017projecting}   \\
\textit{Elasticity of Labor to Perceived Death Risk}                           & $\textcolor{NYUcolor}{\varepsilon_n}$              & $5000$         & \textcolor{NYUcolor}{Empirical Estimates}    \\
\textit{Elasticity of Leisure to Perceived Death Risk}                         & $\textcolor{NYUcolor}{\varepsilon_l}$              & $5000$         & \textcolor{NYUcolor}{Empirical Estimates}    \\  \hline \hline
\end{tabular}
\caption{A SARS-CoV-2 Calibration}
\label{Table:COVID19_parameterization}
\end{small}
\end{center}
\end{table}

The population share of young individuals matches that of individuals younger than $65$ years old in the U.S. Census. I assume that $40\%$ of infected individuals are asymptomatic following \cite{oran2020asymptomatic}, and I assume that the remaining $60\%$ of infections are equally split between mild and severe symptoms. I assume that age does not correlate with the severity of symptoms, following \cite{jung2020clinical}. I keep the assumption that only severe symptomatic individuals can die, and  I choose their infection fatality risks so that the implied unconditional infection fatality risk of each group matches the estimates in \cite{levin2020assessing}. The average incubation period comes from \cite{qin2020incubation}. I set the other lags so that they match the period over which an individual can infect others, as reported by the U.S. Centers for Disease Control and Prevention in October 2020. Specifically, non-severe individuals are on average infectious for $14$ days, while severe individuals stop to be infectious on average after $10$ days since the onset of symptoms.\\

Daily productivity is chosen as follows. I start from daily GDP per-capita in the U.S. and allocate it to each group taking into account that young individuals comprise roughly $84\%$ of the population and that their employment rate is roughly $5$ times higher than that of the old. Contact rates for the U.S. are aggregated from the dataset produced by \cite{prem2017projecting}, which I first correct for non-reciprocity using the pairwise correction suggested by \cite{arregui2018projecting}.\footnote{\cite{acemoglu2020multi} perform a similar calibration exercise for the contact rate matrix, using instead the dataset from \cite{klepac2020contacts}.} Finally, the elasticities to the perceived risk of dying are set equal to the value estimated in the empirical part of the paper.\footnote{I choose the raw estimate for the Google Mobility Data at the state-level since it corresponds precisely to an elasticity. Moreover, I do not pick the estimate from a specification with time fixed-effects, as that would capture a \textit{relative} elasticity, which is not the object of interest.}\\

\begin{figure}[H]
\centerline{\includegraphics[scale=0.55, angle=0]{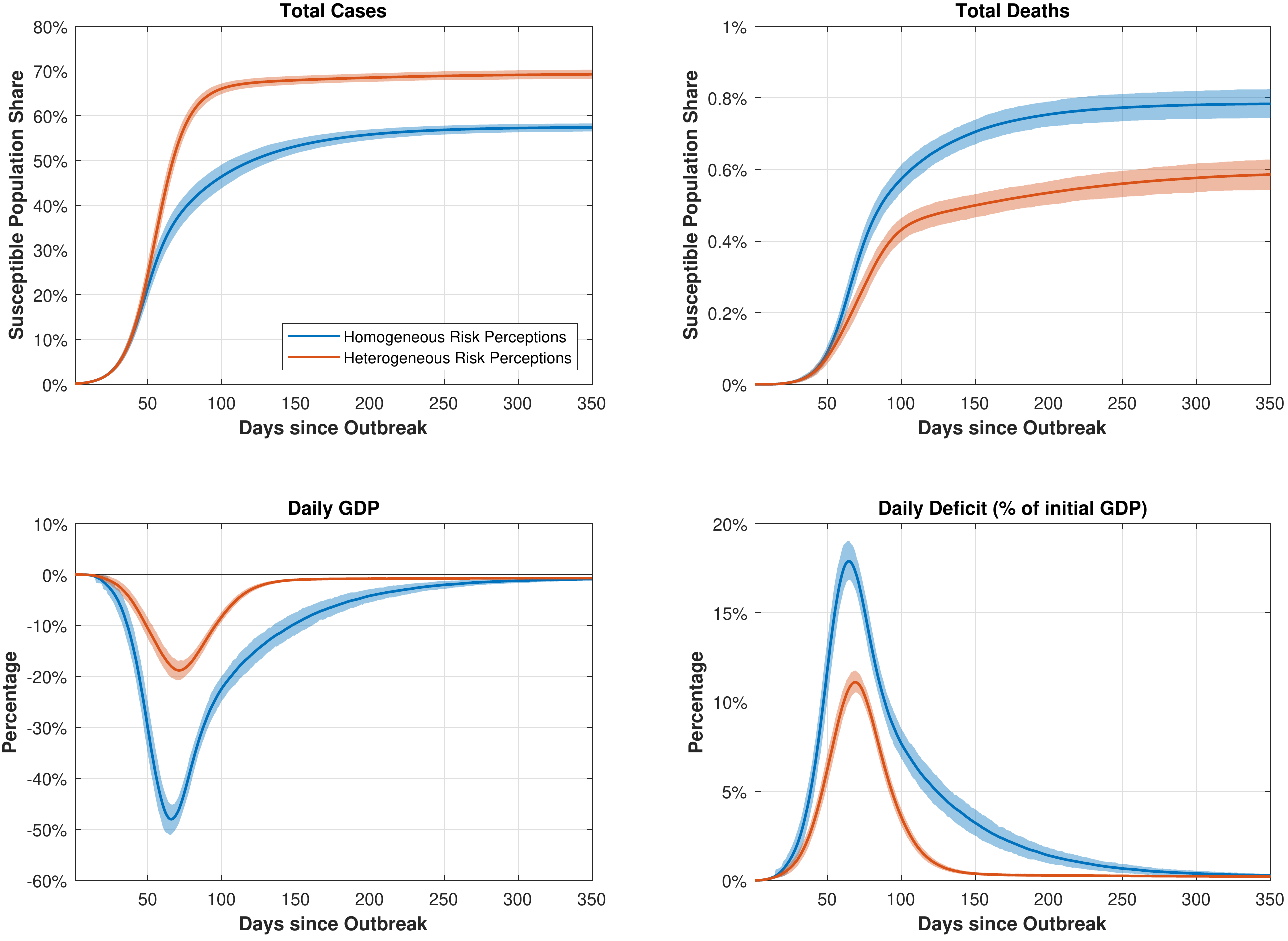}}
\caption{Economic and Health Outcomes for the  SARS-CoV-2 Calibration of the Model}
\label{fig:Aggregate_Disaggregate_COVID19}
\end{figure}

Let's consider the case in which the government does not engage in any additional testing of mild and asymptomatic individuals. \autoref{fig:Aggregate_Disaggregate_COVID19} reports the dynamic evolution of total true latent cases and deaths, GDP and public deficit when the government provides aggregate testing data (in blue) and when it provides disaggregated testing data (in orange). The figure suggests that heterogeneous risk perceptions across the two age groups result in higher total cases, but lower deaths, output losses and budget deficits.\\

\begin{figure}[H]
\centerline{\includegraphics[scale=0.55, angle=0]{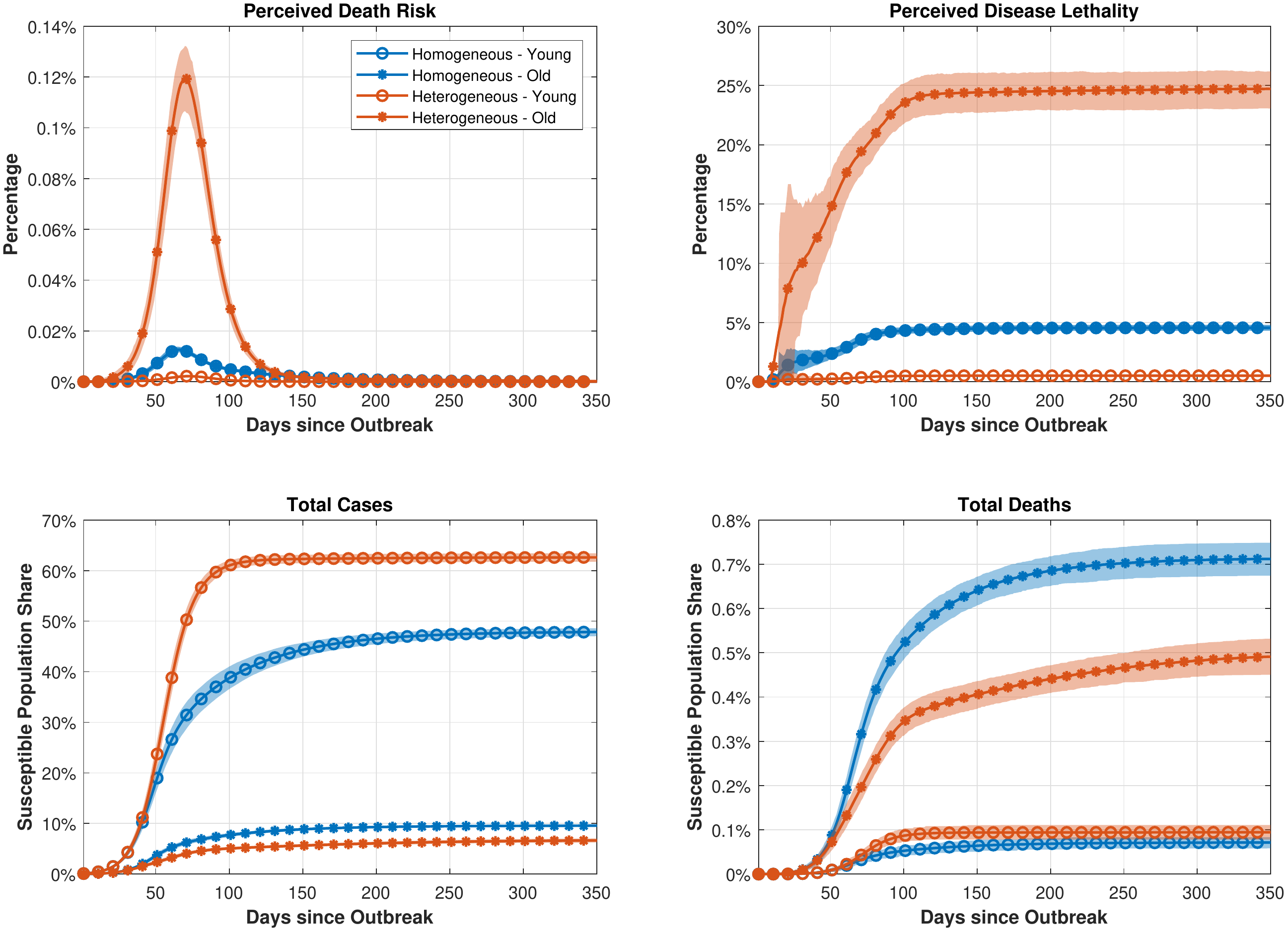}}
\caption{The Effect of Heterogeneous Perceptions on Health Outcomes}
\label{fig:Understanding_Aggregate_Disaggregate_COVID19}
\end{figure}

 \autoref{fig:Understanding_Aggregate_Disaggregate_COVID19} helps understand why. When the government provides aggregate testing data and both the young and the old share the same perceived death risk, young agents \textit{over-estimate} their true risk, while old agents \textit{under-estimate} theirs. This happens because the `aggregate' case fatality rate  estimates the average conditional infection fatality risk across the two groups, which makes the disease  appear more lethal than it actually is to the young, and less lethal to the old. This results in less total true cases, since  the young (who constitute the largest share of the population) `protect' themselves a lot, but more deaths, since the old (whose true risk of dying is higher) do not `protect' themselves enough.\\

\begin{figure}[H]
\centerline{\includegraphics[scale=0.55, angle=0]{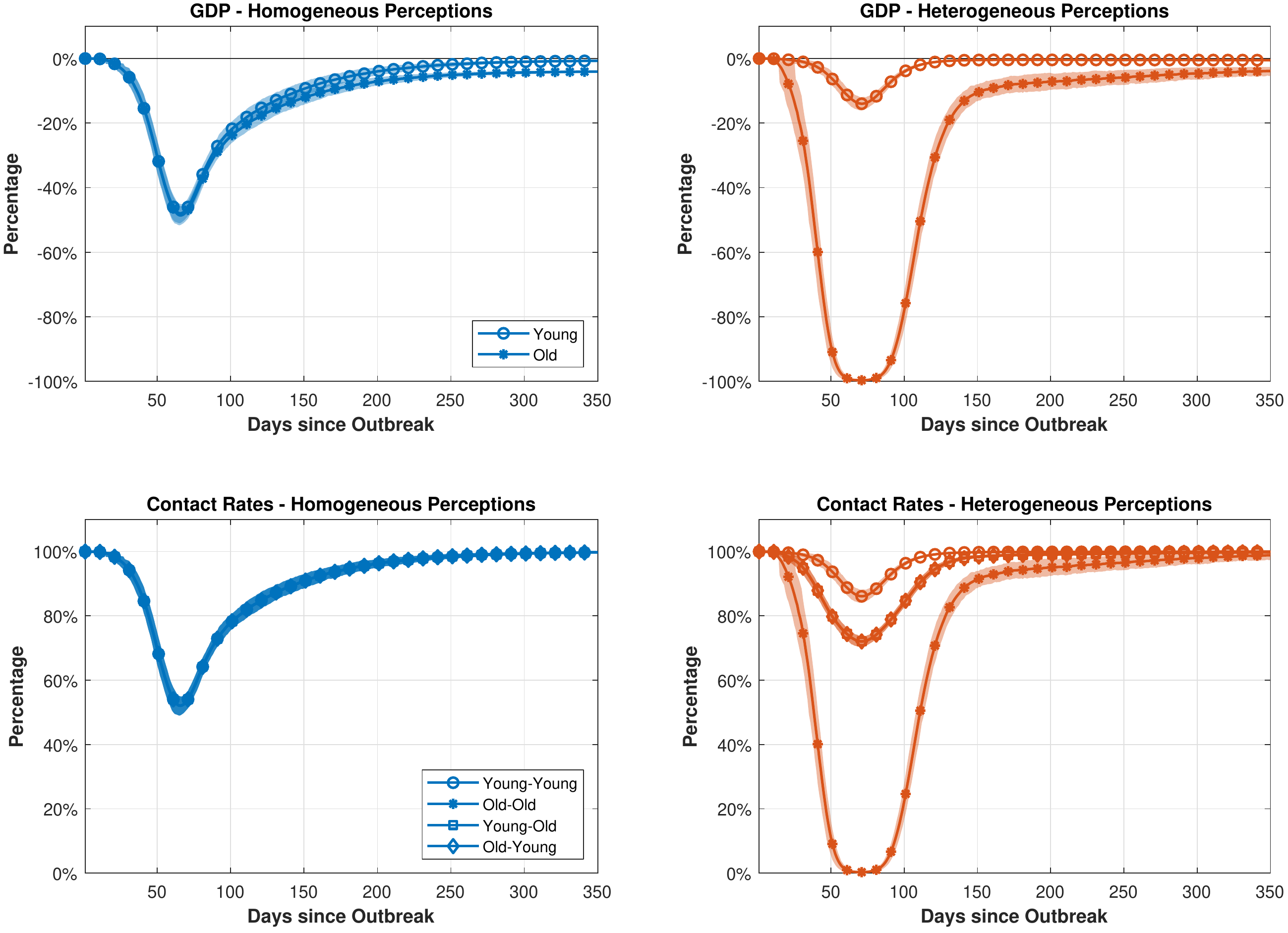}}
\caption{The Effect of Heterogeneous Perceptions on Economic Outcomes}
\label{fig:Understanding_Aggregate_Disaggregate_COVID19_two}
\end{figure}

When the government releases disaggregated  testing data, instead, each group develops a more accurate understanding of the disease. As a result, young agents fear the epidemic disease less and reduce their activity less, which results in more cases and deaths among them. The opposite happens with the old, who now fear the disease more. \autoref{fig:Understanding_Aggregate_Disaggregate_COVID19_two}  shows what happens to contact rates and production across the two groups. \\

With homogeneous risk perceptions, the fall in production and contact rates is homogeneous across groups. With heterogeneous perceptions, instead, production and interactions of the old fall sizeably, but those of the young remain  close to pre-epidemic levels. In other words, heterogeneous perceptions across agents translate into heterogeneous economic responses to the epidemic disease. Since young agents account for most of the population and are the most productive, economic activity falls more when they reduce labor supply. This explains why heterogeneous risk perceptions are associated with better economic outcomes than homogeneous perceptions in this calibration of the model.\\

Importantly, the bottom-right panel of the figure shows the reduced-form infection externality across groups that is at play in the model. Old individuals, who entertain a sizeable share of their interactions with young ones, are unable to reduce the between-group interactions as much as they can reduce their own, and the opposite happens to the young.\footnote{With a little stretch, one can think of heterogeneous risk perceptions as a way to implement an (imperfect) targeted lockdown through behavioral responses. This `endogenous' lockdown is likely to be sub-optimal because of the infection externality.}\\

The importance of heterogeneous risk perceptions generalizes to all testing levels, as clarified by \autoref{fig:Summary_Information_COVID} in the appendix. Across testing levels and relative to the case where all individuals have the same perceptions of risk,  heterogeneous risk perceptions across young and old individuals reduce GDP losses and budget deficits as much as $50\%$, and the total deaths count as much as $30\%$.\\

Importantly, the insight that - relative to homogeneous risk perceptions - heterogeneous  perceptions improve both aggregate economic and health outcomes  is not a general property of the model. In fact,  the overall effect of different risk perceptions across age-groups on aggregate economic and public health outcomes will depend on the population structure, the productivity of the various groups, the pattern of interactions among them, and the characteristics of the disease. In \autoref{Appendix:Pseudo_Spanish}, I illustrate how heterogeneous risk perceptions improve health but not economic outcomes for a disease that is more lethal for the young. \\

\begin{figure}[H]
\centerline{\includegraphics[scale=0.6, angle=0]{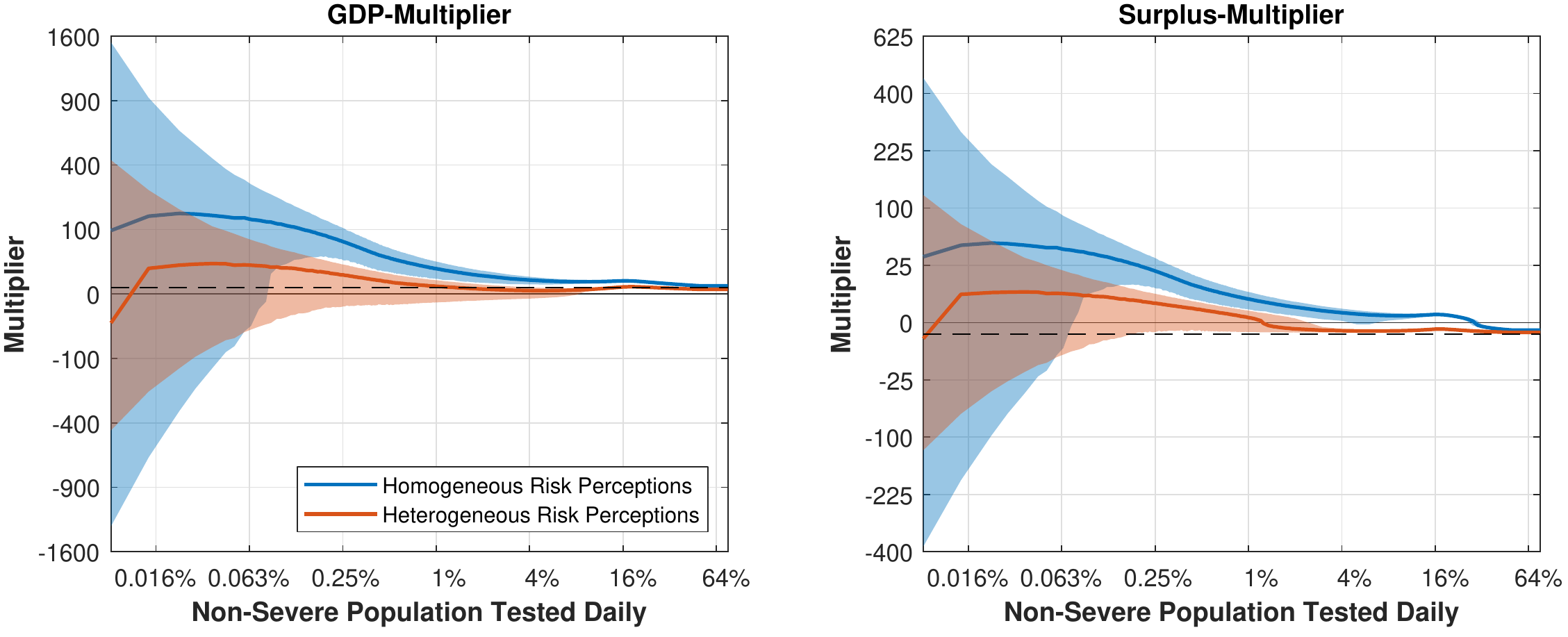}}
\caption{The Testing Multiplier for the SARS-CoV-2 Calibration}
\label{fig:Multiplier_Information_COVID}
\end{figure}

Finally, \autoref{fig:Multiplier_Information_COVID} presents the testing multiplier for the SARS-CoV-2 calibration. In both scenarios, testing appears on average beneficial to the economy and (partially) repays for itself. Interestingly, the multiplier is higher with homogeneous risk perceptions, and this happens partly because the `informational' contribution of additional testing activity is higher in this case.\\

\section{Conclusions}
In this paper, I have proposed a stochastic epidemiological model with realistic testing policies where agents  adjust their behavior in response to a measure of perceived risk that is constructed using testing data produced by the health-care system. I have then used the model to perform counterfactual policy experiments aimed at understanding the economic effects of extending testing to paucisymptomatic and asymptomatic individuals during the outbreak of a novel epidemic disease. \\

My findings suggest that more testing is beneficial to the economy and pays for itself when  performed at a large enough scale, but not necessarily otherwise. Furthermore, they suggest that in a setting with ex-ante heterogeneous agents, heterogeneous perceptions of risk across age-groups  can sizeably affect aggregate economic and public health outcomes.\\

The analysis performed is not aimed at describing any actual epidemic, and its only purpose is to provide insights on the economic effects of testing. It also suffers from several limitations. First, the key ingredient in the analysis is the way agents process information, which in turn determines their behavioral responses. The specification proposed in the model is stylized, and takes a strong stance on the source and the type of information exploited by individuals to form their risk perceptions. As such, it should  only be seen as a first attempt and  more work is needed to understand the details of individual behavior.\\

Second, the analysis abstract from many sources of heterogeneity that could potentially play a role in determining the effects of testing. For example, spatial heterogeneity creates room for geographically-targeted large-scale testing. This could permit epidemic containment without testing a sizeable share of the overall population every day. \\

Finally, the importance of heterogeneous risk perceptions across age-groups  might create incentives for the government to engage in strategic release of information, which might be internalized by the agents in the population, opening up the possibility of a complicated strategic interaction between the government and the agents in the model. This, in turn, might make the information provision by the government more or less relevant for aggregate outcomes.\\

\bibliography{Multiplier_biblio}

\newpage
\appendix
\setcounter{table}{0}
\setcounter{figure}{0}
\renewcommand{\thetable}{A\arabic{table}}
\renewcommand\thefigure{\thesection.\arabic{figure}}
\captionsetup[figure]{font=footnotesize}
\captionsetup[table]{font=footnotesize}

\section{Data Appendix}
\subsection{Data Sources}
\label{Appendix:Data_Sources}
\begin{footnotesize}
The analysis in \autoref{Section:Evidence} is performed combining data from the following data sources:
\begin{itemize}[noitemsep]
    \item \textit{USAFacts}: county-level data on cases, deaths and population
    \item \textit{COVID Tracking Project}: state-level data on cases, deaths and testing
    \item \textit{Google Mobility Report}: state-level data on workplace mobility
    \item \textit{Dallas FED's MEI}: county- and state-level mobility and engagement index
\end{itemize}
\end{footnotesize}

\subsection{Additional Regression Results}
\label{Appendix:Empirical_Additional_Results}

\begin{table}[H]
  \begin{center}
  \begin{scriptsize}
\begin{tabular}{lccccccc}
  \hline \hline
                                     & \multicolumn{3}{c}{\textbf{FED's MEI}}                                                                                                                                                     & \multicolumn{1}{l}{} & \multicolumn{3}{c}{\textbf{Google's Workplace Mobility}}                                                                                                                                   \\ \cline{2-4} \cline{6-8}
\multicolumn{1}{c}{\textbf{}}        & \begin{tabular}[c]{@{}c@{}}(1)\\ OLS\end{tabular}            & \begin{tabular}[c]{@{}c@{}}(2)\\ FE\end{tabular}             & \begin{tabular}[c]{@{}c@{}}(3)\\ FE\end{tabular}             &                      & \begin{tabular}[c]{@{}c@{}}(1)\\ OLS\end{tabular}            & \begin{tabular}[c]{@{}c@{}}(2)\\ FE\end{tabular}             & \begin{tabular}[c]{@{}c@{}}(3)\\ FE\end{tabular}             \\ \hline
\textbf{Spec \#1}                    &                                                              &                                                              &                                                              &                      &                                                              &                                                              & \multicolumn{1}{l}{}                                         \\
\textit{Death Risk ($\chi$)}         & \begin{tabular}[c]{@{}c@{}}-9294.2***\\ (983.5)\end{tabular} & \begin{tabular}[c]{@{}c@{}}-9503.8***\\ (868.9)\end{tabular} & \begin{tabular}[c]{@{}c@{}}-2887.5***\\ (487.4)\end{tabular} &                      & \begin{tabular}[c]{@{}c@{}}-5265.0***\\ (579.3)\end{tabular} & \begin{tabular}[c]{@{}c@{}}-5921.2***\\ (622.1)\end{tabular} & \begin{tabular}[c]{@{}c@{}}-1183.2***\\ (179.2)\end{tabular} \\
\textit{}                            &                                                              &                                                              &                                                              &                      &                                                              &                                                              & \multicolumn{1}{l}{}                                         \\
\textbf{Spec \#2}                    & \multicolumn{1}{l}{}                                         & \multicolumn{1}{l}{}                                         & \multicolumn{1}{l}{}                                         & \multicolumn{1}{l}{} & \multicolumn{1}{l}{}                                         & \multicolumn{1}{l}{}                                         & \multicolumn{1}{l}{}                                         \\
\textit{Lethality (CFR)}             & \begin{tabular}[c]{@{}c@{}}-4.42***\\ (0.74)\end{tabular}    & \begin{tabular}[c]{@{}c@{}}-4.59***\\ (1.01)\end{tabular}    & \begin{tabular}[c]{@{}c@{}}-0.41\\ (0.28)\end{tabular}       &                      & \begin{tabular}[c]{@{}c@{}}-2.66***\\ (0.43)\end{tabular}    & \begin{tabular}[c]{@{}c@{}}-2.94***\\ (0.63)\end{tabular}    & \begin{tabular}[c]{@{}c@{}}-0.20\\ (0.12)\end{tabular}       \\
\textit{Infection Risk (IR)}         & \begin{tabular}[c]{@{}c@{}}-160.6***\\ (44.0)\end{tabular}   & \begin{tabular}[c]{@{}c@{}}-160.5***\\ (46.8)\end{tabular}   & \begin{tabular}[c]{@{}c@{}}-88.7***\\ (15.4)\end{tabular}    &                      & \begin{tabular}[c]{@{}c@{}}-127.5***\\ (16.1)\end{tabular}   & \begin{tabular}[c]{@{}c@{}}-142.3***\\ (15.5)\end{tabular}   & \begin{tabular}[c]{@{}c@{}}-40.3***\\ (5.3)\end{tabular}     \\
                                     &                                                              &                                                              &                                                              &                      &                                                              &                                                              & \multicolumn{1}{l}{}                                         \\ \hline
\multicolumn{1}{l}{State FE}         & N                                                            & Y                                                            & Y                                                            &                      & N                                                            & Y                                                            & Y                                                            \\
\multicolumn{1}{l}{Time FE}          & N                                                            & N                                                            & Y                                                            &                      & N                                                            & N                                                            & Y                                                            \\
\multicolumn{1}{l}{Adj. $R^2$ (Spec \#1)} & 0.15                                                         & 0.17                                                         & 0.96                                                         &                      & 0.20                                                         & 0.25                                                         & 0.97                                                         \\
\multicolumn{1}{l}{Adj. $R^2$ (Spec \#2)} & 0.19                                                         & 0.21                                                         & 0.96                                                         &                      & 0.31                                                         & 0.37                                                         & 0.97                                                         \\
\multicolumn{1}{l}{Obs}              & 1530                                                         & 1530                                                         & 1530                                                         &                      & 1479                                                         & 1479                                                         & 1479                                                         \\ \hline \hline \multicolumn{8}{l}{\textbf{Notes:} Clustered standard errors at the state-level in parenthesis. *$p<0.10$, **$p<0.05$, ***$p<0.01$}
\end{tabular}
\caption{Main Regression Results at State-Level, Non-Standardized Coefficients}
\label{Appendix_A:Table1}
\end{scriptsize}
\end{center}
\end{table}


\begin{table}[H]
\begin{center}
\begin{scriptsize}
\begin{tabular}{lccccc}
  \hline \hline
                              & \multicolumn{5}{c}{\textbf{FED's MEI}}                                                                                                                                                                                                                                                                              \\ \cline{2-6}
\multicolumn{1}{c}{\textbf{}} & \begin{tabular}[c]{@{}c@{}}(1)\\ OLS\end{tabular}            & \begin{tabular}[c]{@{}c@{}}(2)\\ FE\end{tabular}             & \begin{tabular}[c]{@{}c@{}}(3)\\ FE\end{tabular}           & \begin{tabular}[c]{@{}c@{}}(4)\\ FE\end{tabular}            & \begin{tabular}[c]{@{}c@{}}(5)\\ FE\end{tabular}           \\ \hline
\textbf{Spec \#1}             &                                                              &                                                              &                                                            &                                                             &                                                            \\
\textit{Death Risk ($\chi$)}  & \begin{tabular}[c]{@{}c@{}}-1975.4***\\ (182.2)\end{tabular} & \begin{tabular}[c]{@{}c@{}}-1939.8***\\ (153.6)\end{tabular} & \begin{tabular}[c]{@{}c@{}}-910.7***\\ (78.7)\end{tabular} & \begin{tabular}[c]{@{}c@{}}-817.4***\\ (124.7)\end{tabular} & \begin{tabular}[c]{@{}c@{}}-817.4**\\ (323.3)\end{tabular} \\
\textit{}                     &                                                              &                                                              &                                                            &                                                             &                                                            \\
\textbf{Spec \#2}             & \multicolumn{1}{l}{}                                         & \multicolumn{1}{l}{}                                         & \multicolumn{1}{l}{}                                       & \multicolumn{1}{l}{}                                        & \multicolumn{1}{l}{}                                       \\
\textit{Lethality (CFR)}      & \begin{tabular}[c]{@{}c@{}}-1.04***\\ (0.07)\end{tabular}    & \begin{tabular}[c]{@{}c@{}}-1.20***\\ (0.07)\end{tabular}    & \begin{tabular}[c]{@{}c@{}}-0.07***\\ (0.02)\end{tabular}  & \begin{tabular}[c]{@{}c@{}}-0.12***\\ (0.03)\end{tabular}   & \begin{tabular}[c]{@{}c@{}}-0.12***\\ (0.03)\end{tabular}  \\
\textit{Infection Risk (IR)}  & \begin{tabular}[c]{@{}c@{}}-11.98***\\ (2.85)\end{tabular}   & \begin{tabular}[c]{@{}c@{}}-10.45***\\ (2.44)\end{tabular}   & \begin{tabular}[c]{@{}c@{}}-19.79***\\ (3.15)\end{tabular} & \begin{tabular}[c]{@{}c@{}}-14.88***\\ (3.29)\end{tabular}  & \begin{tabular}[c]{@{}c@{}}-14.88**\\ (5.97)\end{tabular}  \\
                              &                                                              &                                                              &                                                            &                                                             &                                                            \\ \hline
County FE                     & N                                                            & Y                                                            & Y                                                          & N                                                           & N                                                          \\
Time FE                       & N                                                            & N                                                            & Y                                                          & N                                                           & N                                                          \\
State-Time FE                 & N                                                            & N                                                            & N                                                          & Y                                                           & Y                                                          \\
SE Clustering                 & County                                                       & County                                                       & County                                                     & County                                                      & State                                                      \\
Adj. $R^2$ (Spec \#1)              & 0.01                                                         & 0.14                                                         & 0.90                                                       & 0.81                                                        & 0.81                                                       \\
Adj. $R^2$ (Spec \#2)              & 0.03                                                         & 0.16                                                         & 0.90                                                       & 0.81                                                        & 0.81                                                       \\
Obs                           & 90599                                                        & 90599                                                        & 90599                                                      & 90599                                                       & 90599                                                      \\ \hline \hline  \multicolumn{6}{l}{\textbf{Notes:} Clustered standard errors in parenthesis. *$p<0.10$, **$p<0.05$, ***$p<0.01$}
\end{tabular}
\caption{Main Regression Results at County-Level, Non-Standardized Coefficients}
\label{Appendix_A:Table2}
\end{scriptsize}
\end{center}
\end{table}

\newpage
\section{Recovering the Deterministic SIR Model}
\label{Appendix:Deterministic_SIR}
\begin{figure}[H]
\centerline{\includegraphics[scale=0.45, angle=0]{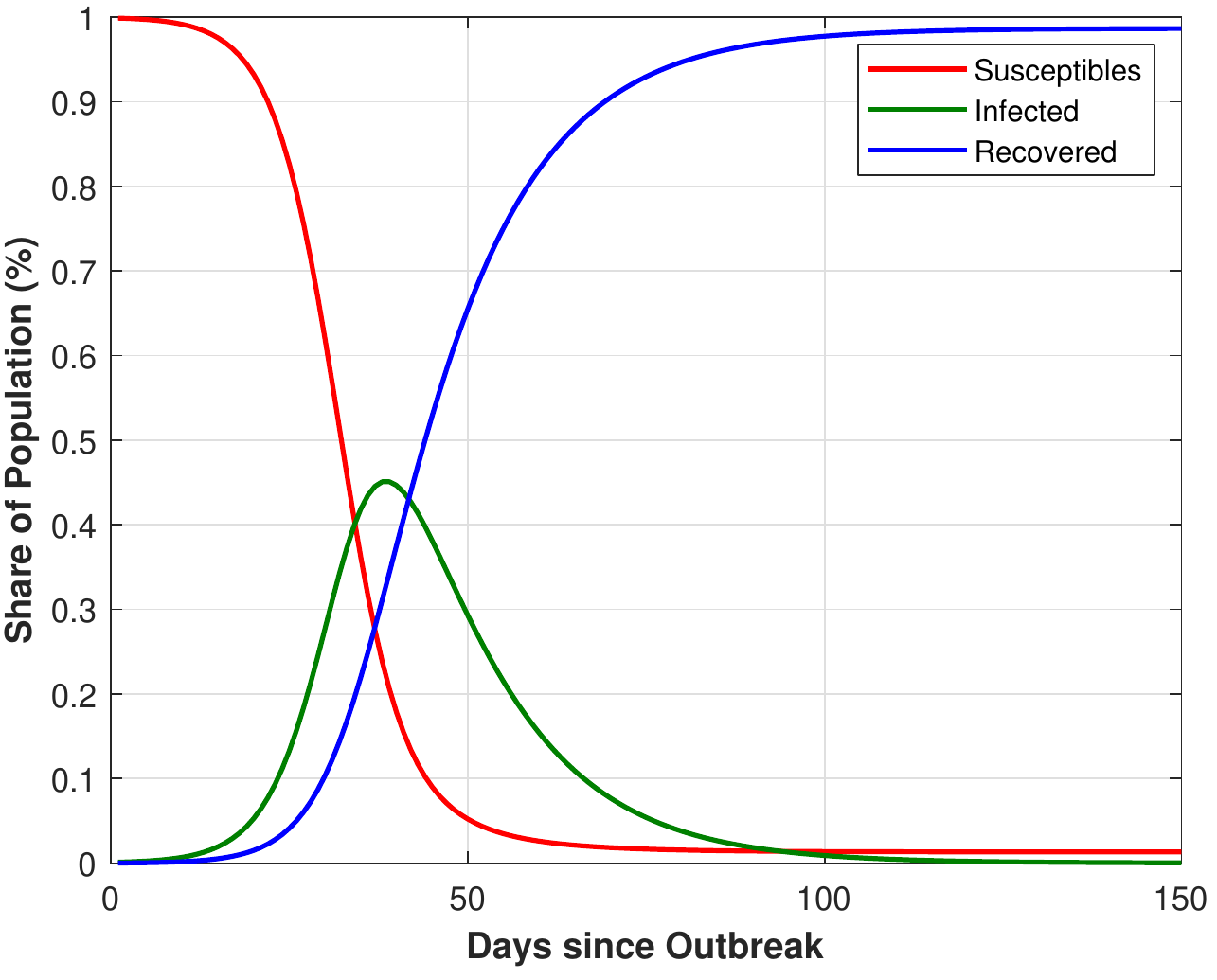}}
\caption{Recovering the SIR Textbook Model}
\label{fig:SIR_Model_Appendix}
\end{figure}

\begin{footnotesize}
It is possible to recover standard textbook epidemiological models by imposing specific restrictions to the model. In this section, I will consider the homogeneous population version of the model and show how to recover the deterministic SIR model. To this end:
\begin{itemize}[noitemsep]
  \item Eliminate the confounding disease by setting $\textcolor{NYUcolor}{\omega^f = \sigma^f = 0}$
  \item Eliminate severe and mild symptomatic states by setting $\textcolor{NYUcolor}{s = m = 0}$
  \item Eliminate the incubation period by setting $\textcolor{NYUcolor}{p(j) = 0}$
  \item Eliminate non-severe testing by setting $\textcolor{NYUcolor}{T_t^{NS} = 0}$
  \item Eliminate behavioral responses by setting $\textcolor{NYUcolor}{\varepsilon_l = \varepsilon_n = 0}$
  \item Eliminate any death risk by setting $\textcolor{NYUcolor}{\phi_s = \phi_m = \phi_a = \phi = 0}$
  \item Assume an exponential form for the time from infection to recovery, i.e. $\textcolor{NYUcolor}{q(j) \sim  Exp(q)}$
  \item Set population size to infinity, i.e. $\textcolor{NYUcolor}{P_0 \rightarrow +\infty}$
\end{itemize}
The resulting aggregate epidemic dynamics will be given by:
\begin{align*}
  \textcolor{NYUcolor}{\Delta X_{t+1}^*} &= - \beta \cdot \frac{I_t^*}{P_0} \cdot X_t^* \\
  \textcolor{NYUcolor}{\Delta I_{t+1}^*} &= \beta \cdot \frac{I_t^*}{P_0}  \cdot X_t^* - \gamma \cdot I_t^* \\
  \textcolor{NYUcolor}{\Delta R_{t+1}^*} &= - \gamma \cdot I_t^*
\end{align*}
where $\textcolor{NYUcolor}{\gamma = \frac{1}{q}}$. Then, define $\textcolor{NYUcolor}{w_t = \frac{W_t}{P_0}}$ for a generic variable $\textcolor{NYUcolor}{W_t}$, and conveniently drop the asterisk denoting latent variables. Divide both sides of each equation by $\textcolor{NYUcolor}{P_0}$ to get:
\begin{align*}
  \textcolor{NYUcolor}{\Delta x_{t+1}} &= - \beta \cdot i_t \cdot x_t \\
  \textcolor{NYUcolor}{\Delta i_{t+1}} &= \beta \cdot i_t  \cdot x_t - \gamma \cdot i_t \\
  \textcolor{NYUcolor}{\Delta r_{t+1}} &= - \gamma \cdot i_t
\end{align*}
The equations above are the same as for a deterministic SIR model in discrete time. \autoref{fig:SIR_Model_Appendix} below provides a visual representation of the aggregate epidemic dynamic under the restrictions above. In the simulation, I set $\textcolor{NYUcolor}{\beta = 0.30}$, $\textcolor{NYUcolor}{\gamma = \frac{1}{14}}$ and increase the population size to $\textcolor{NYUcolor}{P_0 = 1e6}$.\\
\end{footnotesize}

\newpage
\section{More on the Role of Behavioral Responses}
\label{Appendix:Behavioral_responses}
\begin{footnotesize}
It is insightful to dig deeper into the role of behavioral responses in the model. By construction, the model allows to control the intensity of behavioral responses, and, in \autoref{fig:DGP_baseline_noresponses} below, I show what happens with weaker ($\textcolor{NYUcolor}{\varepsilon_l = \varepsilon_n = 500}$) and stronger ($\textcolor{NYUcolor}{\varepsilon_l = \varepsilon_n = 1500}$) behavioral responses.\\
\end{footnotesize}

\begin{figure}[H]
\centerline{\includegraphics[scale=0.40, angle=0]{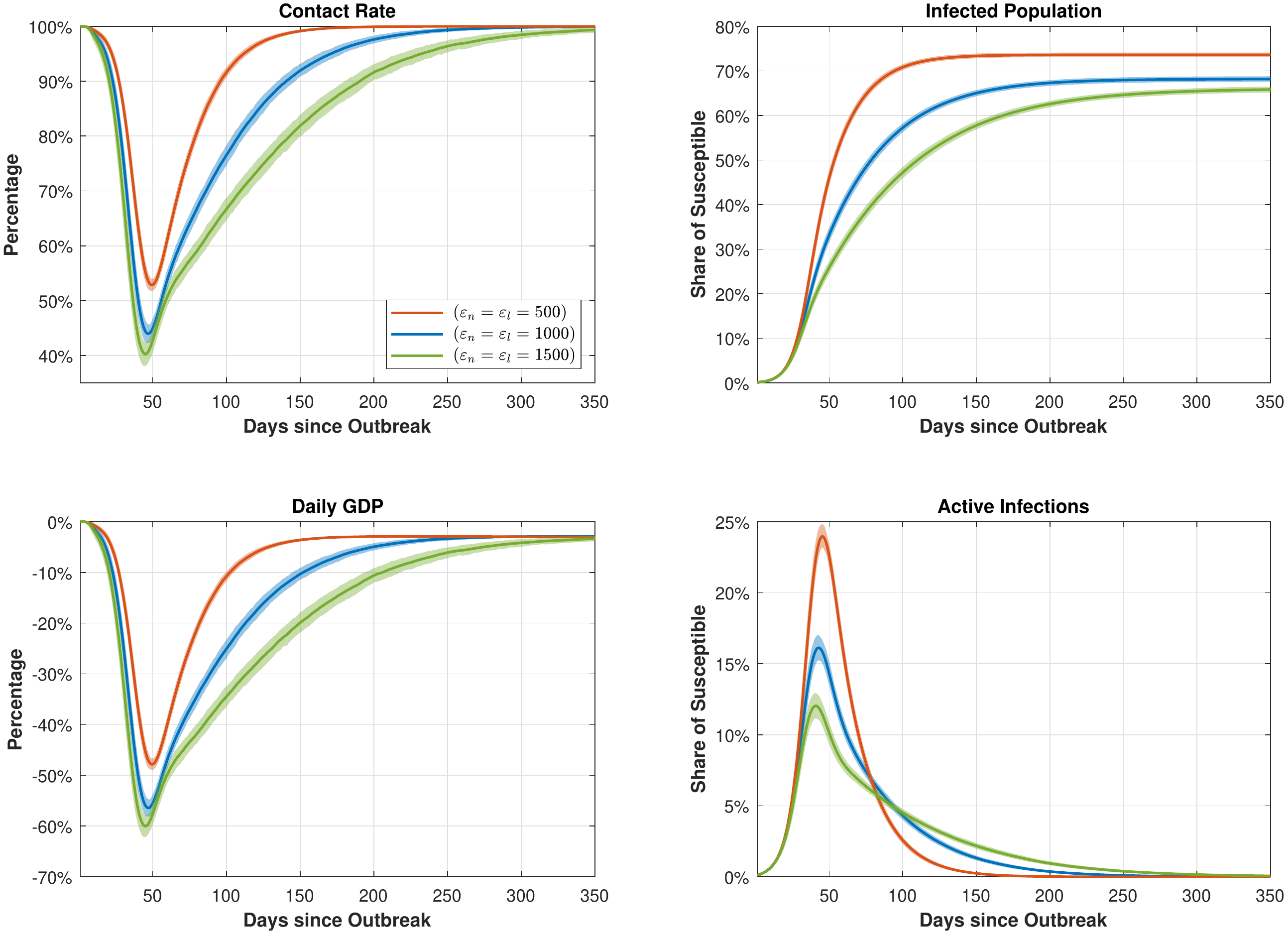}}
\caption{The Importance of Behavioral Responses}
\label{fig:DGP_baseline_noresponses}
\end{figure}
\begin{footnotesize}
Stronger behavioral responses generate a larger fall in labor supply and enjoynment of leisure for the same perceived risk of dying. This translates into a greater reduction of the interactions between agents, and, in turn, into a smaller final epidemic size. At the same time,  lower labor supply causes a sharper contraction of economic activity. Furthermore, stronger behavioral responses ``flatten the curve'' and  lengthen the horizon over which the epidemic disease naturally disappears, as shown in the bottom-right panel.\\
\end{footnotesize}
\newpage
\section{Additional Simulation Results}
\label{Appendix:Additional_Results}

\subsection{The Surplus-Multiplier for Alternative Influenza-Like Diseases}

\begin{figure}[H]
\centerline{\includegraphics[scale=0.4, angle=0]{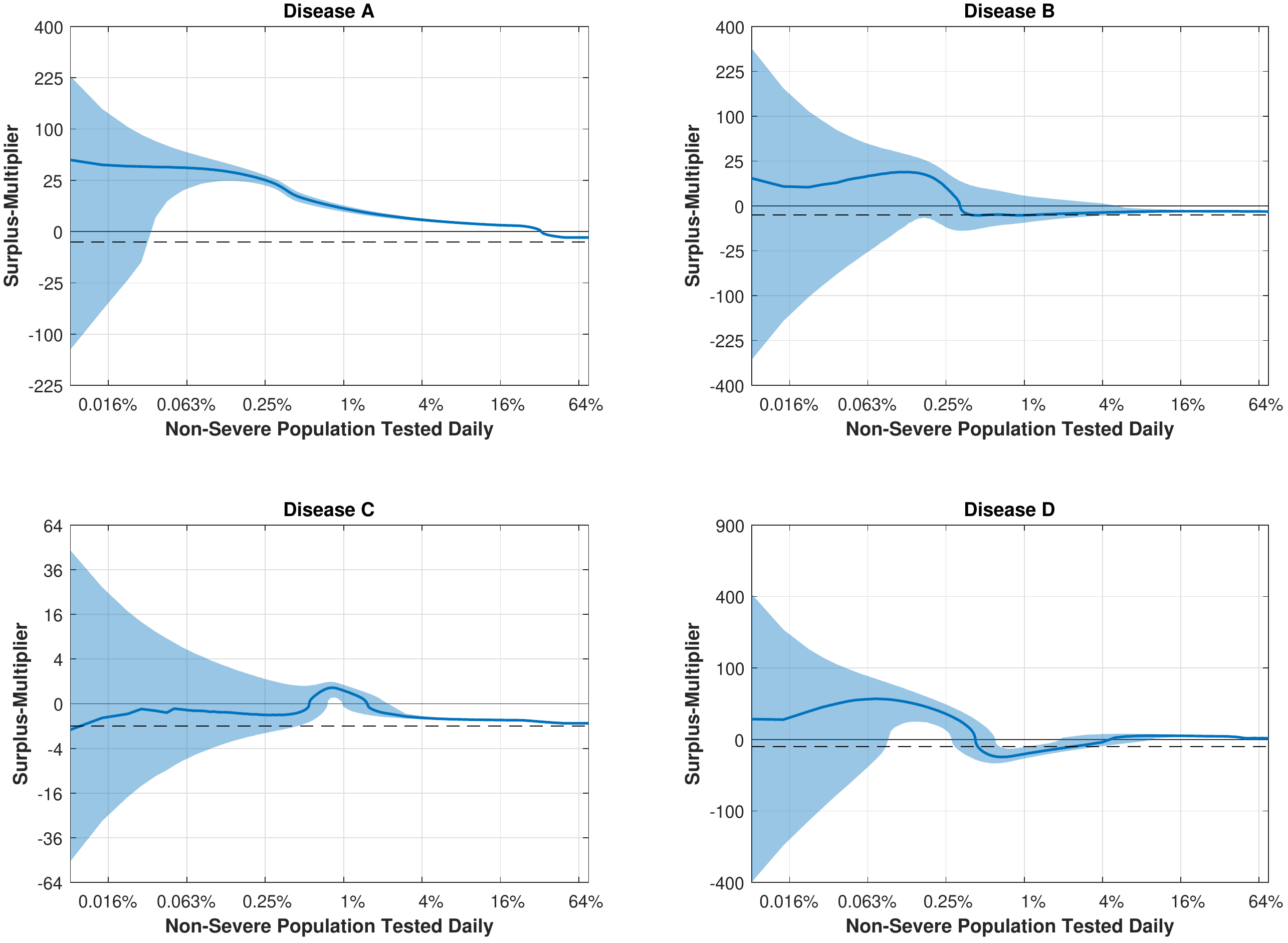}}
\caption{The Surplus-Multiplier for Alternative Diseases}
\label{fig:Appedix_Surplus_Other_Diseases}
\end{figure}

\subsection{Technological Determinants of the Surplus-Multiplier}

\begin{figure}[H]
\centerline{\includegraphics[scale=0.4, angle=0]{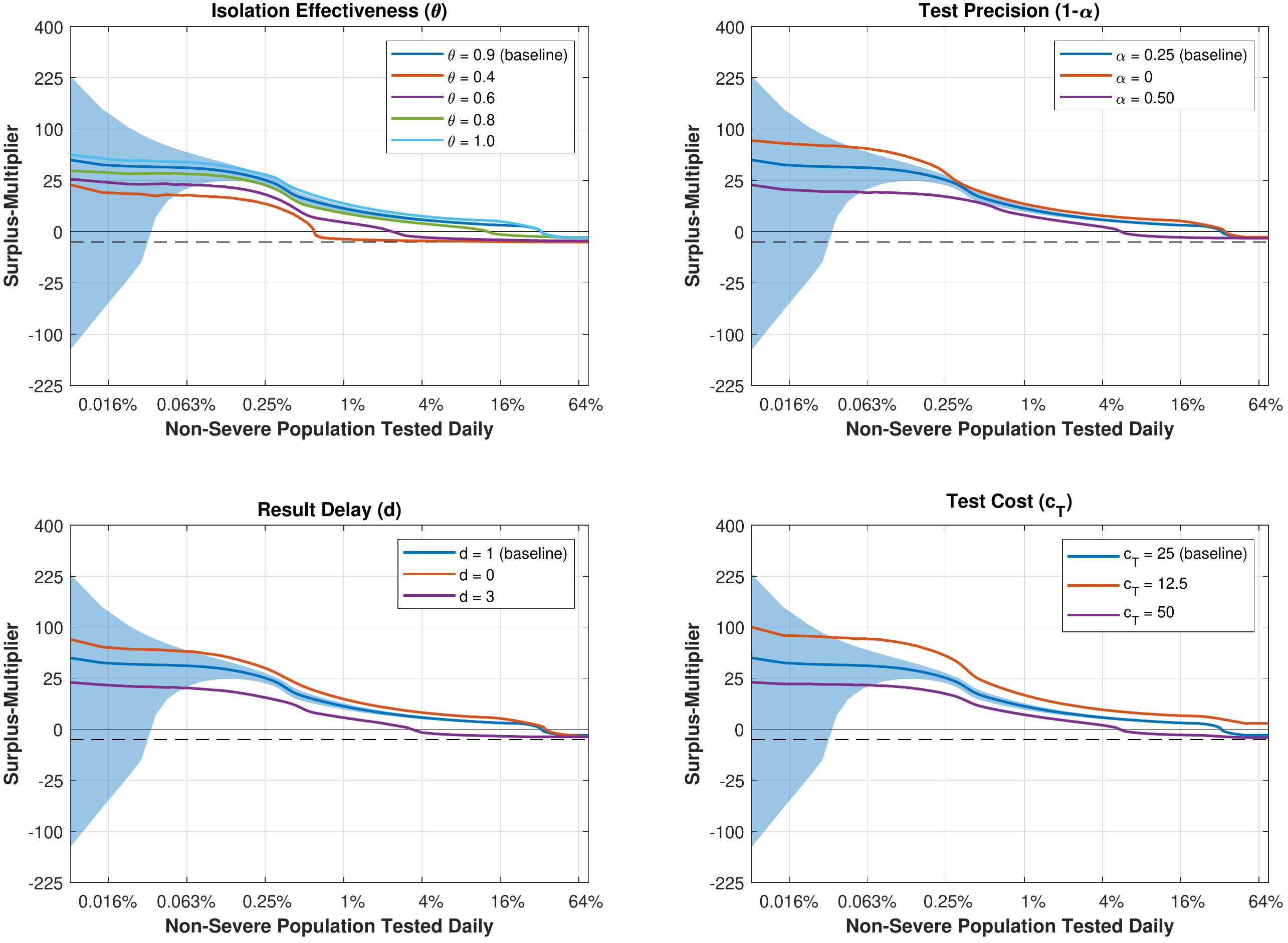}}
\caption{Technological Determinants of the Surplus-Multiplier}
\label{fig:Multiplier_Surplus_Technological_Characteristics}
\end{figure}

\subsection{Testing Multiplier under Alternative Beliefs}

\begin{figure}[H]
\centerline{\includegraphics[scale=0.4, angle=0]{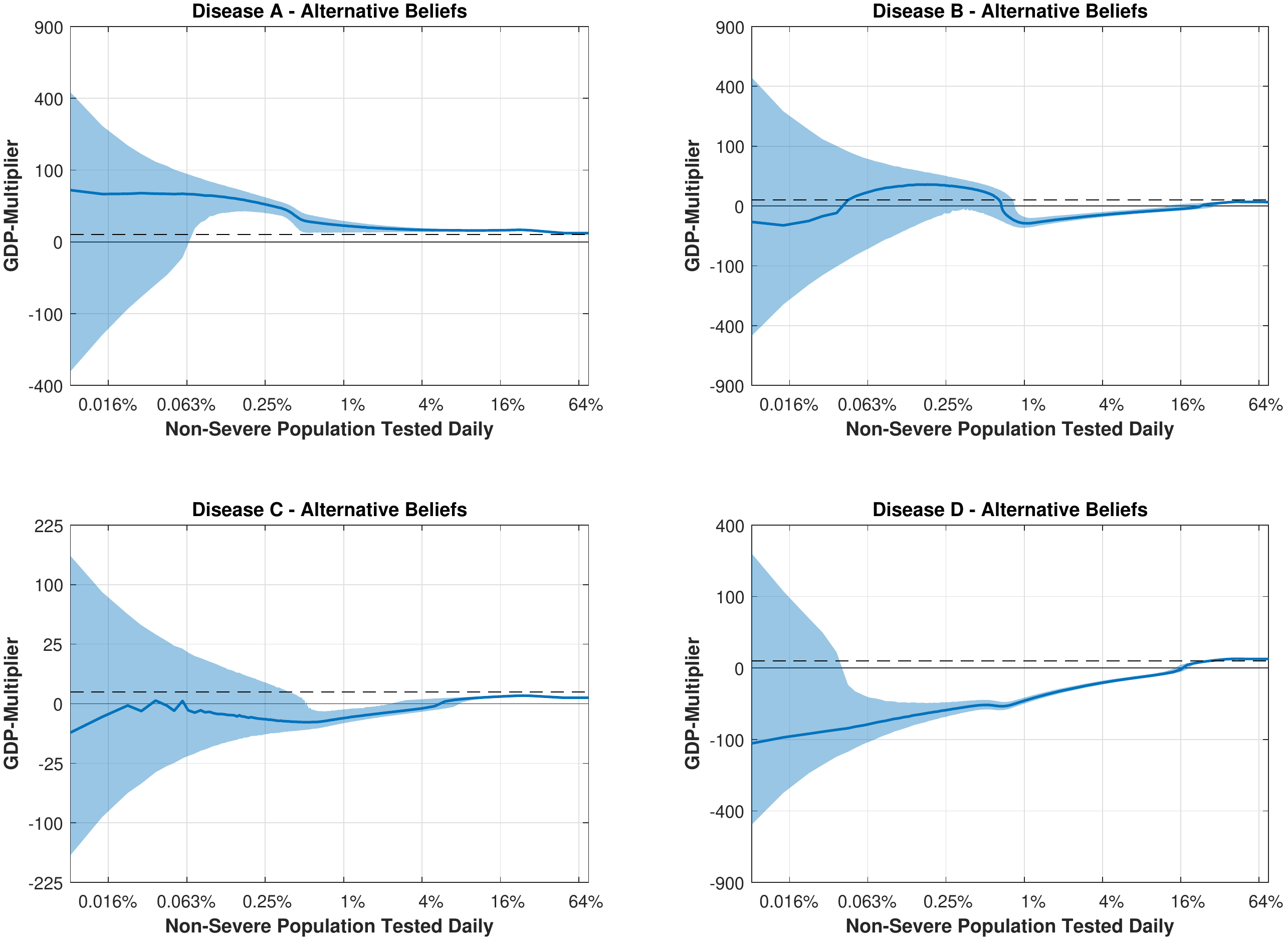}}
\caption{The GDP-Multiplier under Alternative Beliefs}
\label{fig:Multiplier_Alternative_Beliefs}
\end{figure}

\begin{figure}[H]
\centerline{\includegraphics[scale=0.4, angle=0]{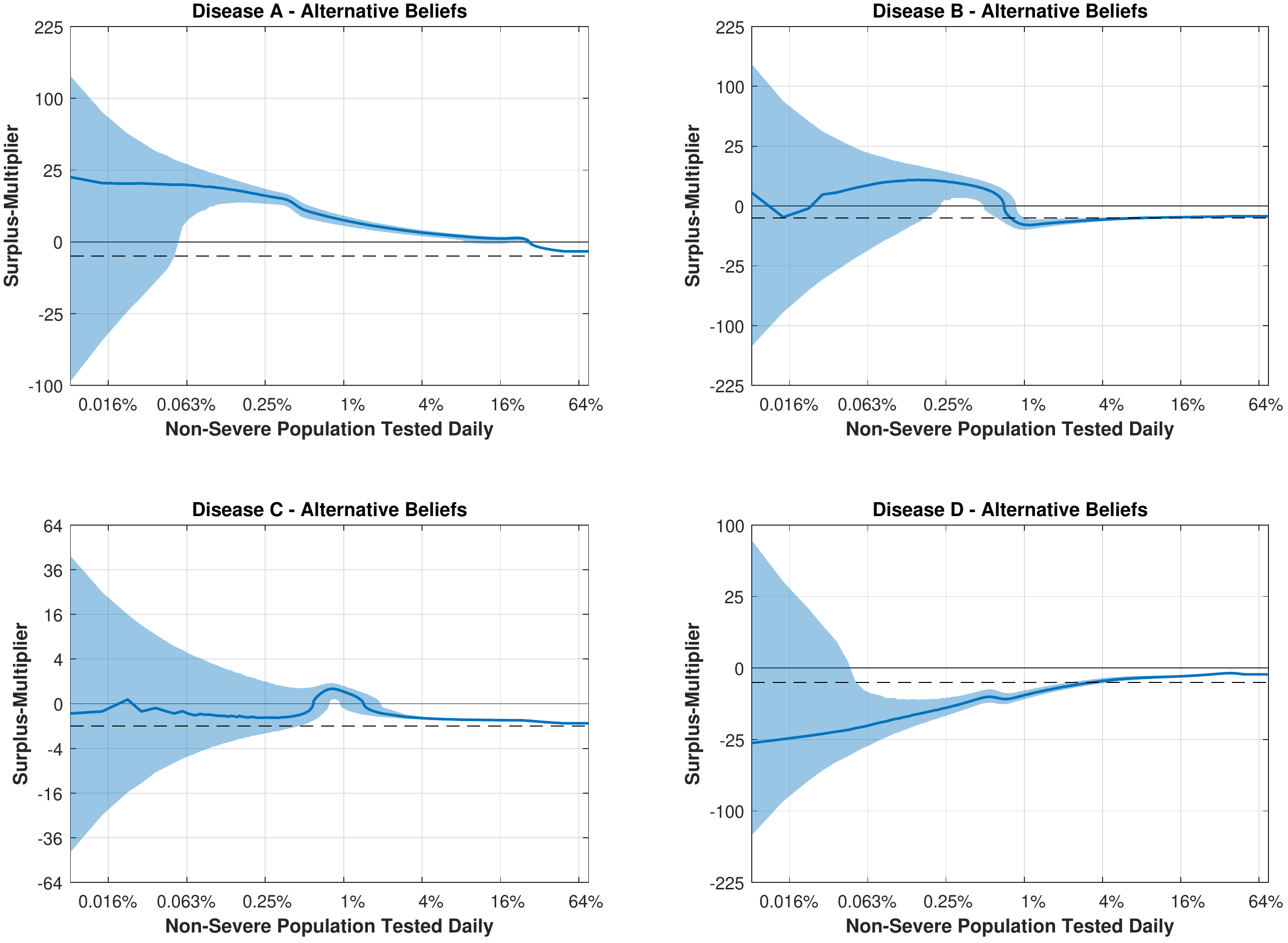}}
\caption{The Surplus-Multiplier under Alternative Beliefs}
\label{fig:Multiplier_Alternative_Beliefs_Surplus}
\end{figure}

\subsection{SARS-CoV-2 Calibration of the Model}

\begin{figure}[H]
\centerline{\includegraphics[scale=0.4, angle=0]{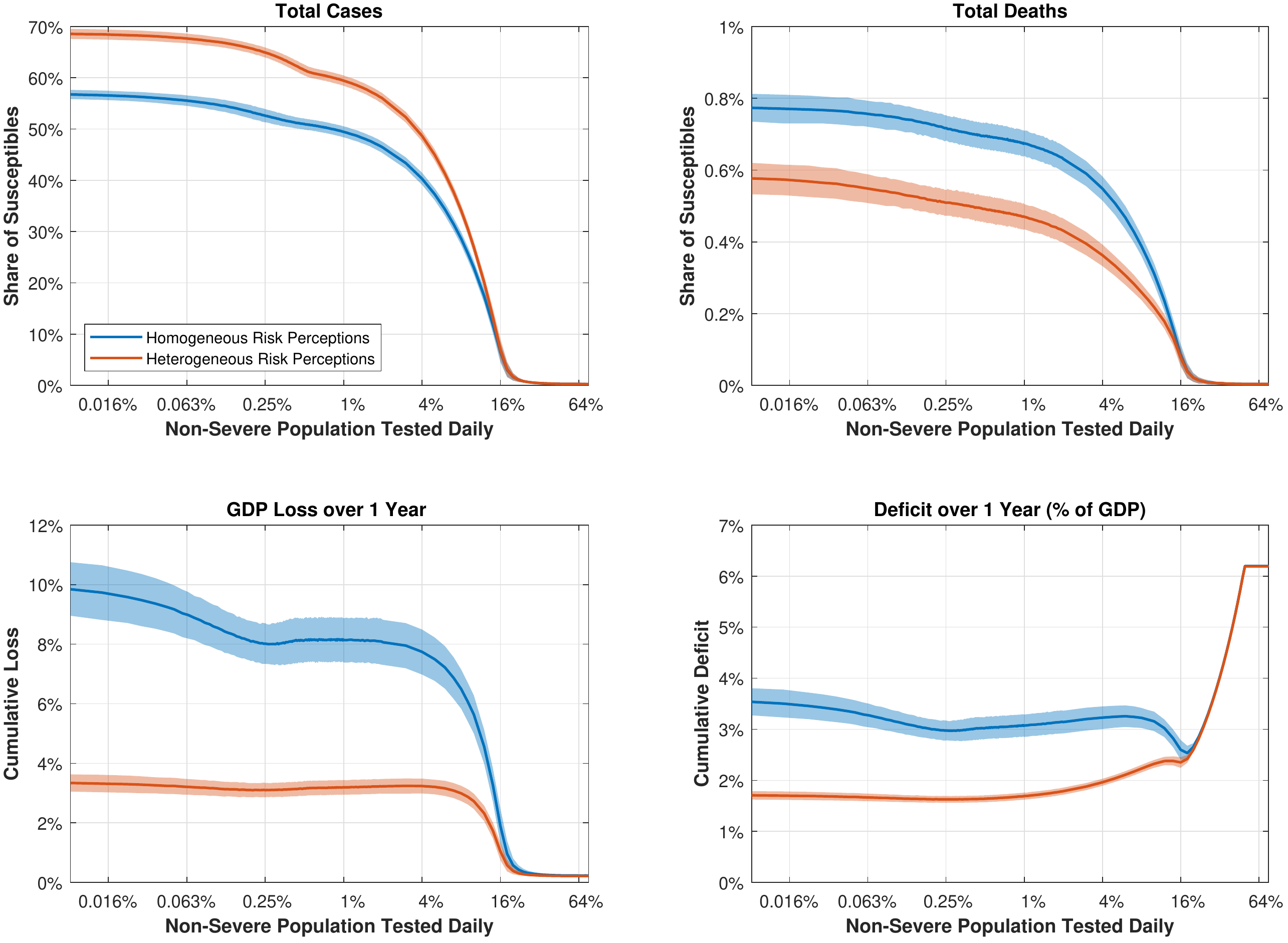}}
\caption{Economic and Health Outcomes for SARS-CoV-2 Across Testing Levels}
\label{fig:Summary_Information_COVID}
\end{figure}

\newpage
\section{A Pseudo-Spanish-SARS-CoV-2 Parameterization}
\label{Appendix:Pseudo_Spanish}

\begin{footnotesize}
The simulations for SARS-CoV-2 suggest that heterogeneous risk perceptions improve both  economic and public health aggregate outcomes, and this happens because high-risk agents  `protect' themselves more and low-risk agents - who contribute the most to economic activity -  `protect' themselves less and return to work. Consider now a disease such that the individuals at high risk are those that contribute the most to economic activity. Interestingly,  the so-called `Spanish Flu' of 1918-1919 is considered to be characterized precisely by this property. \autoref{fig:Spanish_Flu_IFRs} below reports standardized mortality risks across age-groups for the Spanish Flu, as estimated  by  \cite{cilek2018age}.\\

\begin{figure}[H]
\centerline{\includegraphics[scale=0.4, angle=0]{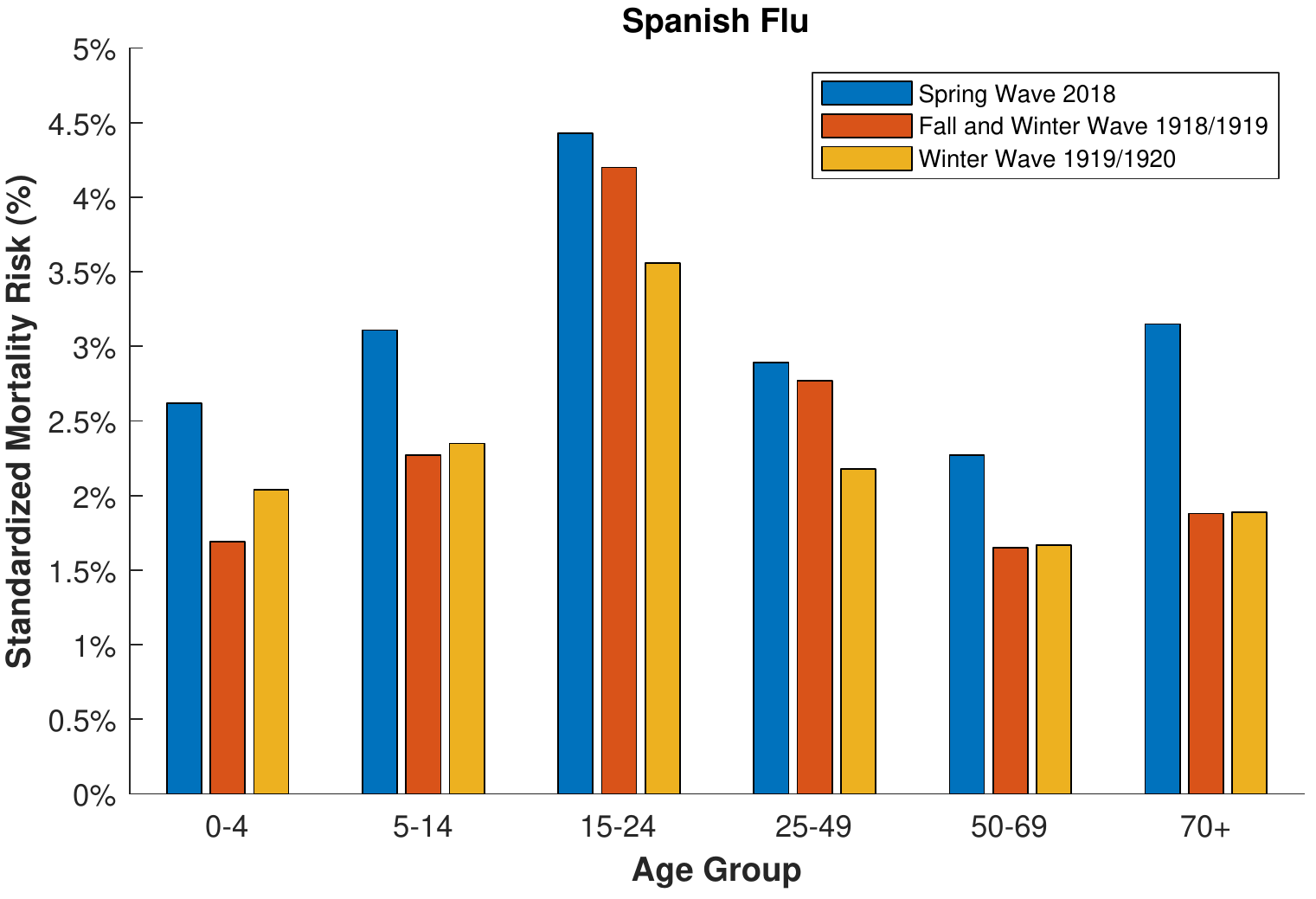}}
\caption{Standardized Mortality Risk Across Age-Groups for the Spanish Flu}
\label{fig:Spanish_Flu_IFRs}
\end{figure}

Given the scarcity of reliable information, calibrating the model to the Spanish Flu would be a daunting task beyond the scope of this paper. I will therefore perform the simplest thought-experiment one could think about: taking the calibration for SARS-CoV-2 and swapping the infection fatality risks across the two groups. More precisely:
\begin{align*}
  \textcolor{NYUcolor}{\phi_s^y} &= 0.248 \\
  \textcolor{NYUcolor}{\phi_s^o} &= 0.005
\end{align*}
All the other parameters stay untouched. \autoref{fig:Pseudo_Spanish} reports economic and health outcomes under this calibration.\\

\begin{figure}[H]
\centerline{\includegraphics[scale=0.4, angle=0]{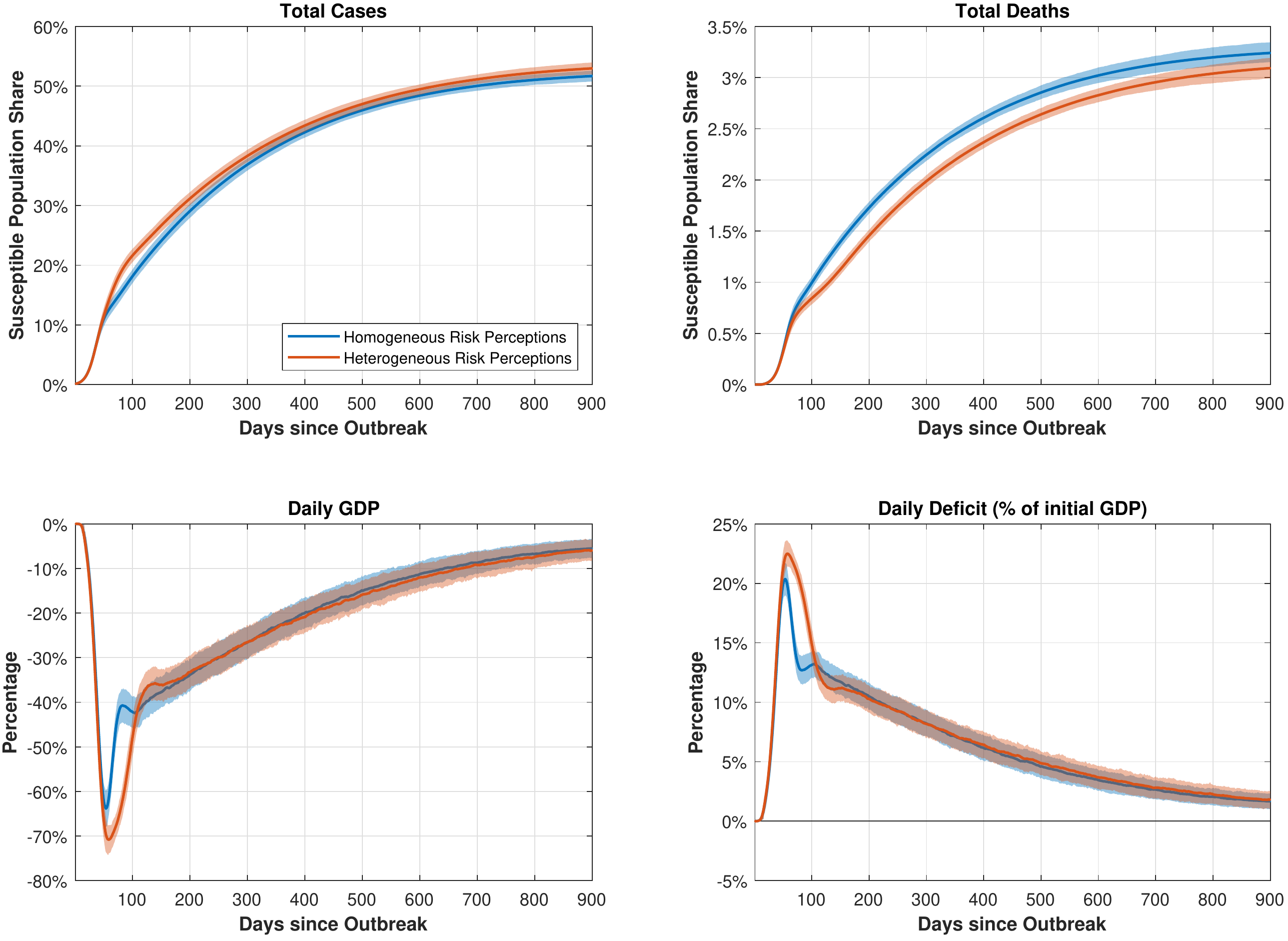}}
\caption{Heterogeneous Risk Perceptions for the Pseudo-Spanish-SARS-CoV-2 Calibration}
\label{fig:Pseudo_Spanish}
\end{figure}

 Heterogeneous risk perceptions still result in less overall deaths, but the economic loss is now slightly higher. This happens because young agents, who generate the vast majority of GDP and are not at high risk, reduce their labor supply more when provided with disaggregated data. This, in turn, produces a larger fall in GDP relative to the scenario in which the government provides aggregate data. \autoref{fig:Appendix_Spanish_Multiplier_Details}  confirms this intuition across all testing levels, whereas \autoref{fig:Appendix_Spanish_Multiplier} displays the testing multiplier.\\

\begin{figure}[H]
\centerline{\includegraphics[scale=0.4, angle=0]{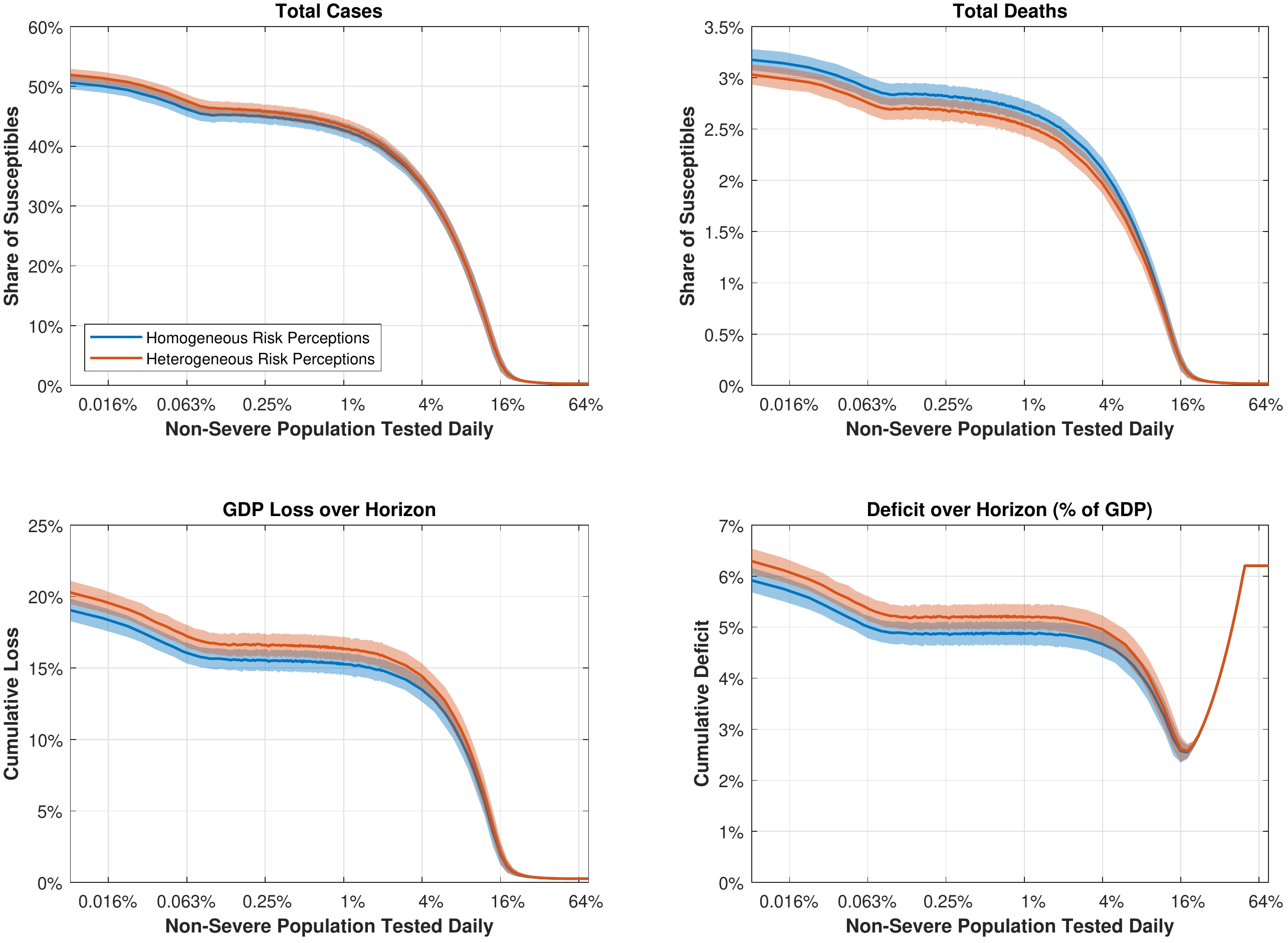}}
\caption{Heterogeneous Risk Perceptions Across Testing Levels for the Pseudo-Spanish-SARS-CoV-2 Calibration}
\label{fig:Appendix_Spanish_Multiplier_Details}
\end{figure}

\begin{figure}[H]
\centerline{\includegraphics[scale=0.4, angle=0]{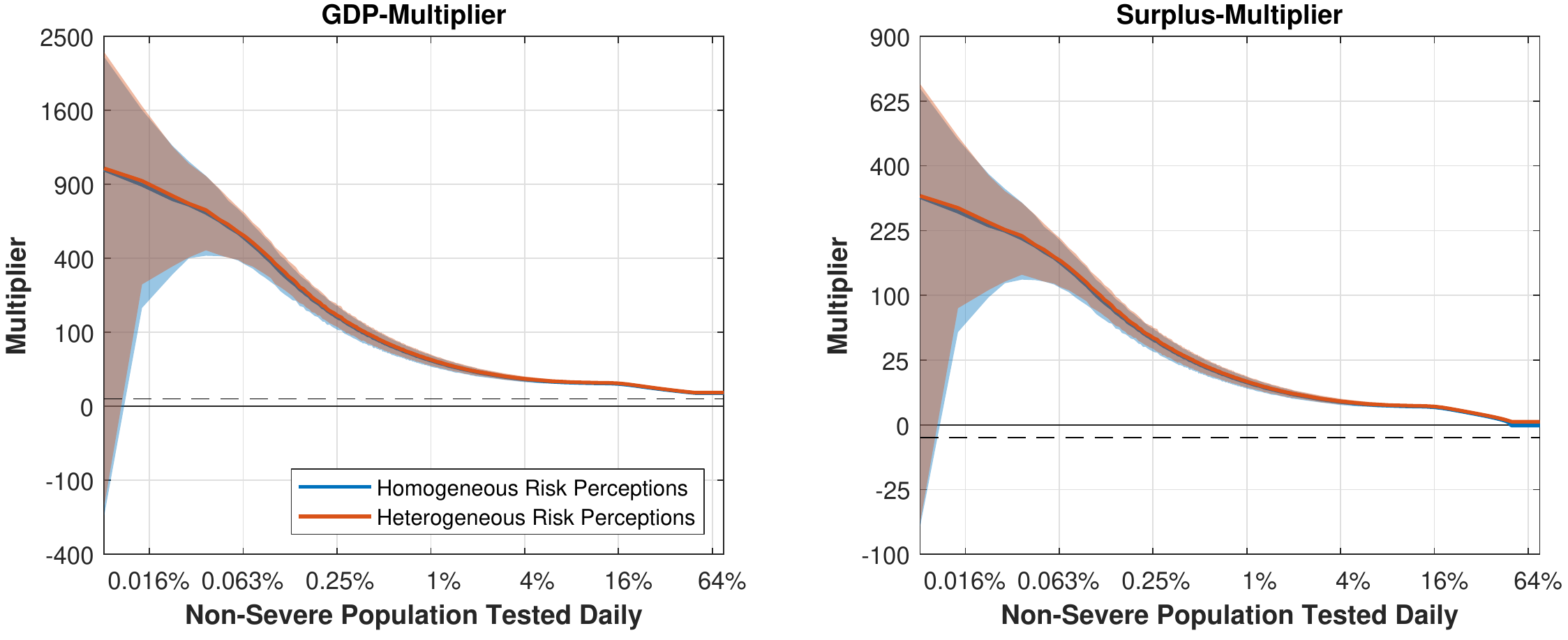}}
\caption{The Testing Multiplier for the Pseudo-Spanish-SARS-CoV-2 Calibration}
\label{fig:Appendix_Spanish_Multiplier}
\end{figure}

It is interesting to notice how powerful the behavioral responses of the young are in slowing down epidemic transmission. Because the young population comprises most of the population, the case fatality rate is very high with both  aggregate and disaggregated testing data. As a result, in an attempt to  `protect' themselves, the young produce a catastrophic collapse of economic activity and sizeably increase the time necessary for the population to acquire herd immunity.\footnote{Because of this, I am forced to increase the time horizon considered from $350$ to $900$.}

\end{footnotesize}

\end{document}